\documentclass{pasj01}

\Received{$\langle$reception date$\rangle$}
\Accepted{$\langle$acception date$\rangle$}
\Published{$\langle$publication date$\rangle$}

\usepackage{framed}
\usepackage{txfonts}
\usepackage{xcolor}
\usepackage{rotating}
\usepackage{xspace}
\usepackage{ulem}
\usepackage{url}
\usepackage{comment}
\usepackage{mathrsfs}
\usepackage{multirow}

\newcommand{\be}{\begin{eqnarray}}
\newcommand{\ee}{\end{eqnarray}}
\newcommand{\dfrac}[2]{\frac{\displaystyle #1}{\displaystyle #2}}

\newcommand{\kny}{\, k_\mathrm{Ny}}
\newcommand{\Lbox}{\, L_\mathrm{box}}
\newcommand{\hMpci}{\, h\,\mathrm{Mpc}^{-1}}
\newcommand{\hiMpc}{\, h^{-1}\,\mathrm{Mpc}}
\newcommand{\hiGpc}{\, h^{-1}\,\mathrm{Gpc}}

\newcommand{\Om}{\Omega_\mathrm{m}}
\newcommand{\om}{\omega_\mathrm{m}}
\newcommand{\Ob}{\Omega_\mathrm{b}}
\newcommand{\ob}{\omega_\mathrm{b}}
\newcommand{\Oc}{\Omega_\mathrm{c}}
\newcommand{\oc}{\omega_\mathrm{c}}
\newcommand{\Ode}{\Omega_\mathrm{de}}
\newcommand{\ode}{\omega_\mathrm{de}}
\newcommand{\Ok}{\Omega_\mathrm{k}}
\newcommand{\ok}{\omega_\mathrm{k}}
\newcommand{\As}{A_\mathrm{s}}
\newcommand{\lnAs}{\ln(10^{10}\As)}
\newcommand{\ns}{n_\mathrm{s}}

\newcommand{\CDM}{\mathrm{CDM}}
\newcommand{\LCDM}{\Lambda\CDM}
\newcommand{\wCDM}{w_0 \CDM}
\newcommand{\wnCDM}{w_0\nu\CDM}
\newcommand{\wwnCDM}{w_0 w_a \nu\CDM}
\newcommand{\wwnoCDM}{w_0 w_a \nu o\CDM}

\newcommand{\DE}{\textsc{DarkEmulator2}\xspace}
\newcommand{\EE}{\textsc{EuclidEmulator2}\xspace}
\newcommand{\BCE}{\textsc{BaccoEmu}\xspace}
\newcommand{\AEMU}{\textsc{Aemulus}$\nu$\xspace}
\newcommand{\MTU}{\textsc{Mira-Titan Univ. IV}\xspace}
\newcommand{\FEMU}{\textsc{FrankenEmu}\xspace}
\newcommand{\CEMU}{\textsc{CosmicEmu}\xspace}
\newcommand{\PA}{\textsc{PkANN}\xspace}
\newcommand{\GOKU}{\textsc{GokuEmu}\xspace}
\newcommand{\GOKUN}{\textsc{GokuNEmu}\xspace}
\newcommand{\ALETHEIA}{\textsc{Aletheia}\xspace}
\newcommand{\CSST}{\textsc{CsstEmulator}\xspace}

\newcommand{\strv}[1]{\textcolor{magenta}{#1}}

\newcommand{\ykrv}[1]{\textcolor{blue}{#1}}

\begin{document}
\title{Dark Quest II: A Wide-Coverage Neural Network Emulator of the Nonlinear Matter Power Spectrum Across Extended Cosmologies}


\author{Satoshi \textsc{Tanaka},\altaffilmark{1,3}
Takahiro \textsc{Nishimichi},\altaffilmark{2,3,4,5} and Yosuke \textsc{Kobayashi}\altaffilmark{2,6}}

\altaffiltext{1}{Graduate School of Social Data Science, Hitotsubashi University, 2-1, Naka, Kunitachi 186-8601, Tokyo, Japan}
\altaffiltext{2}{Department of Astrophysics and Atmospheric Sciences, Faculty of Science, Kyoto Sangyo University, Motoyama, Kamigamo, Kita-ku, Kyoto 603-8555, Japan}
\altaffiltext{3}{Center for Gravitational Physics and Quantum Information, Yukawa Institute for Theoretical Physics, Kyoto University,
Kyoto 606-8502, Japan}
\altaffiltext{4}{Kavli Institute for the Physics and Mathematics of the Universe (WPI),
The University of Tokyo Institutes for Advanced Study (UTIAS),
The University of Tokyo, Kashiwa, Chiba 277-8583, Japan}
\altaffiltext{5}{RIKEN Center for Advanced Intelligence Project, 1-4-1 Nihonbashi, Chuo-ku, Tokyo 103-0027, Japan}
\altaffiltext{6}{Department of Astronomy and Steward Observatory, The University of Arizona}

\email{stanaka.cosmos@gmail.com}



\maketitle

\begin{abstract}
\DE\ is a neural network emulator of the nonlinear matter power spectrum in a nine-dimensional $w_0 w_a \nu o \mathrm{CDM}$ parameter space, developed as the emulator component of the \textsc{Dark Quest II} (DQ2) program.
It is trained on simulations generated with the \textsc{Ginkaku} code, whose numerical implementation, accuracy tests, and post-processing pipeline are described in the companion paper.
The design follows a unified strategy: in addition to the cosmological parameter vector, we supplement the neural network's inputs with three families of physically motivated auxiliary quantities --- the linear matter power spectrum, descriptors of the simulation resolution, and a low-dimensional summary of the initial Gaussian random field --- that are expected to improve generalization across the parameter space.
Training a single network jointly across three simulation resolution tiers allows the emulator to exploit a small number of high-resolution simulations while retaining broad coverage from lower-resolution simulations.
For a $L_{\mathrm{box}}=1\,\hiGpc$ box with $N=3000^{3}$ particles, the emulator
reproduces the simulated matter power spectrum to subpercent accuracy up to
the particle Nyquist scale, $k_{\mathrm{Ny}}\simeq 10\,\hMpci$.
The emulator remains accurate over the calibrated wavenumber range, while its highest-$k$ predictions depend on the simulation resolution and shot noise.
We validate the emulator on independent test suites and, through a cross-comparison with several public emulators and widely used fitting formulas, characterize the inter-model consistency and the parameter-dependent trends in their residuals.
\end{abstract}

\section{Introduction}
\label{sec:intro}

A reliable forward model is essential for cosmological parameter inference from current and forthcoming high-precision large-scale structure surveys such as SDSS, KiDS, DES, HSC-SSP, the Subaru Prime Focus Spectrograph (PFS) survey, DESI, Euclid, and LSST \citep{sdss,de-Jong2015-vr,Abbott2016-uy,2018PASJ...70S...4A,2014PASJ...66R...1T,DESI-Collaboration2022-pa,Euclid-Collaboration2025-pt,Ivezic2019-ux}.
Precision analyses increasingly require cosmological models beyond the minimal
$\LCDM$ framework.
In particular, the total neutrino mass, dark energy dynamics beyond a cosmological constant, and possible deviations from spatial flatness remain observationally relevant.
Recent DESI DR2 baryon acoustic oscillation (BAO) measurements, in combination with cosmic microwave background and type-Ia supernova data, hint at a dynamical dark energy direction with $w_{0}>-1$ and $w_{a}<0$ relative to the cosmological-constant limit~\citep{DESI-DR2}, while ongoing cosmic shear analyses from HSC, KiDS, and DES continue to probe a possible $S_{8}$ tension with cosmic microwave background constraints~\citep{Hikage19,Hamana2020-bh,Dalal2023-HSCY3,Li2023-HSCY3,More2023-HSCY3,KiDS1000:Heymans,Asgari2021-KiDS,Amon2022-DESY3,Secco2022-DESY3,DESY3:3x2pt}.

Distance-based probes such as BAO, type-Ia supernovae, and the CMB primarily constrain the background expansion history and the geometry of the universe. Once $\LCDM$ is extended with additional degrees of freedom---for example dynamical dark energy or non-zero spatial curvature---model classes that are similarly favored by these distance-based data can nevertheless predict visibly different nonlinear matter clustering. Clustering and weak-lensing observables therefore carry information that is complementary to distance-based probes, and can help separate model classes that remain degenerate at the level of geometric measurements. Exploiting this complementarity requires forward models that accurately cover an extended cosmological parameter space, including dynamical dark energy, non-zero curvature, and massive neutrinos simultaneously, rather than being restricted to the minimal six-parameter $\LCDM$ model.

Although one could in principle combine high-resolution $N$-body simulations with Bayesian parameter inference, often implemented with Markov chain Monte Carlo (MCMC) methods~\citep{Lewis2002-zi,Trotta2008-qt}, to obtain accurate constraints from nonlinear observables, directly embedding such costly simulations in inference pipelines is computationally infeasible.
A single high-resolution run remains too expensive for repeated evaluation within such pipelines.
Analytical approximations, including linear theory, perturbative treatments, and fitting formulas, are computationally much cheaper, but they do not maintain uniformly high accuracy over the full range of cosmologies and scales relevant for current survey analyses.

Simulation-based emulators provide a practical alternative by learning fast predictions from suites of cosmological simulations.
The Cosmic Calibration and Coyote programs have established the feasibility of accurate emulation for the nonlinear matter power spectrum \citep{Heitmann06,Habib07,Schneider08,Coyote1,Coyote2,Coyote3}. 
PkANN provided an early neural network (NN)-based extension of this line of work~\citep{PkANN1,PkANN2}.
Later public emulators, including the EuclidEmulator series~\citep{EuclidEmu1,EuclidEmu2}, the Mira-Titan series~\citep{MiraTitan1,MiraTitan2,MiraTitan4}, BACCO~\citep{BACCO}, and Aemulus-$\nu$~\citep{Aemulus1,Aemulusnu}, have expanded the accessible cosmological domain and refined emulator design through larger simulation suites, updated target parameterizations, and more flexible surrogate models. 
More recent approaches, such as GokuN~\citep{Goku,GokuN} and Aletheia~\citep{Aletheia}, represent a further diversification of emulator methodology through multi-fidelity designs, alternative parameterizations, and updated regression strategies. 
These developments reflect a broader progression from early proof-of-concept emulators toward higher-dimensional, more accurate, and more computationally efficient nonlinear matter power spectrum emulation.

Despite this progress, recent emulators still disagree at the percent level over a non-trivial range of scales and cosmologies. The supported cosmological parameter ranges also differ substantially between emulators, with most public emulators occupying only the central region of a broader physically motivated domain rather than reaching the parameter directions probed by current dark energy and curvature analyses. Different emulators further adopt different strategies for mitigating cosmic variance in their training data, including paired-and-fixed initial conditions~\citep{Angulo2016-FP,Pontzen2016-FP}, control-variate techniques~\citep{Chartier2021-uu,Kokron2022-mh,DeRose2023-wn}, and hierarchical multi-fidelity designs that combine separate models trained at distinct resolutions~\citep{Ho2021-nu,Ho2023-ls,GokuN}. These differences in coverage, cosmic variance treatment, and modeling strategy translate into emulator-dependent systematics that motivate both a fresh look at design choices and a direct cross-comparison among existing emulators.

Although the nonlinear matter power spectrum is not itself a directly observed quantity, it remains a fundamental building block for a wide range of cosmological predictions.
In particular, it enters theoretical predictions for weak lensing and related projected statistics, while also serving as a survey-agnostic characterization of nonlinear gravitational clustering. 
For this reason, emulating the nonlinear matter power spectrum remains valuable even as recent emulator efforts increasingly target observables that are more directly connected to survey measurements. 
In parallel, other emulator programs have been developed for halo abundances, halo bias, and galaxy clustering observables~\citep{MiraTitan3,Aemulus2,Aemulus3,Aemulus4,Aemulus5,Yuan2022-op,2020MNRAS.492.2872W,CSST-emu-II,CSST-emu-III}; see also~\citet{Moriwaki2023-sy}.
In particular, \textsc{DarkEmulator}, based on the \textsc{Dark Quest I} (DQ1) simulation suite, was developed primarily for halo and galaxy applications rather than for direct emulation of the nonlinear matter power spectrum~\citep{DQI,2020PhRvD.102f3504K,Cuesta-Lazaro2023-pq}.

Within the \textsc{Dark Quest} series, the first emulator in the DQ1 project was based on Gaussian processes (GP), whereas later extensions adopted neural networks for additional observables~\citep{DQI,2020PhRvD.102f3504K,Cuesta-Lazaro2023-pq}.
Motivated by these later results and by the larger, higher-dimensional training set considered here, we adopt a neural network approach for DQ2.
In this paper, we present \DE, a neural network emulator for the nonlinear matter power spectrum in a nine-dimensional $\wwnoCDM$ cosmological parameter space whose support is designed to cover, in a single emulator, the parameter directions probed by current dark energy and curvature analyses. The companion paper describes \textsc{Ginkaku}, including the $N$-body solver, numerical accuracy tests, and post-processing pipeline used to generate the DQ2 simulations analyzed here~\citep{Ginkaku26}; the present paper focuses on the DQ2 simulation design, emulator construction, and validation of \DE.

Emulating the nonlinear matter power spectrum amounts to a high-dimensional-to-high-dimensional regression that maps a nine-dimensional cosmological parameter vector to a continuous function of wavenumber and redshift, sampled with the limited training budget afforded by full $N$-body simulations. How efficiently this mapping can be learned, and how well it generalizes beyond the densely sampled core of the parameter design, depend critically on the way the inputs to the regressor are organized and on how the underlying simulation data are exposed to it. We exploit the flexibility of a feed-forward neural network to address this problem in a single coherent architecture, in which the cosmological parameter vector is supplemented by three families of physically motivated auxiliary inputs, each targeting a distinct source of residual structure in the mapping.

First, rather than learning a multiplicative boost factor relative to the linear matter power spectrum, we supply the sampled linear power spectrum as an explicit input feature to the network, so that scale-dependent information not captured by the parameter vector alone is propagated into the regression target. Second, instead of training a separate emulator at each simulation resolution and combining them with an external correction, as in hierarchical multi-fidelity designs~\citep{Ho2021-nu,Ho2023-ls,GokuN}, we treat the simulation resolution as an additional input feature of a single network, so that low-, middle-, and high-resolution $N$-body simulations are jointly used to constrain one mapping. Third, rather than relying on paired-and-fixed initial conditions or external control-variate schemes to suppress cosmic variance in the training data~\citep{Angulo2016-FP,Pontzen2016-FP,Chartier2021-uu,Kokron2022-mh,DeRose2023-wn}, we expose the realization dependence directly to the network through a low-dimensional summary of the initial Gaussian random field, enabling cosmic-variance-aware predictions at evaluation time.

To assess the emulator beyond the construction stage, we cross-compare \DE\ with widely used fitting formulas and with a representative set of public matter power spectrum emulators on common parameter volumes, and we propagate the predictions to the cosmic-shear convergence power spectrum through a Limber-type line-of-sight integration.
The wide cosmological coverage of \DE\ further enables side-by-side predictions of the nonlinear matter power spectrum and the convergence power spectrum for extended-cosmology benchmarks motivated by recent survey results. As a concrete example, we take posterior-mean cosmologies for several extended-$\LCDM$ model classes reported in the DESI DR2 cosmological analysis~\citep{DESI-DR2}. Because these cosmologies are inferred primarily from background-expansion and geometric probes (BAO combined with CMB, and in some cases with type-Ia supernova data), they are not strongly differentiated by such distance-based measurements, while their predicted clustering and weak-lensing signals can nevertheless differ visibly. This is presented as an illustration of the emulator's reach across model classes suggested by current survey data, rather than as a parameter-inference analysis.
Although the present implementation focuses on the matter power spectrum, the same training strategy can be extended directly to other summary statistics, including halo and galaxy clustering and their cross-correlations.

The paper is organized as follows. 
Section~\ref{sec:data} describes the numerical setup, the cosmological parameter design, the multi-resolution simulation suite, the power spectrum measurement procedure, and the preprocessing applied to the training targets. 
Section~\ref{sec:emulation} presents the emulator framework, including the network architecture, the training and hyperparameter-tuning procedures, the independent test suites, and the accuracy metrics.
Section~\ref{sec:design} develops the three design choices that distinguish the present implementation from existing public emulators: using the linear power spectrum as an explicit input feature, jointly training on simulations at multiple resolutions, and conditioning on a reduced description of the initial Gaussian random field to enable cosmic-variance-aware predictions.
Section~\ref{sec:apps} compares our emulator with public emulators and widely used fitting formulas for the nonlinear matter power spectrum, and then demonstrates applications to lensing convergence observables and extended cosmologies.
Section~\ref{sec:est_time} compares the estimation time of \DE\ with those of public emulators and fitting formulas.
Section~\ref{sec:summary} summarizes the main results and concludes the paper.
Appendices~\ref{append:lin_pk_emu}--\ref{append:cmaes} describe the auxiliary amplitude and linear power spectrum emulators, the conversion from $P_{\mathrm{cb}}$ to $P_{\mathrm{tot}}$, the sensitivity diagnostics of the trained emulator, the behavior outside the calibrated domain, and the CMA-ES hyperparameter optimization procedure.

\section{Simulation suite and datasets}
\label{sec:data}

This section describes the simulation suite and measurement pipeline used to
train and assess the emulator.

\subsection{Numerical setup}
\label{subsec:setup}

We perform cosmological $N$-body simulations with the TreePM code \textsc{Ginkaku}~\citep{Ginkaku26},
which is designed for large ensembles on modern MPI/OpenMP architectures. \textsc{Ginkaku}
is built on \textsc{Fdps}\footnote{\url{https://github.com/FDPS/FDPS}}~\citep{2016PASJ...68...54I,2018PASJ...70...70N},
which provides domain decomposition, load balancing, and tree construction. 
Long-range forces are evaluated with the \textsc{GreeM} particle-mesh (PM) solver~\citep{Yoshikawa_2005,2009PASJ...61.1319I,Ishiyama2012-ey,Ishiyama2022-uf}.
Short-range particle-particle (PP) forces are computed using \textsc{Phantom-GRAPE}\footnote{\url{https://bitbucket.org/kohji/phantom-grape/}}
kernels~\citep{Nitadori2006-ek,2012NewA...17...82T,2013NewA...19...74T}, which
employ SIMD vectorization and software pipelining.
Time integration is performed with a kick-drift-kick (KDK) leapfrog scheme.
We adopt the ``fiducial'' accuracy setting of \textsc{Ginkaku} described in the companion paper~\citep{Ginkaku26}, which corresponds to a Plummer softening length $\epsilon_\mathrm{p} = 2.14\%$ of the mean inter-particle spacing, time-step accuracy parameters $\eta_\mathrm{tree} = 0.70$ and $\eta_\mathrm{PM} = 0.125$ with a maximum step $\Delta\tau_\mathrm{max} = 0.03$, tree opening angle $\theta = 0.5$, and a PM grid with $N_\mathrm{PM}^{1/3} = 2\,N_\mathrm{p}^{1/3}$ (i.e., $\epsilon_\mathrm{g} = r_\mathrm{s}/H_\mathrm{g} = 3$). The accuracy of this setting is demonstrated in~\citet{Ginkaku26} to control the matter power spectrum to within $\sim 1\%$ up to the particle Nyquist wavenumber at $z = 0$.

Initial conditions are generated using second-order Lagrangian perturbation theory (2LPT) with transfer functions from \textsc{Class}~\citep{class1,class2} in the $N$-body gauge~\citep{2015PhRvD..92l3517F}, starting at an initial redshift $z_{\mathrm{ini}}$.
The value of $z_{\mathrm{ini}}$ is chosen separately for each cosmology by requiring that the one-dimensional rms Zel'dovich displacement remains smaller than $0.25$ times the mean inter-particle spacing.
For each model, we adopt the lowest redshift $z_{\mathrm{ini}}$ that satisfies this condition.
For the fiducial cosmology in the $L_\mathrm{box} = 1\,\hiGpc$ box, the resulting starting redshifts are approximately $z_\mathrm{ini} \simeq 30$, $70$, and $100$ for the LR, MR, and HR resolutions, respectively, and they vary across the parameter space according to the cosmological growth history and the linear amplitude. For larger boxes such as $L_\mathrm{box} = 2\,\hiGpc$, the mean inter-particle spacing doubles, so $z_\mathrm{ini}$ is correspondingly lower (during this epoch, the Universe is close to Einstein--de Sitter, so doubling the displacement at fixed mass resolution roughly halves $1 + z_\mathrm{ini}$).

We evolve CDM and baryons as a single collisionless component (``cb'').
Massive neutrinos, radiation, and dark energy perturbations are incorporated via the linear response method~\citep{2013MNRAS.428.3375A}, ensuring background and linear-level consistency with \textsc{Class}. In practice, this is implemented by adding an external source term $S_\mathrm{ext}$ to the Poisson equation solved on the PM grid, where $S_\mathrm{ext}$ encodes the linear contribution of the non-cb species evaluated from the \textsc{Class} transfer functions in the $N$-body gauge. The cb particles thus respond to gravity from their own nonlinear clustering and to the linear-level potential generated by neutrinos, radiation, and dark energy perturbations; see~\citet{Ginkaku26} for details. 
When total matter spectra are required, we map $P_{\mathrm{cb}}$ to $P_{\mathrm{tot}}$ using the linear theory total-to-cb ratio $\mathcal{R}^{\mathrm{lin}}_{\mathrm{tot/cb}}(k,z)$ computed with \textsc{Class}. Appendix~\ref{append:cb_to_tot} provides the definition and validation of this mapping.

\subsection{Parameter space and sampling design}
\label{subsec:sampling}

Our fiducial cosmology follows the \textit{Planck}~2015 results~\citep{planck-collaboration:2015fj}.
The emulator targets a nine-parameter $\wwnoCDM$ space that extends the six-parameter $\wCDM$ set of DQ1. Here the label ``$o$'' indicates that we allow nonzero spatial curvature ($\Ok\neq 0$), encompassing both open and closed geometries.
Compared to DQ1, the parameter domain is widened to cover the $\Om$-$\sigma
_{8}$ degeneracy region suggested by the HSC-Y1 analysis~(\cite{Hamana2020-bh}; $3\sigma$ region).
We take the nine independent parameters to be 
\be 
\boldsymbol\theta=(\Om,\ \ob,\ \sigma_8,\ n_s,\ h,\ M_\nu,\ w_0,\ w_a,\ \Ok), 
\ee 
with $\ob\equiv \Ob h^{2}$. 
The dependent quantities are $\oc=\Oc h^{2}$, $\Ode =1-\Om-\Ok$, the primordial amplitude $A_{s}$ (or $\ln(10^{10}A_{s})$), and $S_{8}=\sigma_{8}\sqrt{\Om/0.3}$. Table~\ref{table:params} summarizes the bounds and fiducial values at $z=0$. 
We fix the pivot scale to $k_{\mathrm{pivot}}=0.05 \, \mathrm{Mpc}^{-1}$ and the dark energy sound speed to $c_{s}^{2}=1$. 
Radiation is not listed among the inputs but is included consistently via the linear response treatment in the simulations. For neutrinos, we adopt the equal mass three-species approximation; $M_{\nu}$ denotes the total mass.

Given a parameter vector $\boldsymbol{\theta}$, we denote the matter power spectrum by $P(k,z;\boldsymbol{\theta})$, but for brevity we henceforth suppress the explicit $\boldsymbol{\theta}$-dependence and simply write $P(k,z)$.

\begin{table}[htbp]
  \caption{Parameter ranges and fiducial values at $z=0$ (fiducial values follow
  \textit{Planck}~2015~\citep{planck-collaboration:2015fj}). Independent parameters
  are $(\Om,\ob,\sigma_{8},n_{s},h,M_{\nu},w_{0},w_{a},\Ok)$; dependent ones
  are $(\oc,\Ode,\ln(10^{10}A_{s}),S_{8})$.}
  \vspace{0.3em}
  \label{table:params}
  \centering
  \begin{tabular}{ccc}
    \hline
    \hline
    Parameter           & Range            & Fiducial value \\
    \hline
    $\Om$               & [0.05, 0.62]     & 0.3156         \\
    $\ob$               & [0.015, 0.03]    & 0.02225        \\
    $\sigma_{8}$        & [0.47, 1.23]     & 0.831          \\
    $n_{s}$             & [0.916, 1.012]   & 0.9645         \\
    $h$                 & [0.5, 0.9]       & 0.674          \\
    $M_{\nu}$ [eV]      & [0.0, 0.5]       & 0.06           \\
    $w_{0}$             & [-1.5, -0.5]     & -1             \\
    $w_{a}$             & [-0.5, 0.5]      & 0              \\
    $\Ok$               & [-0.1, 0.1]      & 0              \\
    \hline
    $\oc$               & [0.01, 0.30]     & 0.1198         \\
    $\Ode$              & [0.4125, 0.9385] & 0.6843         \\
    $\ln(10^{10}A_{s})$ & [1.0412, 5.7548] & 3.0911         \\
    $S_{8}$             & [0.60, 0.95]     & 0.85233        \\
    \hline
    \hline
  \end{tabular}
\end{table}

We construct the DQ2 cosmology design with weighted sequential designs (WSDs), rather than a Latin hypercube design, and apply a cut based on a linear spectrum distance metric.
Specifically, we define a linear spectrum distance from the fiducial cosmology at $z=0$ by
\be
D=\frac{1}{N_{k}}\sum_{i=1}^{N_k}
\frac{\big|P_{\mathrm{lin,fid}}(k_{i})-P_{\mathrm{lin}}(k_{i})\big|}{\min\!\left[P_{\mathrm{lin,fid}}(k_{i}),\,P_{\mathrm{lin}}(k_{i})\right]}\Bigg|_{z=0},
\label{eq:p_dist}
\ee
and retain draws with $D<5$.
We adopted this threshold after confirming that the retained domain still broadly covers the region motivated by the HSC-Y1 cosmic shear constraints. As a one-dimensional reference, the adopted ranges correspond to approximately $-2.9\sigma$ and $+5.8\sigma$ in $\Om$, $-3.3\sigma$ and $+3.8\sigma$ in $\sigma_8$, relative to the corrected HSC-Y1 flat $\LCDM$ marginalized 68\% constraints. The threshold therefore removes models with extreme linear power spectrum shapes while preserving the intended broad coverage.

WSDs proceed as follows:
\begin{enumerate}
  \item We first define a curved support region in the $\Om$--$\sigma_8$ plane using lower and upper spline boundaries motivated by the HSC-Y1 cosmic shear constraints. The sampling therefore does not assume a single linear degeneracy slope. Within this support, $\Om$ is drawn from a Gaussian distribution centered on the fiducial value, $\Om=0.3156$, with standard
  deviation 0.13.
  At fixed $\Om$, $\sigma_8$ is drawn from a Gaussian distribution with standard deviation 0.09 and mean $0.52[\sigma_{8,\mathrm{lower}}(\Om)+\sigma_{8,\mathrm{upper}}(\Om)]$.
  Samples outside the lower and upper spline boundaries are rejected. This procedure follows the curved $\Om$--$\sigma_8$ degeneracy while concentrating samples around the fiducial cosmology.

  \item $(\oc,h,M_{\nu})$ are drawn uniformly within bounds; $\ob$ is then
    inferred to satisfy 
    \be 
    \Om=\frac{\oc}{h^{2}}+\frac{\ob}{h^{2}}+\frac{M_{\nu}}{\alpha h^{2}}, 
    \ee 
    where $\alpha\simeq 93.14\,\mathrm{eV}$ is the standard conversion factor defined by $\omega_\nu \equiv \Omega_\nu h^2 = M_\nu/\alpha$; in the actual calculation, we use the exact conversion consistent with the adopted background cosmology.

  \item $n_{s}$ is drawn uniformly within its bounds.

  \item $\Ok$ is drawn uniformly, and we set $\Ode=1-\Om-\Ok$.

  \item For the dark energy equation of state, we use the form
    $w(a)=w_0+w_a(1-a)$. We draw the two quantities $w_0$ and
    $w_0+w_a$ uniformly from $[-1.5,-0.5]$, and retain only samples
    satisfying $-0.5<w_a<0.5$. This construction keeps $w(a)$ within
    $[-1.5,-0.5]$ over $a\in[0,1]$.

  \item Finally, $A_{s}$ and $P_{\mathrm{lin}}(k,z)$ are computed with \textsc{Class};
    the draw is accepted if $D<5$.
\end{enumerate}

\begin{figure*}[htbp]
  \centering
  \includegraphics[width=0.8\linewidth]{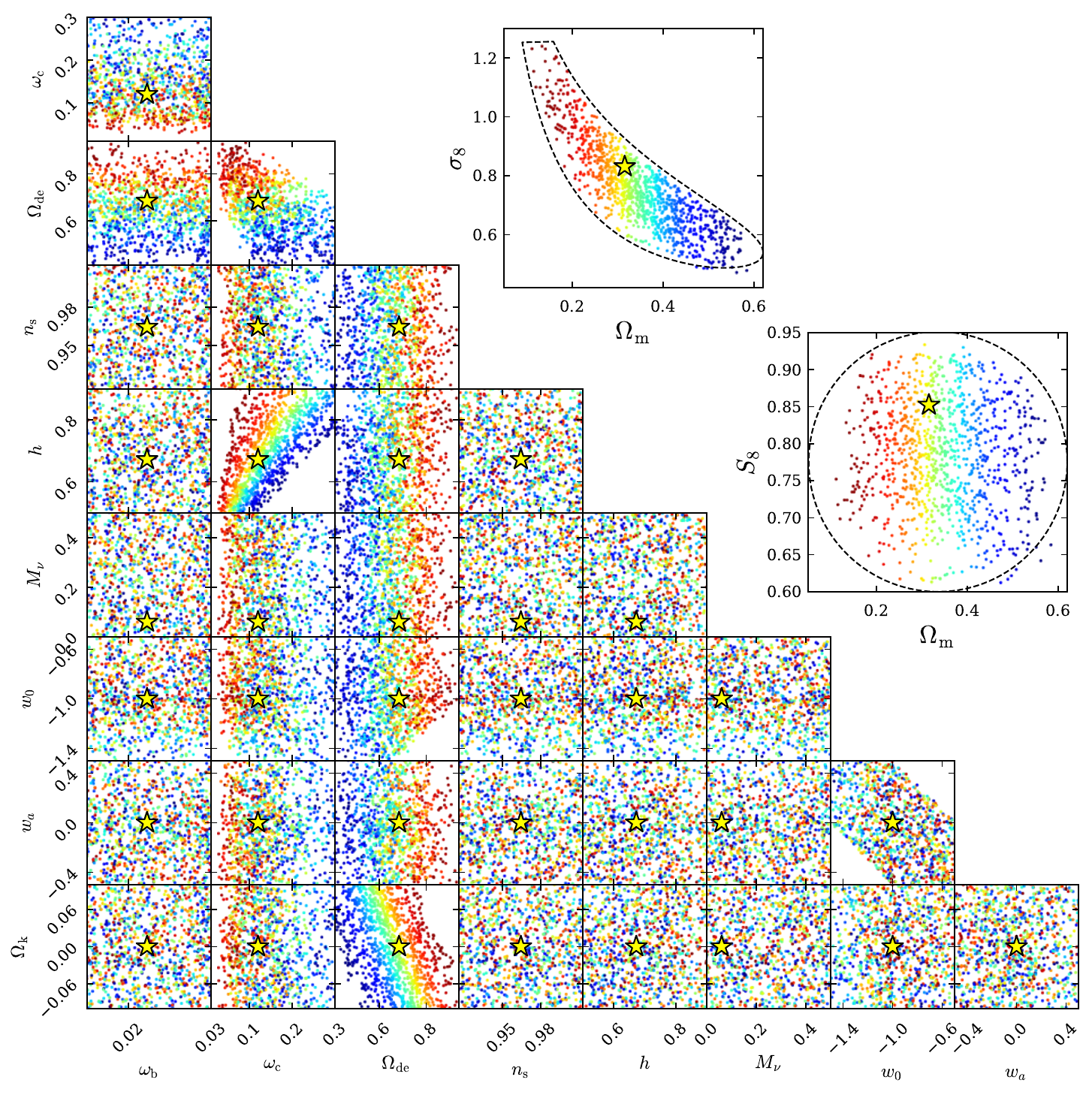}
  \caption{Distribution of 1,000 cosmologies in the nine-dimensional parameter
  space. Colors encode $\Om$; the star marks the fiducial model. The independent
  set is $(\Om,\ob,\sigma_{8}, \ns, h, M_{\nu},w_{0},w_{a},\Ok)$; dependent
  quantities $(\oc,\Ode,S_{8})$ are shown for reference.
  {Alt text: Corner plot with one-dimensional parameter distributions on the diagonal and pairwise projections off the diagonal.}
  }
  \label{fig:9d_params}
\end{figure*}

\begin{figure*}[htbp]
  \centering
  \includegraphics[width=\linewidth]{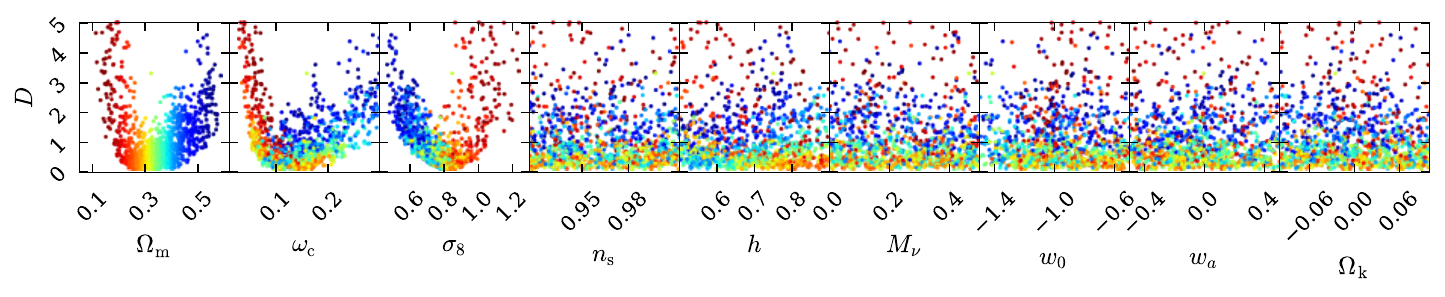}
  \caption{Same sample as figure~\ref{fig:9d_params}, colored by the linear
  spectrum distance $D$ defined in equation~\ref{eq:p_dist}.
  {Alt text: Row of scatter panels comparing the distance metric with individual cosmological parameters. The strongest visible trends occur for matter density, cold dark matter density, and sigma eight.}
  }
  \label{fig:9d_params_D}
\end{figure*}

Figure~\ref{fig:9d_params} displays the 1,000 cosmologies sampled with WSDs in the nine-dimensional space.
The WSD sampling density is a design distribution for emulator training, not an inference prior.
In parameter inference, users may adopt any science-motivated prior, but samples must remain within the emulator domain defined by Table~\ref{table:params}, the non-rectangular $\Om$--$\sigma_8$ support in Fig.~\ref{fig:9d_params}, and the cut $D<5$. Samples inside the rectangular parameter bounds but outside this domain should be treated as extrapolations and interpreted with caution.
Figure~\ref{fig:9d_params_D} shows the same points colored by $D$.
The stronger $D$-dependence of $\Om$, $\oc$, and $\sigma_{8}$ reflects the intentional overweighting along the empirical $\Om$-$\sigma_{8}$ degeneracy.
The sampling density in $D$ is approximately Gaussian, with fewer samples at larger $D$; this nonuniformity is accounted for when interpreting the independent-test performance.

We adopt a base design of 1,000 cosmologies.
In a nine-dimensional space, the curse of dimensionality \citep{Bellman1961-vb} makes the required training set size difficult to determine a priori, so we treat this as an initial design and expand it as needed.
The specific choice of 1,000 cosmologies is in part ad hoc, but it was chosen to be at least on par with the simulation suites underlying recent public emulators reviewed in Section~\ref{subsubsec:emu_overview} so that the design is comparable in size to established benchmarks. The final accuracy of the emulator reported below indicates that this size is sufficient for our target precision in the 9-dimensional space considered here.
WSDs naturally support such sequential augmentation, whereas Latin hypercube designs (LHDs; \cite{McKay1979-tk}) are typically optimized for a fixed sample size and are less convenient to extend without compromising stratification and space-filling.

\subsection{Simulation dataset}
\label{subsec:dataset}

Filling the entire nine-dimensional cosmological parameter space with uniformly
high-resolution runs is computationally prohibitive.
We therefore adopt a tiered, multi-resolution simulation suite, \textsc{Dark Quest II} (DQ2).
We first sample 1,000 cosmologies with $L_{\mathrm{box}}=1024 \hiMpc$ and $N_{\mathrm{p}}=1024^{3}$, which can be run efficiently and provides broad coverage of the space (hereafter, the \textit{low-resolution} configuration; corresponding to
the sampling points shown in figure~\ref{fig:9d_params}).
To better match the requirements of current large-scale structure surveys, we additionally run smaller \textit{middle-resolution} and \textit{high-resolution} subsets with $L_{\mathrm{box}}=1024 \hiMpc$ and $N_{\mathrm{p}}=2048^{3}$ (comparable to the resolution of DQ1) and with $L_{\mathrm{box}}=1000 \hiMpc$ and $N_{\mathrm{p}}=3000^{3}$, respectively.

We denote each simulation set by a two-letter label: L, M, and H indicate low-, middle-, and high-resolution tiers, respectively, while F and R indicate fixed and random initial phases. Thus, \textsc{LF/MF/HF} are the fixed-phase runs at the three resolution tiers, and \textsc{LR/MR/HR} are their random-phase counterparts.
The slight inconsistency between $L_\mathrm{box}=1024\,\hiMpc$ (LF/LR, MF/MR) and $L_\mathrm{box}=1000\,\hiMpc$ (HF/HR) is historical: the high-resolution suite originates from the $N_\mathrm{p}=3{,}000^3$ runs at $L_\mathrm{box}=1{,}2{,}4\,\hiGpc$ presented in the companion paper~\citep{Ginkaku26}, while the lower-resolution suites were generated independently for emulator training. No physical significance should be attached to this difference; all three tiers are used jointly in training and validation.
The suite is designed so that the low-resolution runs provide dense coverage while the higher-resolution subsets supply small-scale information and strengthen the emulator's inductive bias at high $k$. All resolutions are used jointly in training and validation. For the middle- and high-resolution subsets, we sample 50 and 20 cosmologies, respectively, and assess whether this is sufficient in section~\ref{subsec:mixed_emu}.
The 50 \textsc{MF}/\textsc{MR} cosmologies are selected from the 1,000 \textsc{LF}/\textsc{LR} cosmologies using a maximin-like space-filling criterion based on averaged nearest-neighbor distances in the normalized
cosmological-parameter space. The 20 \textsc{HF}/\textsc{HR} cosmologies are then selected from the 50 \textsc{MF}/\textsc{MR} cosmologies using the same criterion. This produces nested subsets in which each higher-resolution cosmology has matching lower-resolution counterparts, while retaining broad coverage at each tier.

To disentangle cosmological dependence from realization scatter and to improve robustness to cosmic variance, we control the Gaussian random fields (GRFs) used in the initial conditions. The fixed-phase runs isolate purely cosmological responses, whereas the random-phase runs expose the emulator to realization scatter at fixed cosmological parameters.
In addition, we run \textsc{LRfid} with 100 realizations at the fiducial cosmology to quantify variance and improve generalization at fixed parameters; this set serves as the basis for the cosmic variance emulation strategy developed in Section~\ref{subsec:cv_emu}.
All configurations and counts are summarized in table~\ref{table:ldataset}.
The value $k_\mathrm{max}=100\,\hMpci$ listed in Table~\ref{table:ldataset} indicates the upper wavenumber retained in the measured spectra (cf. Section~\ref{subsec:pkmeas}); it is set roughly an order of magnitude above the typical reach of Stage-IV weak-lensing analyses ($k\sim 10\,\hMpci$, where baryonic effects in any case dominate), and is also kept generous to provide a comfortable buffer for the Hankel transforms used in the hybrid $\xi$--$P(k)$ estimator.
We store particle data at 11 snapshots with redshifts $z=3$, $2.482$, $2.031$, $1.639$, $1.297$, $1$, $0.741$, $0.516$, $0.320$, $0.149$ and $0$, for measuring summary statistics such as the matter power spectrum.

\begin{table*}[htbp]
  \caption{Simulation models used to construct the training dataset for the
  emulator. Columns list the number of realizations (Runs), box size
  $L_{\mathrm{box}}\,[\hiMpc]$, particle number $N_{\mathrm{p}}$, PM grid size
  $N_{\mathrm{PM}}$, Nyquist wavenumber $k_{\mathrm{Ny}}$ and maximum supported
  wavenumber $k_{\max}$ (both in units of $\hMpci$), and the treatment of the
  initial phases.}
  \label{table:ldataset}
  \centering
  \scalebox{1.0}{
  \begin{tabular}{lccccccl}
    \hline
    \hline
    Model          & Runs   & $L_{\mathrm{box}}$ & $N_{\mathrm{p}}$ & $N_{\mathrm{PM}}$ & $k_{\mathrm{Ny}}$ & $k_{\max}$ & Phases / Notes                               \\
    \hline
    \textsc{LF}    & $1000$ & $1024$             & $1024^{3}$       & $2048^{3}$        & $3.14$            & $100$      & Fixed GRFs                                   \\
    \textsc{LR}    & $1000$ & $1024$             & $1024^{3}$       & $2048^{3}$        & $3.14$            & $100$      & Random GRFs; same cosmologies as \textsc{LF} \\
    \textsc{LRfid} & $100$  & $1024$             & $1024^{3}$       & $2048^{3}$        & $3.14$            & $100$      & Random GRFs at the fiducial cosmology        \\
    \textsc{MF}    & $50$   & $1024$             & $2048^{3}$       & $4096^{3}$        & $6.28$            & $100$      & Fixed GRFs; subset of \textsc{LF}            \\
    \textsc{MR}    & $50$   & $1024$             & $2048^{3}$       & $4096^{3}$        & $6.28$            & $100$      & Random GRFs; same cosmologies as \textsc{MF} \\
    \textsc{HF}    & $20$   & $1000$             & $3000^{3}$       & $6000^{3}$        & $9.42$            & $100$      & Fixed GRFs; subset of \textsc{MF}            \\
    \textsc{HR}    & $20$   & $1000$             & $3000^{3}$       & $6000^{3}$        & $9.42$            & $100$      & Random GRFs; same cosmologies as \textsc{HF} \\
    \hline
    \hline
  \end{tabular}}
\end{table*}

\subsection{Power spectrum measurement}
\label{subsec:pkmeas}

The nonlinear power spectrum $P_{\mathrm{sim}}(k,z)$ is measured with the cosmic-variance-reducing estimator developed in the companion paper~\citep{Ginkaku26}, of which we summarize only the underlying idea.

The estimator works in configuration space, where each of three correlation-function templates for the same realization is empirically a good approximation over a distinct range of $x$: the simulation estimate (sim), itself a hybrid of an FFT-based measurement on large separations and direct pair counting below the FFT grid spacing, on small scales where nonlinearity dominates; the linear-theory prediction (lin) on the largest scales; and a propagator-damped linear template ($G^{2}\cdot$\,lin) across the BAO region, with $G$ the nonlinear propagator~\citep{crocce06b,crocce:2006uq,crocce08} that captures the smearing of the BAO peak. The three are stitched smoothly in $x$, and the round trips between $\xi(x)$ and $P(k)$ are implemented with FFTLog~\citep{Hamilton00}. Section~5.3 of~\citet{Ginkaku26} provides the explicit transition windows and validates the procedure.

Because the direct pair-counting branch of the sim estimator resolves separations well below the FFT grid spacing, the measurement of $P_{\mathrm{sim}}(k,z)$ extends beyond the FFT mesh Nyquist up to $k \simeq 100\,\hMpci$, with the shot-noise contribution faithfully included at the high-$k$ end.
The particle Nyquist wavenumber,
\be
\kny = \pi\,\frac{N_{\mathrm{p}}^{1/3}}{L_{\mathrm{box}}} \, , \label{eq:kny}
\ee
serves as a resolution-dependent reference scale throughout this paper. At sufficiently small scales the matter distribution loses memory of its grid pre-initial configuration and becomes asymptotically Poisson, so $P_{\mathrm{sim}}(k,z)$ approaches the resolution-dependent floor
\be
P_\mathrm{sn} \equiv \frac{L_{\mathrm{box}}^{3}}{N_{\mathrm{p}}} \, , \label{eq:pksn}
\ee
which vanishes as $N_\mathrm{p}$ is increased. The particle discreteness is part of the dynamical state that sources the subsequent nonlinear evolution, and because the relaxation just described is itself scale-dependent, the discreteness contribution to $P_{\mathrm{sim}}(k,z)$ carries a non-trivial $k$ dependence that does not lend itself to a clean a~priori subtraction~\citep{POWMES}. We therefore do not subtract any analytical noise model from the measurement: $P_{\mathrm{sim}}$ itself --- including its resolution-dependent high-$k$ asymptote --- is the quantity the emulator is trained to reproduce, with the $N_p$ dependence learned by the resolution-conditioned network of Section~\ref{subsec:mixed_emu}.

The emulation targets are interpolated onto 301 fixed, logarithmically spaced
$k$ nodes spanning $k_{\min}=10^{-3}\,\hMpci$ to
$k_{\max}=100\,\hMpci$, matching the $k$ range covered by the emulator, rather
than onto bin centers defined from bin edges.
Unless otherwise noted, $P_{\mathrm{sim}}(k,z)$ denotes the ``cb'' component.
Total matter spectra are obtained by converting $P_{\mathrm{cb}}$ to $P_{\mathrm{tot}}$ using the linear theory ratio $\mathcal{R}^{\mathrm{lin}}_{\mathrm{tot/cb}}(k,z)$ (Appendix~\ref{append:cb_to_tot}).

\subsection{Preprocessing}
\label{subsec:preprocessing}

To stabilize training across redshifts and resolutions, we reduce the dynamic range of the targets and the leverage of high-$k$ fluctuations by applying the following target transformations.

\noindent
\textit{Preprocessing steps:}
\begin{enumerate}
  \item \textbf{Linear normalization.}

    Dividing the simulated nonlinear power spectrum by the fiducial linear power
    spectrum removes the bulk redshift trend while retaining the scale dependence.
    Specifically, we define the nonlinear boost factor with respect to the fiducial
    cosmology as \be B_\mathrm{fid}(k,z) \equiv P_\mathrm{sim}(k,z)/P_\mathrm{lin,fid}(k,z)
    \, . \label{eq:Bfid} \ee

    Note that this fiducial-normalized boost factor differs slightly from the
    convention adopted in several existing emulators~\citep{EuclidEmu1,EuclidEmu2,BACCO}, where the nonlinear correction
    is commonly defined as \be B(k,z) \equiv P_\mathrm{nl}(k,z)/P_\mathrm{lin}(k,z)
    \, , \label{eq:B} \ee for each cosmology.
    We adopt the fiducial-normalized definition in equation~(\ref{eq:Bfid}) so that all training targets share a common, cosmology-independent denominator. This keeps the scale-dependent features at fixed $k$ comparable across cosmologies, which is convenient when treating $B_\mathrm{fid}(k,z)$ as a regression target jointly with the linear power spectrum input (Section~\ref{subsec:9d_param_linpk}).

  \item \textbf{Log-ratio target.}

    We then take the base-10 logarithm, \be Y(k,z)=\log_{10} B_\mathrm{fid}(k,z)
    \, , \label{eq:pk_out} \ee which compresses the dynamic range and yields an
    approximately additive target across redshifts.

  \item \textbf{High-$k$ taper.}

    At high wavenumbers, the training targets are more sensitive to particle shot noise, finite-resolution effects, and the details of the small-scale power spectrum estimator. To reduce the leverage of these high-$k$ fluctuations in the regression target, we apply a high-$k$ taper,
    \be \tilde{Y}(k,z)
    = \left\{
    \begin{array}{ll}
      Y(k,z)                                                          & (k<k_0)    \\
      \log_{10}\left[B_\mathrm{fid}(k,z) \times (k/k_0)^{-3/4}\right] & (k\ge k_0) \\
    \end{array}
    \right. \label{eq:tilde_pk_out} \ee with $k_{0}=0.1\,h\,\mathrm{Mpc}^{-1}$.
    This transformation makes the residuals closer to homoscedastic with respect to an $L_{2}$ loss. The transition scale $k_0$ and the slope $-3/4$ are empirical choices that we found to work well in practice; they are not derived from any specific noise model. At prediction time, the inverse transformation is applied to recover $P(k,z)$ from $\tilde{Y}(k,z)$.
\end{enumerate}

\begin{figure}[htbp]
  \centering
  \includegraphics[width=\linewidth]{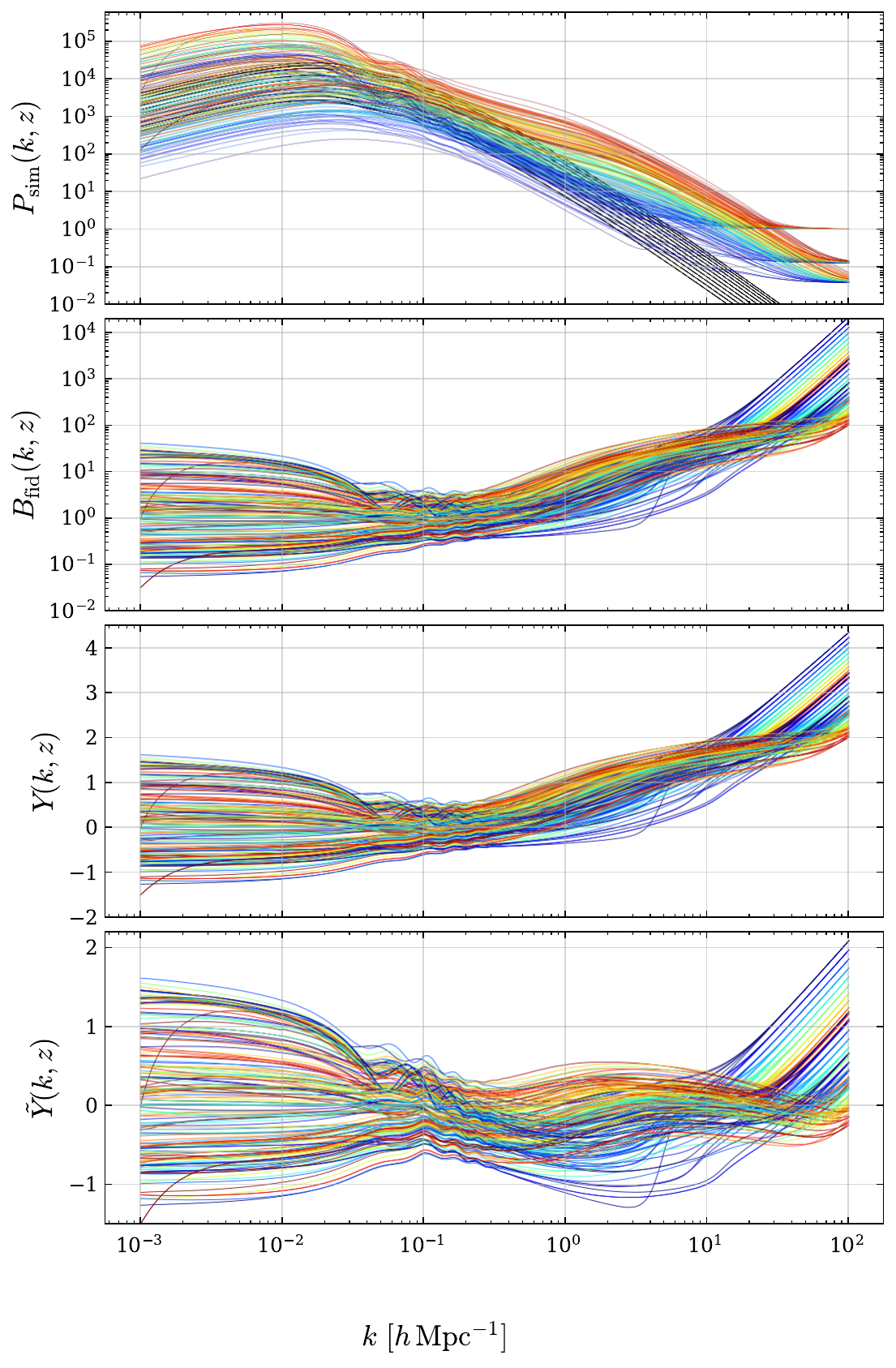}
  \caption{ Random samples of training targets across cosmologies, redshifts, and
  resolutions. Colors encode the linear power spectrum distance $D$ (blue:
  small, red: large). 
  The top, second, third, and bottom panels show the original nonlinear ``cb'' power spectrum $P_{\mathrm{sim}}(k,z)$, the nonlinear boost factor $B_{\mathrm{fid}}(k,z)$, the log-ratio $Y(k,z)$, and the tapered target $\tilde{Y}(k,z)$, respectively.
  {Alt text: Four stacked line plots in which the curves become progressively more compressed and uniform after normalization, logarithmic scaling, and tapering at high wavenumber.}
  }
\label{fig:teaching_form}
\end{figure}

\section{Emulator framework}
\label{sec:emulation}

This section describes the overall architecture of the \DE\ emulator, the training and hyperparameter-tuning procedures, the independent test suites, and the accuracy metrics used to quantify emulator performance. The three design choices that distinguish the present implementation from existing public emulators --- the use of the linear power spectrum as an input feature, the joint treatment of multiple simulation resolutions, and the conditioning on realization-dependent inputs --- are then developed in section~\ref{sec:design}.

\subsection{Emulator overview}
\label{subsec:emu_overview}

DQ1 adopted Gaussian-process (GP) regression~\citep{NIPS1995_7cce53cf,Ambikasaran2016-vv}
as the main emulation technique. 
In \DE, however, the cosmological parameter space is nine-dimensional, and the number of training cosmologies is substantially larger than in DQ1. In this regime, standard GP regression with dense covariance matrices is computationally expensive~\citep{Ambikasaran2016-vv}, whereas neural networks (NNs) scale more favorably with both dimensionality and training set size. We therefore adopt NN as our primary emulation method.

We construct a set of fully connected feed-forward neural networks (FFNN), implemented with the \textsc{PyTorch}\footnote{\url{https://pytorch.org/}} framework, to emulate the nonlinear matter power spectrum. 
For each cosmology, the networks are trained to predict the preprocessed targets $\tilde{Y}(k,z)$ defined in equation~\ref{eq:tilde_pk_out}. 
The emulator predicts these compressed spectra, which are then mapped back to the nonlinear power spectrum through the inverse preprocessing pipeline described in section~\ref{subsec:preprocessing}.

The emulator takes as input nine independent cosmological parameters, $(\Om,\ob,\sigma_{8},\ns,h,M_{\nu},w_{0},w_{a},\Ok)$, as defined in section~\ref{subsec:sampling}.
Internally, these degrees of freedom can be provided through several equivalent parameter combinations. For example, the matter content can be specified either in terms of physical densities $(\ob,\oc)$ together with $h$ (with $\Om$ then implied), or by $(\Om,\oc,h)$. Likewise, the total energy budget can be specified by either $\Ode$ or $\Ok$, and the fluctuation amplitude by one of $(\sigma_{8},\As,\lnAs)$. Any supported input specification is mapped to a unique internal parameter vector before being passed to the networks.

The overall architecture of \DE\ is modular. 
We train four fully connected neural networks: (i) a nonlinear ``cb'' power spectrum emulator, (ii) a linear ``cb'' power spectrum emulator, (iii) an amplitude emulator that maps between $\As$ and $\sigma_{8}$, and (iv) a linear theory total-to-cb ratio emulator, $\mathcal{R} ^{\mathrm{lin}}_{\mathrm{tot/cb}}(k,z)$. 
The linear modules (ii)--(iv) provide fast internal evaluations of $P_{\mathrm{lin,cb}}(k,z)$, the $\As\leftrightarrow \sigma_{8}$ conversion, and the scale- and redshift-dependent cb-to-total mapping. 
When total matter spectra are required, we approximate the nonlinear total matter spectrum as $P_{\mathrm{tot}}(k,z) \approx \mathcal{R}^{\mathrm{lin}}_{\mathrm{tot/cb}}(k,z)\,P_{\mathrm{cb}}(k,z)$ (see appendix~\ref{append:cb_to_tot}). 
Their targets and preprocessing are summarized in section~\ref{subsec:preprocessing}. 
Further details of the linear spectrum and amplitude emulators are provided in appendix~\ref{append:lin_pk_emu}, and the definition and validation of $\mathcal{R}^{\mathrm{lin}}_{\mathrm{tot/cb}}$ are given in appendix~\ref{append:cb_to_tot}.

\begin{figure}[htbp]
  \centering
  \includegraphics[width=\linewidth]{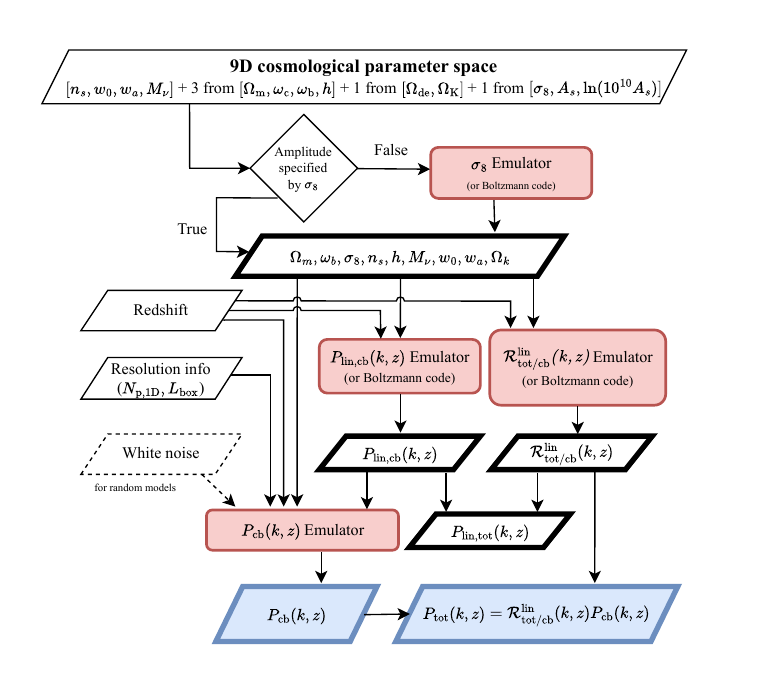}
  \caption{Overview of the \DE\ emulator architecture. 
  {Alt text: Flowchart showing connected emulator modules that combine cosmological inputs, redshift, resolution information, and optional white noise variables to produce matter power spectra.}
  }
  \label{fig:dq2_emu}
\end{figure}

The training setup, hyperparameter tuning, and test suites are described in section~\ref{subsec:train_test}, and the accuracy metrics used in later performance assessments are defined in section~\ref{subsec:acc_metrics}. The design choices specific to \DE\ are discussed separately in section~\ref{sec:design}.

\subsection{Training setup and test suites}
\label{subsec:train_test}

The emulator training data are drawn from the simulation suite described in
section~\ref{subsec:dataset} and summarized in table~\ref{table:ldataset}. We use
simulations with multiple mass and force resolutions and with both fixed (\textsc{LF},
\textsc{MF}, and \textsc{HF} models) and random initial phases (\textsc{LR}, \textsc{LRfid},
\textsc{MR}, and \textsc{HR} models), as indicated in the table.

Within this pool of emulator samples, we adopt Monte Carlo cross-validation (MCCV) with five repeated train/validation splits~\citep{Kohavi1995-zp}. Each split is constructed using 90\% of the cosmological models for training and the remaining 10\% for validation. The 10\% subset in each MCCV split is a validation set used during training and should not be confused with the independent test suite described below.
The splits are defined at the level of cosmological models: each cosmology is assigned exclusively to either the training set or the validation set.

We train the network by minimizing the root mean square error (RMSE) of the compressed targets $\tilde{Y}(k,z)$ defined in equation~\ref{eq:tilde_pk_out}. For a minibatch containing $N_{\mathrm{batch}}$ samples, the loss function is
\be
\mathrm{RMSE}
=
\sqrt{
\frac{1}{N_{\mathrm{batch}}N_k}
\sum_{b=1}^{N_{\mathrm{batch}}}
\sum_{i=1}^{N_k}
\left(\hat{y}_{b,i}-y_{b,i}\right)^2
},
\label{eq:rmse}
\ee
where $y_{b,i}=\tilde{Y}_{b}(k_i,z)$ is the compressed target for the $b$-th sample at the $i$-th wavenumber node, $\hat{y}_{b,i}$ is the corresponding network prediction, and $N_k$ is the number of wavenumber nodes. For each split, we monitor the validation RMSE and retain the checkpoint that
achieves the lowest validation RMSE. For a given hyperparameter configuration, we define its MCCV validation score as the minimum of these split-wise best-validation RMSE values across the five splits, and we adopt the configuration that minimizes this score.

For each emulator, network hyperparameters (e.g., numbers of hidden layers and
neurons, learning rate, and weight decay) are tuned based on this MCCV validation score.
For the nonlinear $P_{\mathrm{cb}}(k)$ emulator, we use the covariance matrix adaptation evolution strategy (CMA-ES) optimizer~\citep{Hansen1996-me,Hamano2022-fr} to explore the hyperparameters of both the network architecture and the training setting.
In practice, this survey is conducted for the mixed-resolution emulator and the associated cosmic variance correction network, and we adopt the resulting optimal configuration. 
The CMA-ES runs indicate that relatively shallow but wide fully connected networks (3
hidden layers with 2,000 neurons each) yield the best validation accuracy for this
emulator; therefore, we adopt ELU activation functions. 
For the remaining emulators, i.e., the linear $P_{\mathrm{lin,cb}}(k)$ emulator, the $\sigma_{8}\leftrightarrow \As$ amplitude mapping, and the $P_{\mathrm{lin,tot}}/P_{\mathrm{lin,cb}}$ correction factor, we perform a simple grid search over a smaller set of hyperparameters and find that networks with GELU activations and a moderate number of neurons per layer achieve sufficient accuracy. 
The resulting network architectures and training settings are summarized in table~\ref{table:net_params}, and the detailed CMA-ES setup and the optimal hyperparameter set for the nonlinear emulator are given in Appendix~\ref{append:cmaes}.
In table~\ref{table:net_params}, $N_{\mathrm{train}}$ denotes the total number of samples in the full training pool for the nonlinear emulator before the MCCV train/validation split. 
$N_{\mathrm{test}}$ denotes the number of distinct test cosmologies, not the total number of test realizations.

\begin{table*}[htbp]
  \caption{Network hyperparameters.}
  \label{table:net_params}
  \centering
  \begin{tabular}{ccccc}
    \hline
    \hline
    Parameter             & $P_{\mathrm{cb}}(k,z)$ & $P_{\mathrm{lin,cb}}(k,z)$ & Amplitude ($\sigma_{8}\leftrightarrow \As$) & $\mathcal{R}^{\mathrm{lin}}_{\mathrm{tot/cb}}$ \\
    \hline
    $N_{\mathrm{hidden}}$ & 3                      & 5                          & 7                                           & 3                                              \\
    $N_{\mathrm{neuron}}$ & 2000                   & 400                        & 1000                                        & 400                                            \\
    $N_{\mathrm{epoch}}$  & 5000                   & 10000                      & 10000                                       & 10000                                          \\
    $N_{\mathrm{batch}}$  & 50                     & 100                        & 100                                         & 100                                            \\
    activation func.      & ELU                    & GELU                       & GELU                                        & GELU                                           \\
    optimizer             & LAMB                   & LAMB                       & LAMB                                        & LAMB                                           \\
    loss func.            & RMSE                   & RMSE                       & RMSE                                        & RMSE                                           \\
    scheduler             & StepLR                 & StepLR                     & StepLR                                      & StepLR                                         \\
    \hline
    $N_{\mathrm{train}}$  & 2240$^{\dagger}$       & 49000                      & 49000                                       & 49000                                          \\
    $N_{\mathrm{test}}$   & 80$^{\ddagger}$        & 1000                       & 1000                                        & 1000                                           \\
    Hyperparameter tuning & CMA-ES                 & grid search                & grid search                                 & grid search                                    \\
    \hline
    \hline
  \end{tabular}

  \begin{minipage}{0.96\linewidth}
    { \footnotesize $^{\dagger}$For the nonlinear $P_{\mathrm{cb}}(k)$ emulator, the training set comprises 2,240 simulation runs drawn from 1,000 distinct cosmological models:
    $N_{\mathrm{train}} = 1000(\textsc{LF}) + 1000(\textsc{LR}) + 100(\textsc{LRfid}) + 50(\textsc{MF}) + 50(\textsc{MR}) + 20(\textsc{HF}) + 20(\textsc{HR})$.
    }\\
    { \footnotesize $^{\ddagger}$For the nonlinear $P_{\mathrm{cb}}(k)$ emulator, the test set comprises 80 independent test cosmologies:
    $N_{\mathrm{test}} = 30(\textsc{TLF}) + 30(\textsc{TLR}) + 10(\textsc{TMF}) + 10(\textsc{TMR})$.
    }\\
  \end{minipage}
\end{table*}

In addition to the training and validation data, we prepare a completely independent test suite, summarized in table~\ref{table:tdataset}. These test simulations are never used for training, validation, or hyperparameter tuning, and are reserved exclusively for assessing the emulator accuracy using the metrics defined in section~\ref{subsec:acc_metrics}.

\begin{figure}[htbp]
  \centering
  \includegraphics[width=\linewidth]{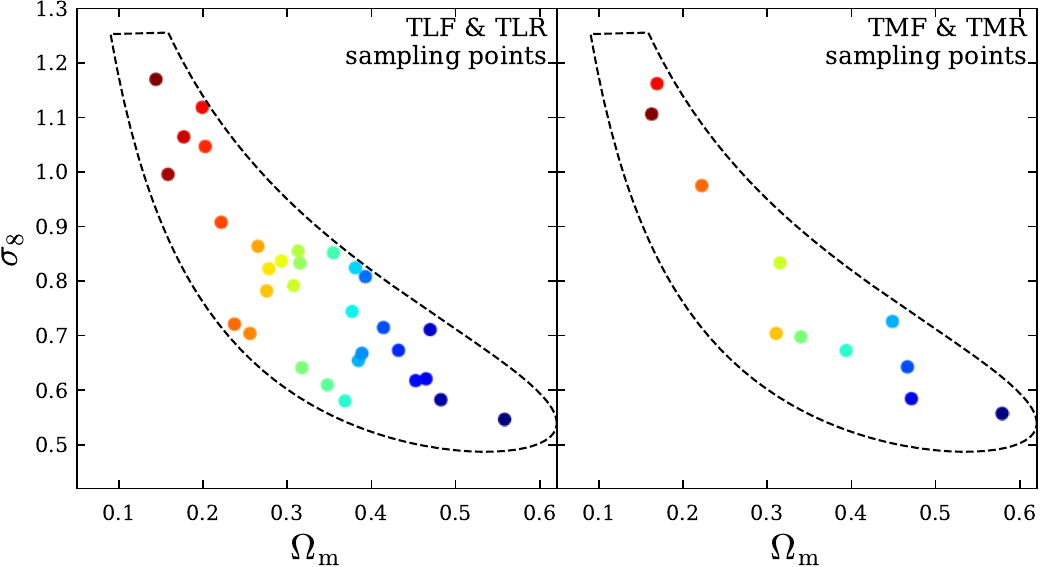}
  \caption{Distribution of test cosmologies in the $\Om$-$\sigma_{8}$ plane. The
  left panel shows the low-resolution test sets (\textsc{TLF}/\textsc{TLR}) and
  the right panel shows the middle-resolution test sets (\textsc{TMF}/\textsc{TMR}).
  The color coding, with red indicating smaller $\Om$ and blue larger $\Om$,
  is reused in the figures comparing emulator predictions to these test sets.
  {Alt text: Points lie within a dashed, banana-shaped support region and follow a decreasing trend in sigma eight with increasing matter density. The right panel is sparser than the left panel.}
  }
  \label{fig:testset}
\end{figure}

The distribution of the test cosmologies in the $\Om$-$\sigma_{8}$ plane for
the low- and middle-resolution models is shown in figure~\ref{fig:testset}.
The remaining seven cosmological parameters are chosen to sample the allowed
ranges as evenly as possible. A sufficiently large number of test cosmologies is
essential for a reliable performance assessment, so we select 30 sampling
points for the low-resolution (\textsc{TLF}/\textsc{TLR}) setting and 10
sampling points for the middle-resolution (\textsc{TMF}/\textsc{TMR}) setting,
as summarized in table~\ref{table:tdataset}. Although additional test sets at
high resolution would be desirable, constructing an independent test suite comparable
in size to the training sample would be computationally prohibitive. We
therefore restrict the independent tests in this work to the low- and middle-resolution
suites and leave a systematic validation against high-resolution simulations for
future work. The color coding of the points in figure~\ref{fig:testset} is
reused in later figures that compare emulator predictions to these test sets, with
red indicating smaller $\Om$ and blue larger $\Om$.

For the random-phase test models (\textsc{TLR} and \textsc{TMR}), we generate ten realizations at each test cosmology.  Because our primary comparison target is the ensemble-averaged power spectrum rather than individual realizations, we compare the emulator predictions with the mean power spectrum over these realizations at each test cosmology.

\begin{table*}[htbp]
  \caption{Summary of test data sets.}
  \label{table:tdataset}
  \centering
  \scalebox{1.0}{
  \begin{tabular}{lccccccl}
    \hline
    \hline
    Model        & Runs         & $L_{\mathrm{box}}\,[\hiMpc]$ & $N_{\mathrm{p}}$ & $N_{\mathrm{PM}}$ & $k_{\mathrm{Ny}}$ & $k_{\max}[\hMpci]$ & Phases / Notes                            \\
    \hline
    \textsc{TLF} & $30$         & $1024$                       & $1024^{3}$       & $2048^{3}$        & $3.14$            & $100$              & Fixed                                     \\
    \textsc{TLR} & $30\times10$ & $1024$                       & $1024^{3}$       & $2048^{3}$        & $3.14$            & $100$              & Random, 10 realizations for 30 parameters \\
    \textsc{TMF} & $10$         & $1024$                       & $2048^{3}$       & $4096^{3}$        & $6.28$            & $100$              & Fixed                                     \\
    \textsc{TMR} & $10\times10$ & $1024$                       & $2048^{3}$       & $4096^{3}$        & $6.28$            & $100$              & Random, 10 realizations for 10 parameters \\
    \hline
    \hline
  \end{tabular}}
\end{table*}

\subsection{Accuracy metrics}
\label{subsec:acc_metrics}

We assess the final emulator performance in physical units by comparing the
nonlinear power spectra from the simulations, $P_{\mathrm{sim}}(k,z)$, with the
corresponding emulator outputs, $P_{\mathrm{DE2}}(k,z)$, for the independent
test suites summarized in table~\ref{table:tdataset}.
Because $P(k)$ spans a wide dynamic range across both $k$ and $z$, we adopt the
mean absolute percentage error (MAPE), i.e., the mean absolute relative error
expressed in percent, as our primary accuracy metric.
For a single test cosmology, we define
\be
\mathrm{MAPE}
=
100\times
\frac{1}{N_k}
\sum_{i=1}^{N_k}
\left|
\frac{\hat{y}_{i}-y_{i}}{y_{i}}
\right|
\quad [\%],
\label{eq:mape}
\ee
where $y_i\equiv P_{\mathrm{sim}}(k_i,z)$ and
$\hat{y}_i\equiv P_{\mathrm{DE2}}(k_i,z)$.
We further summarize the performance over $N_{\mathrm{test}}$ independent test
cosmologies by averaging the MAPE values,
\be
\overline{\mathrm{MAPE}}
=
\frac{1}{N_{\mathrm{test}}}
\sum_{j=1}^{N_{\mathrm{test}}}
\mathrm{MAPE}_{j}
\quad [\%].
\label{eq:mean_mape}
\ee
Throughout, we quote both $\mathrm{MAPE}$ and
$\overline{\mathrm{MAPE}}$ in the relevant $k$ ranges as our primary figures of
merit for the nonlinear power spectrum emulator.

\section{Design choices for the nonlinear emulator}
\label{sec:design}

Building on the framework of section~\ref{sec:emulation}, this section presents three design choices that distinguish the \DE\ nonlinear matter power spectrum emulator from existing publicly available emulators. These are not three independent additions but three instantiations of a single underlying principle: in addition to the cosmological parameter vector, we supplement the neural network's inputs with three families of physically motivated auxiliary quantities that are expected to improve the accuracy across the parameter space. The three axes considered here are the linear power spectrum (which simplifies the regression target and generalizes the long-standing practice of emulating a multiplicative boost factor relative to it), descriptors of the simulation resolution, and a low-dimensional summary of the initial Gaussian random field; the latter two carry information about the actual spectrum realized in each finite-resolution, single-realization simulation that is not contained in the cosmological parameters alone.

Each axis is introduced below through a contrast with the dominant alternative in the literature, motivated by the underlying physical or numerical structure of the problem, and validated against our independent test suite. Section~\ref{subsec:9d_param_linpk} discusses the use of the linear power spectrum as an explicit input feature, generalizing the more common practice of emulating a multiplicative boost factor relative to it. Section~\ref{subsec:mixed_emu} introduces a resolution-conditioned mixed-resolution training scheme that allows simulations at different mass and force resolutions to contribute jointly to a single emulator. Section~\ref{subsec:cv_emu} develops a cosmic-variance-aware emulator that conditions on a reduced description of the initial Gaussian random field and enables efficient ensemble-mean predictions at evaluation time.

\subsection{Linear power spectrum as an input feature}
\label{subsec:9d_param_linpk}

Conventional emulators of the nonlinear matter power spectrum typically learn the multiplicative boost factor $B(k,z;\boldsymbol{\theta}) \equiv P_\mathrm{nl}(k,z;\boldsymbol{\theta})/P_\mathrm{lin}(k,z;\boldsymbol{\theta})$ as their regression target~(e.g., \cite{EuclidEmu2, BACCO}). Because the linear spectrum $P_\mathrm{lin}(\boldsymbol{\theta})$ already absorbs a large part of the cosmology dependence, the residual variation of $B$ with $\boldsymbol{\theta}$ is comparatively mild and is well suited to a relatively shallow neural network. The network is, however, restricted to predict $B$ from $\boldsymbol{\theta}$ alone, without direct access to the scale-dependent linear features that drive much of the cosmology dependence; in extended cosmologies such as those including massive neutrinos, dynamical dark energy, or non-zero spatial curvature, where the cosmology dependence does not factor cleanly through $P_\mathrm{lin}$, $B$ retains non-trivial structure that must be learned from $\boldsymbol{\theta}$ alone.

We take a different route: rather than dividing by $P_\mathrm{lin}$ and emulating $B$, we keep $P_\mathrm{nl}$ (or a fiducial-normalized form thereof) as the regression target and supply the sampled linear power spectrum directly as an additional input feature of the network. The conventional boost-factor approach is recovered as the special case in which the learned mapping is linear in $\mathbf{p}_\mathrm{lin}$ with a coefficient depending only on $\boldsymbol{\theta}$, but the network is also free to express additive or otherwise nonlinear dependencies on $P_\mathrm{lin}$ that cannot be captured by a purely multiplicative combination. This added flexibility is particularly relevant for the extended cosmologies considered here.

\begin{figure*}[htbp]
  \centering
  \includegraphics[width=0.8\linewidth]{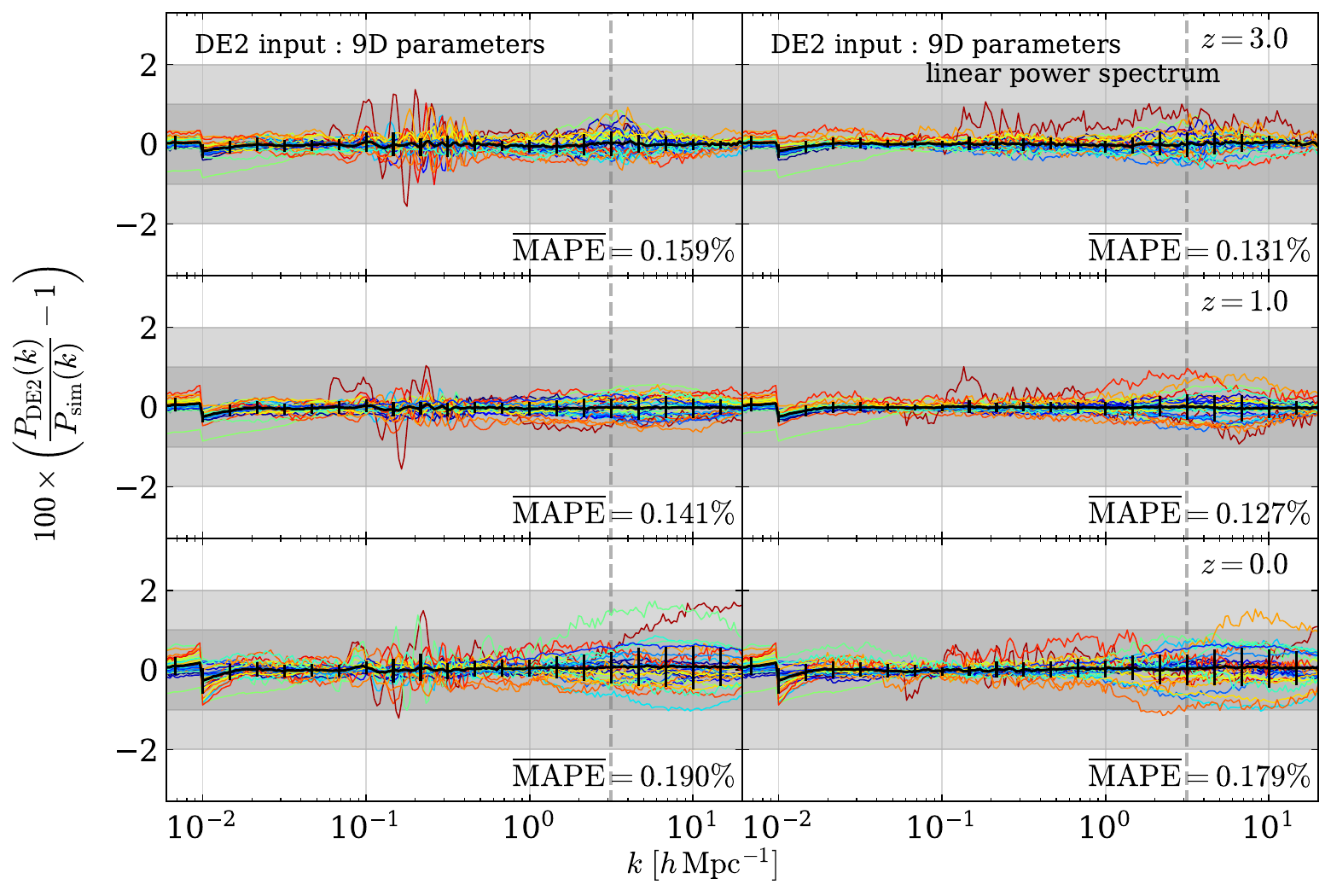}
  \caption{Accuracy of the emulator trained on the 1,000 \textsc{LF}
  cosmologies, evaluated on the 30 test cosmologies. The left panels use only
  the nine-dimensional cosmological parameter vector as input, whereas the right
  panels additionally include the linear power spectrum as input features. 
  For $k < 0.01 \, \hMpci$, the residual is evaluated against the linear-theory spectrum predicted by the linear emulator, whereas for $k \geq 0.01 \, \hMpci$ it is evaluated against the nonlinear emulator.
  The black curve with error bars shows the mean and standard deviation over the 30 test cosmologies, the light and dark shading indicate the $2\%$ and $1\%$ error bands, and the vertical dashed lines mark $\kny$ of the low-resolution simulations. 
  {Alt text: Residual curves are more coherent and less oscillatory in the right column than in the left column, especially around baryon acoustic oscillation scales.}
  }
  \label{fig:pk_p9vsp9lp}
\end{figure*}

To assess this design choice empirically, we compare two NN emulators trained on the low-resolution (\textsc{LF}) simulations of table~\ref{table:ldataset} that differ only in their inputs.
Let $\boldsymbol{\theta} \in\mathbb{R}^{9}$ denote the nine-dimensional cosmological parameter vector.
We consider the following two input representations, 
\be 
\mathbf{x}^{(\theta)} = (\boldsymbol{\theta},\, z), \qquad \mathbf{x}^{(\theta+\mathrm{lin})} = \left(\boldsymbol{\theta},\,
z,\,\mathbf{p}_\mathrm{lin}\right), \label{eq:x_lin} 
\ee 
where $z$ is redshift,
and 
\be 
\mathbf{p}_\mathrm{lin} = \mathbf{p}_\mathrm{lin}(\boldsymbol{\theta},z)
= \left\{ P_\mathrm{lin,cb}(k_i;\boldsymbol{\theta},z) \right\}_{i=1}^{N_\mathrm{lin}},
\label{eq:p_x_lin}
\ee 
denotes the sampled linear power spectrum evaluated at a set of wavenumbers $\{k_{i}\}_{i=1}^{N_\mathrm{lin}}$. 
The nine cosmological parameters used to compute the input linear spectrum are identical to those fed directly into the NN. In constructing the nonlinear-emulator training inputs, we use both linear spectra computed directly with the Boltzmann solver and spectra predicted by the auxiliary linear emulator described in appendix~\ref{append:lin_pk_emu}, evaluated for the same nine-dimensional cosmological parameter vectors. This choice makes the nonlinear emulator consistent with the input representation used in the final prediction pipeline, where the linear spectrum may be supplied either by a direct Boltzmann calculation or by the auxiliary linear emulator. Consequently, small interpolation errors in the auxiliary linear spectra are already represented in the training inputs used by the nonlinear emulator, rather than being introduced only at inference time.
In principle, a sufficiently expressive network could extract all relevant information from $\boldsymbol{\theta}$ alone, and the linear spectrum would then be redundant. In practice, however, the finite number of neurons and hidden layers limits the effective expressivity of the model. 
Providing $\mathbf{p}_{\mathrm{lin}}$ as an additional, scale-dependent input supplies the NN with $k$-bin summary information and is therefore expected to improve the emulator accuracy. 
In our implementation we set $N_{\mathrm{lin}}=150$ and choose the sampling wavenumbers $k_{i}$ to be logarithmically spaced between $0.01$ and $10 \,\hMpci$.

Figure~\ref{fig:pk_p9vsp9lp} summarizes the emulator accuracy on the 30 test cosmologies.
Each colored line corresponds to one of the test cosmologies shown in figure~\ref{fig:testset}.
For large-scale modes ($k<0.01 \, \hMpci$), we evaluate the accuracy with respect to the linear theory spectrum $P_{\mathrm{lin,cb}}(k)$, because the nonlinear spectrum measured from the simulations suffers from additional scatter due to cosmic variance in the finite-volume realizations. 
The residual mismatch at the junction between the linear and nonlinear regimes is explicitly modeled and corrected by the cosmic variance emulator introduced in section~\ref{subsec:cv_emu}.
When the emulator input consists only of the nine cosmological parameters, the fractional errors exhibit residual oscillations around the BAO scales, especially for test cosmologies with smaller $\Om$. 
By contrast, when the linear power spectrum is included in the inputs, these $k$-dependent errors are significantly suppressed. 
At $z=0$, this input augmentation reduces $\overline{\mathrm{MAPE}}$ by about 5.8\% relative to the parameter-only input.
Nevertheless, the errors remain somewhat larger for the most extreme test cosmologies near the boundaries of the prior volume, where the sampling density of the training data is relatively low. 
At wavenumbers higher than those displayed in figure~\ref{fig:pk_p9vsp9lp}, the measured power spectra of the \textsc{LF} simulations are already dominated by the shot-noise level $P_{\mathrm{sn}}$ and become nearly flat. 
On such strongly shot noise dominated scales, the fractional error is less informative for cosmological applications, so we do not include these modes in the accuracy assessment in this subsection.

These tests confirm that including the linear power spectrum alongside the nine cosmological parameters improves the emulator accuracy around the BAO scales.
In all subsequent validation and application results, we therefore adopt this augmented input configuration as the default.
Sensitivity diagnostics for this fixed-phase low-resolution emulator, including permutation feature importance and partial dependence analyses, are presented in Appendix~\ref{append:sensitivity}.

\subsection{Resolution as a continuous input: integrated multi-fidelity learning}
\label{subsec:mixed_emu}

Hierarchical multi-fidelity emulators in cosmology~\citep{Ho2021-nu, Ho2023-ls, GokuN} address the cost of high-resolution training data by maintaining separate models for each fidelity level (or each resolution) and combining them with an explicitly learned correction $\Delta P$ between adjacent levels. The architecture itself encodes a discrete fidelity hierarchy: the user must commit to a specific number of levels and define how they connect.

Our approach is different in spirit. Rather than building a separate model per resolution and combining them with an explicit correction term, we treat the simulation resolution as an additional input feature of a single network. The network learns the dependence on $(N_\mathrm{p,1D}, L_\mathrm{box})$ as a continuous function alongside the cosmological parameters, in the same way it learns the dependence on $\boldsymbol{\theta}$. The \textsc{LF}, \textsc{MF}, and \textsc{HF} fixed-phase models listed in table~\ref{table:ldataset} therefore contribute jointly to a single end-to-end emulator, with the network internally interpolating between resolution levels rather than relying on a predefined hierarchy. This integrated treatment leverages the flexibility of the neural network architecture and avoids committing to a fixed fidelity structure at the design stage. We refer to this approach as resolution-conditioned training.

The technique known as ``super-resolution'' reconstructs high-resolution outputs from low-resolution inputs and has been widely studied in image processing and inference from volumetric data~\citep{Dong2014-nx,Kodi_Ramanah_2020,Li2021-iw}.
In contrast, our ``mixed-resolution'' emulation does not upsample any individual low-resolution realization. 
Instead, we train the emulator on a small set of high-resolution simulations and a much larger set of low-resolution simulations that densely sample the cosmological parameter space. Building on the input representation $\mathbf{x}^{(\theta+\mathrm{lin})}$ in equation~\ref{eq:x_lin}, we augment the neural network input vector with resolution descriptors: the one-dimensional particle number $N_{\mathrm{p,1D}}\equiv N_{\mathrm{p}}^{1/3}$ and the box size $L_{\mathrm{box}}\,[\hiMpc]$. 
In this way, the emulator can represent power spectra over a range of numerical resolutions while being constrained toward the highest-resolution behavior wherever such training data are available.
For a cosmology $\boldsymbol{\theta}$, redshift $z$, particle number
$N_{\mathrm{p,1D}}$, and box size $L_{\mathrm{box}}$, the mixed-resolution
emulator takes the augmented input vector
$\left(\boldsymbol{\theta},\,z,\,\mathbf{p}_{\mathrm{lin}},\,N_{\mathrm{p,1D}},\,L_{\mathrm{box}}\right)$.
In section~\ref{sec:apps}, we evaluate the emulator only at our highest-resolution configuration.

Low-resolution simulations cannot reliably represent small-scale clustering because of shot noise, so higher-resolution runs are essential. 
However, simulating more than 1,000 cosmological models at higher resolution (e.g., $N_{\mathrm{p}}=2000^{3}$ or $3000^{3}$) is computationally prohibitive in both wall-clock time and storage.
We therefore test whether the emulator can reproduce the power spectra of such high-resolution simulations when only a small number of high-resolution models are added to a large low-resolution training set. 
If successful, this strategy substantially reduces the computational cost and storage required to construct high-resolution training data.

\begin{figure}[htbp]
  \centering
  \includegraphics[width=\linewidth]{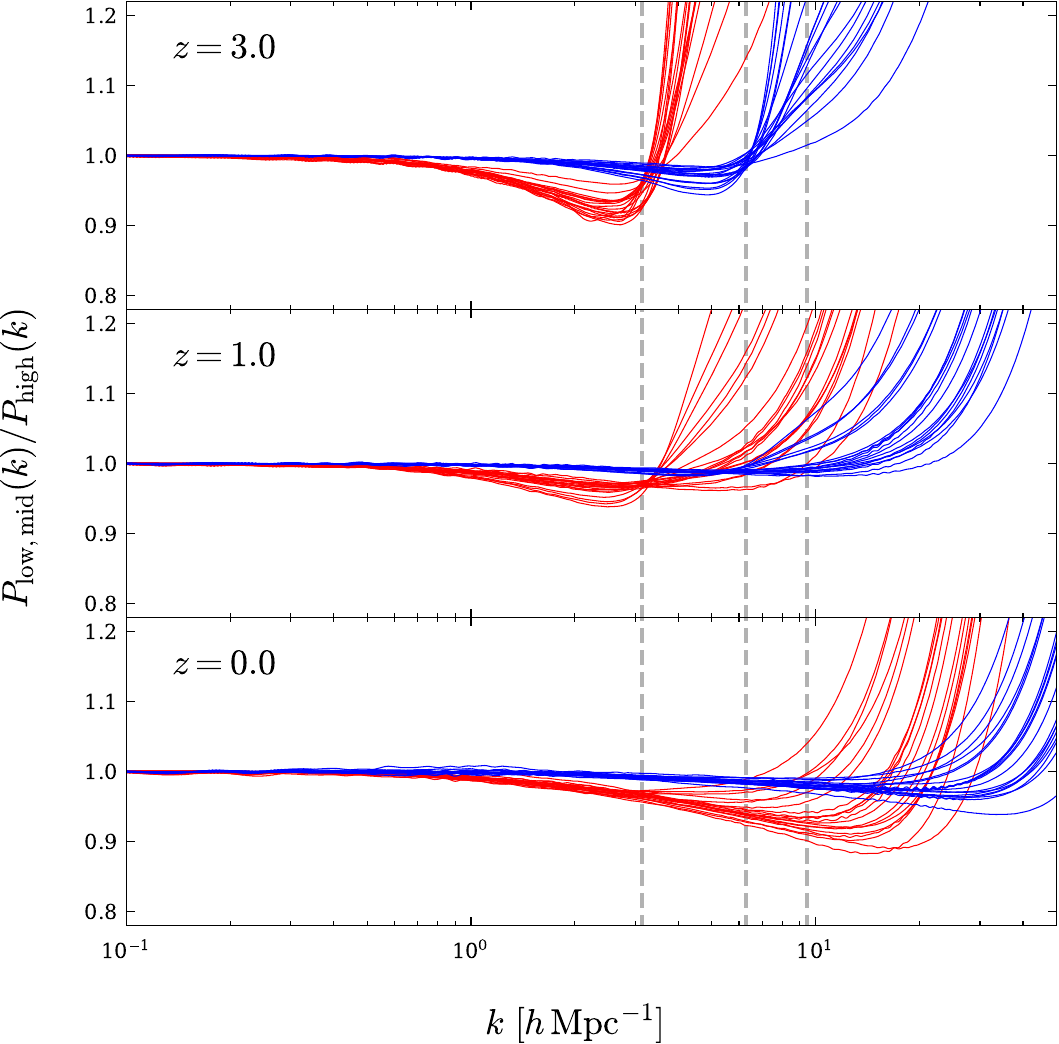}
  \caption{Ratios of the \textsc{LF} (red) and \textsc{MF} (blue) matter power
  spectra to the corresponding \textsc{HF} results. Each curve corresponds to one
  of the 20 cosmological models with matched \textsc{LF}, \textsc{MF}, and \textsc{HF}
  simulations. Vertical dashed lines mark $\kny$ for the \textsc{LF}, \textsc{MF},
  and \textsc{HF} simulations (left to right). 
  {Alt text: Lower-resolution spectra depart from the high-resolution reference before their Nyquist scales, with larger departures at high wavenumber and clearer resolution dependence toward lower redshift.}
  }
  \label{fig:pk_reso_check}
\end{figure}

To make the network sensitive to resolution, we need training examples that differ only in numerical resolution while sharing the same cosmological parameters. 
As described in section~\ref{subsec:dataset}, the \textsc{MF} and \textsc{HF} subsets are chosen using a maximin-like averaged nearest-neighbor criterion, rather than a strict maximin rule. 
This selection preserves broad coverage of the weighted \textsc{LF} design while keeping the subsets nested across resolution levels. 
We first select 50 \textsc{MF} cosmologies from the 1,000 \textsc{LF} cosmologies and then select 20 \textsc{HF} cosmologies from these 50 \textsc{MF} cosmologies using the same criterion. For each of these 20 cosmologies, we have three simulations at low, middle, and high resolution.
In this way, the network sees low-, middle-, and high-resolution realizations at identical cosmological parameters and can learn systematic resolution trends. 
Figure~\ref{fig:pk_reso_check} shows the resolution dependence of the measured power spectra for these matched simulations. 
Compared with the high-resolution results, the lower-resolution spectra begin to deviate well below their respective Nyquist wavenumbers, already at approximately one-third of $k_{\mathrm{Ny}}$, and shot noise dominates near $k_{\mathrm{Ny}}$.

As noted above, we encode the simulation resolution by supplying the one-dimensional particle number $N_{\mathrm{p,1D}}\equiv N_{\mathrm{p}}^{1/3}$ and the box size $L_{\mathrm{box}}$ as additional input variables. 
Because we have not yet constructed an independent highest-resolution (\textsc{HF}) test set, we first validate a mixed-resolution network trained on the 1,000 \textsc{LF} and 50 \textsc{MF} simulations using the \textsc{TLF} and \textsc{TMF} test sets.

\begin{figure}[htbp]
  \centering
  \includegraphics[width=\linewidth]{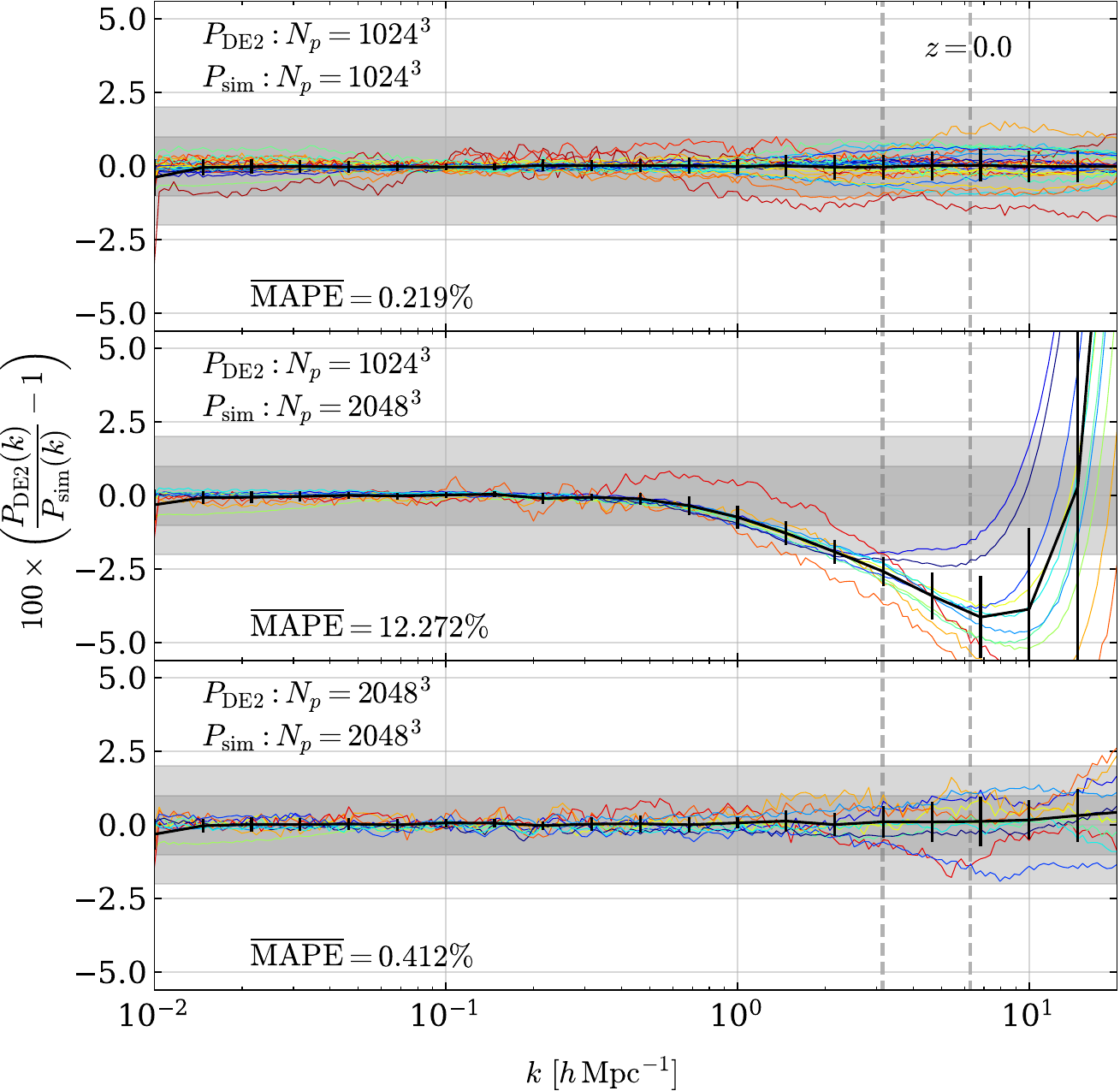}
  \caption{ Accuracy of the mixed-resolution emulator. The top panel compares
  an emulator trained only on the 1,000 \textsc{LF} simulations with the \textsc{TLF}
  test set at low resolution. The middle panel compares the same LF-only emulator
  with the \textsc{TMF} test set at middle resolution. The bottom panel compares
  an emulator trained on a mixed set of 1,000 \textsc{LF} and 50 \textsc{MF} simulations
  with the \textsc{TMF} test set at middle resolution. Black curves with error
  bars show the mean and standard deviation over the test cosmologies. The
  shaded bands indicate the 2\% and 1\% error ranges, and vertical dashed lines
  mark $\kny$ for the low- and middle-resolution simulations (left to right). 
  {Alt text: The upper two panels show that an emulator trained only on low resolution data remains accurate for low-resolution tests but fails for middle-resolution tests, while the bottom panel shows recovery after adding middle-resolution training data.}
  }
  \label{fig:pk_mixed_emu}
\end{figure}

Figure~\ref{fig:pk_mixed_emu} shows the performance of the mixed-resolution emulator.
In the top and middle panels, the emulator is trained only on the 1,000 \textsc{LF} training simulations and evaluated on the \textsc{TLF} and \textsc{TMF} test sets, respectively. 
In the bottom panel, the emulator is trained on 1,000 \textsc{LF} plus 50 \textsc{MF} training simulations and evaluated on the \textsc{TMF} test set. 
The network trained only on the \textsc{LF} training set achieves high accuracy on the \textsc{TLF} test set (top panel), as expected. 
However, when the same model is applied to the \textsc{TMF} test set, the accuracy degrades at small scales (middle panel).
By contrast, augmenting the training set with only 50 \textsc{MF} models (5\% of the \textsc{LF} sample size) substantially improves the overall accuracy and restores high accuracy for the \textsc{TMF} test cosmologies (bottom panel).

\begin{figure}[htbp]
  \centering
  \includegraphics[width=\linewidth]{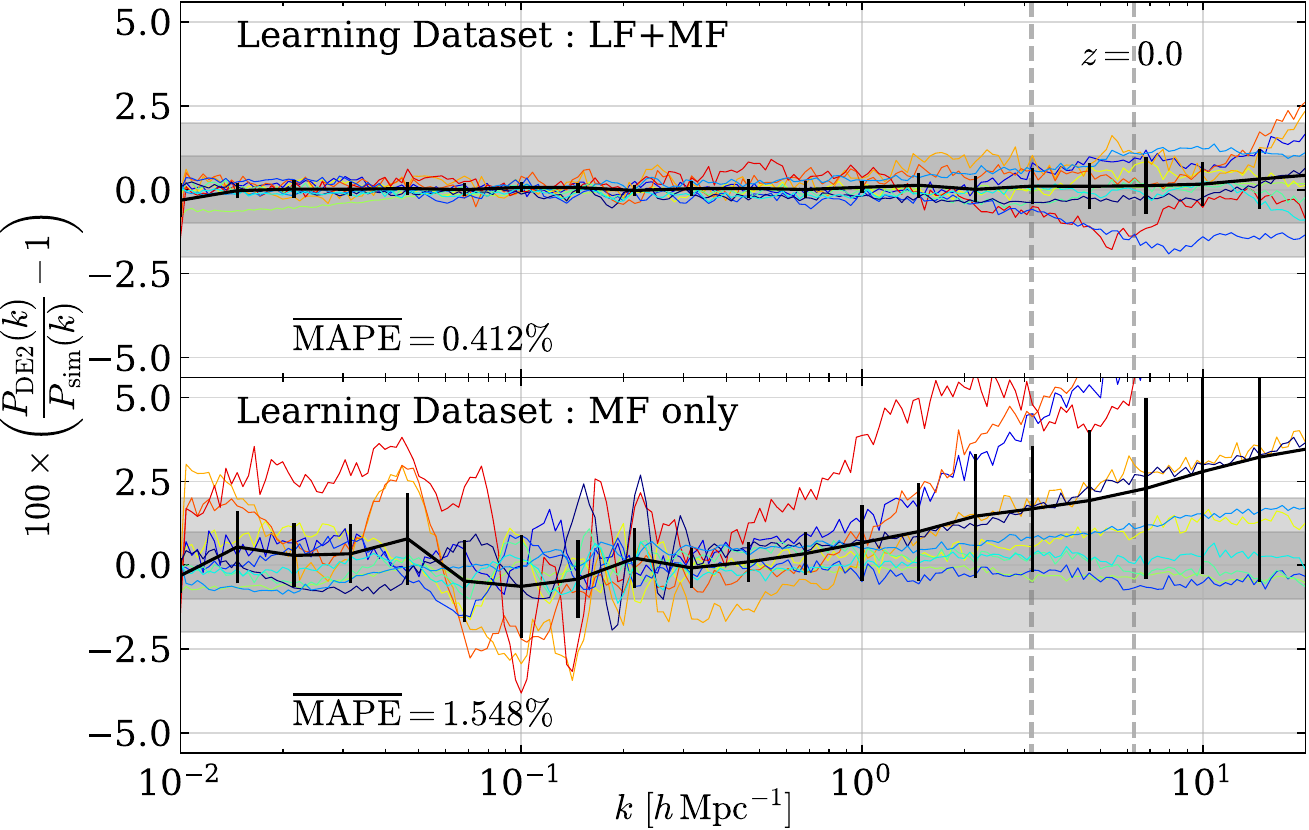}
  \caption{Accuracy on the \textsc{TMF} test set. The upper panel shows the emulator trained on 1,000 \textsc{LF} + 50 \textsc{MF} models, and the lower panel shows the emulator trained on 50 \textsc{MF} models only. Plotting conventions are the same as in figure~\ref{fig:pk_mixed_emu}.
  {Alt text: The mixed low- and middle-resolution training set gives residuals clustered near zero, whereas the model trained only on middle resolution data shows larger scatter and systematic deviations over much of the wavenumber range.}
  }
  \label{fig:pk_only_mid}
\end{figure}

Figure~\ref{fig:pk_only_mid} further shows that training on the 50 \textsc{MF} models alone does not yield sufficiently accurate predictions for the \textsc{TMF} test set. 
Thus, the middle-resolution (\textsc{MF}) training data alone are not sufficient. 
They become effective only when combined with the large \textsc{LF} ensemble that densely samples the cosmological parameter space.

\begin{figure}[htbp]
  \centering
  \includegraphics[width=\linewidth]{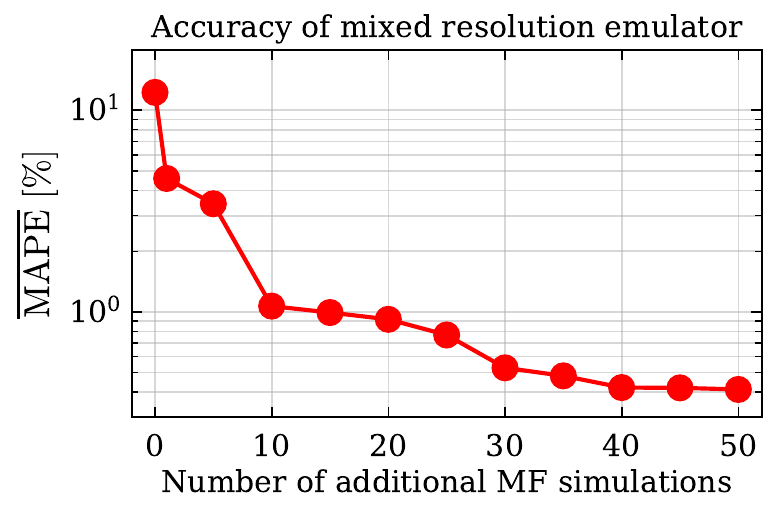}
  \caption{ $\overline{\mathrm{MAPE}}$ on the \textsc{TMF} test set at $z=0$
  as a function of the number of \textsc{MF} simulations added to the baseline
  of 1,000 \textsc{LF} simulations. 
  {Alt text: The error drops steeply when the first several middle-resolution simulations are added, then decreases more slowly and approaches a plateau after roughly forty added simulations.}
  }
  \label{fig:mr_mape}
\end{figure}

To quantify the impact of the middle-resolution training data, we train a series of mixed-resolution networks in which we increase the number of \textsc{MF} simulations while keeping the 1,000 \textsc{LF} simulations fixed. 
We evaluate the $\overline{\mathrm{MAPE}}$ on the \textsc{TMF} test set at $z=0$. 
Figure~\ref{fig:mr_mape} shows that the error decreases rapidly once $\sim 10$ \textsc{MF} simulations are added, and then declines more gradually as $N_{\mathrm{MF}}$ increases. 
It levels off at $\overline{\mathrm{MAPE}}\simeq 0.4\%$ for $N_{\mathrm{MF}}\simeq 40$--$50$. Beyond $N_{\mathrm{MF}}\sim 40$, the gains from adding further \textsc{MF} simulations are modest, indicating that the accuracy is close to saturation. 
These results indicate that including about 50 \textsc{MF} simulations is sufficient to achieve subpercent mean errors at middle resolution without a large additional computational cost.

To extend the mixed-resolution approach to our highest-resolution configuration, we add 20 \textsc{HF} simulations to the training set. 
These \textsc{HF} runs are available for 20 of the 50 \textsc{MF} cosmologies, which provide matched realizations at low, middle, and high resolution for identical cosmological parameters.
This matching helps the network learn resolution trends while reducing degeneracies with cosmology. 
These \textsc{HF} simulations also serve as high-resolution anchor points and are expected to improve the emulator fidelity at \textsc{HF} resolution, at least up to the \textsc{HF} Nyquist wavenumber.
For the 20 training cosmologies, the emulator predictions evaluated at \textsc{HF} resolution closely track the measured \textsc{HF} power spectra without obvious bias or unphysical features. 
This indicates that the mixed-resolution model can interpolate the training examples smoothly at \textsc{HF} resolution, although an independent \textsc{HF} test set is still required for a systematic validation.
Because constructing an independent \textsc{HF}-resolution test suite requires a substantial computational investment, a fully systematic validation is deferred to future work.

Table~\ref{table:mr_cost} summarizes the resource cost of the mixed-resolution suite under simple storage- and compute-based proxies, comparing it with the hypothetical all-\textsc{HF} suite of equal cosmological coverage. The storage cost per realization scales as $(N_\mathrm{p,1D}/N_{\mathrm{p,1D}}^{\textsc{LF}})^{3}$, giving relative factors of $8\times$ for \textsc{MF} and $27\times$ for \textsc{HF}. The compute cost is further multiplied by the typical increase in the number of time steps with resolution, taken here as $2\times$ for \textsc{MF} and $3\times$ for \textsc{HF}. Under these proxies, the mixed-resolution strategy reduces both storage and compute by more than an order of magnitude relative to a hypothetical all-\textsc{HF} training suite ($7.2\%$ and $4.2\%$, respectively).

\begin{table*}[htbp]
\caption{Storage and compute cost of the mixed-resolution training suite, expressed in units of one \textsc{LF} realization (i.e., the cost of a single $L_\mathrm{box}=1024\,\hiMpc$, $N_\mathrm{p}=1024^{3}$ simulation). ``Storage'' uses $N_\mathrm{p}$-proportional cost factors ($1$:$8$:$27$ for \textsc{LF}:\textsc{MF}:\textsc{HF}); ``Compute'' multiplies the storage factor by an additional time-step factor ($1$:$2$:$3$). The all-\textsc{HF} reference row assumes $1{,}000$ \textsc{HF} realizations covering the same cosmological models as the actual training suite.}
\label{table:mr_cost}
\centering
\begin{tabular}{lccc}
\hline\hline
Suite & \#runs (\textsc{LF}/\textsc{MF}/\textsc{HF}) & Storage [one-\textsc{LF}-run units] & Compute [one-\textsc{LF}-run units] \\
\hline
DQ2 mixed-resolution & $1{,}000$ / $50$ / $20$ & $1{,}940$ & $3{,}420$ \\
All-\textsc{HF} reference & $0$ / $0$ / $1{,}000$ & $27{,}000$ & $81{,}000$ \\
\hline
Fractional cost & --- & $7.2\%$ & $4.2\%$ \\
\hline\hline
\end{tabular}
\end{table*}

\subsection{Cosmic-variance-aware emulation via white-noise conditioning}
\label{subsec:cv_emu}

Most matter power spectrum emulators reduce the impact of cosmic variance in their training data by averaging over many realizations per cosmology, or by combining paired-and-fixed initial conditions~\citep{Pontzen2016-FP, Angulo2016-FP, Villaescusa-Navarro2018-FP, Klypin2020-FP} to suppress fluctuations from individual large-scale modes. The training target is then a (nearly) ensemble-averaged power spectrum, and information about the specific realization underlying each training simulation is effectively discarded.

We instead exploit the flexibility of the neural network architecture to include a cosmic variance summary of the initial conditions as an additional input feature: the binned white-noise power $W(k)$ of the initial Gaussian random field is supplied alongside the cosmological parameters, so the network can learn how the nonlinear power spectrum varies with both the cosmology and the realization-specific $\mathbf{W}$. This treatment keeps the training requirement at one finite-volume realization per cosmology across the training sets summarized in table~\ref{table:ldataset}, while still allowing realization scatter to be controlled at inference time. Section~\ref{subsubsec:cv_lrfid} characterizes the residual realization scatter that motivates the approach, Section~\ref{subsubsec:cv_whitenoise} defines the white-noise input, Section~\ref{subsubsec:cv_validation} validates the resulting cosmic variance emulator, and Section~\ref{subsec:ave_cv} develops a calibrated single-call approximation to the ensemble mean.

\subsubsection{Realization scatter in the fiducial model}
\label{subsubsec:cv_lrfid}

\begin{figure*}[htbp]
  \centering
  \includegraphics[width=0.4\linewidth]{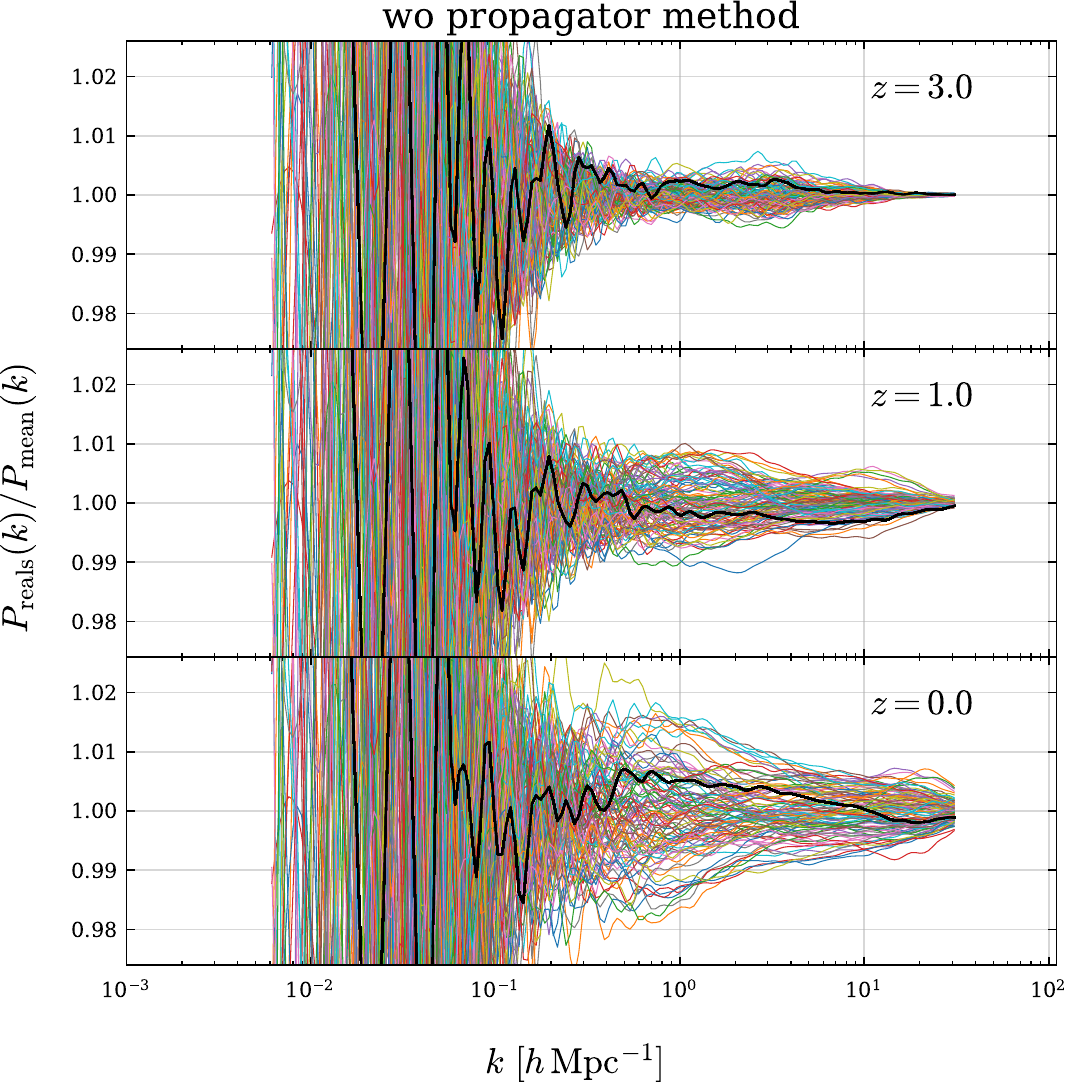}
  \includegraphics[width=0.4\linewidth]{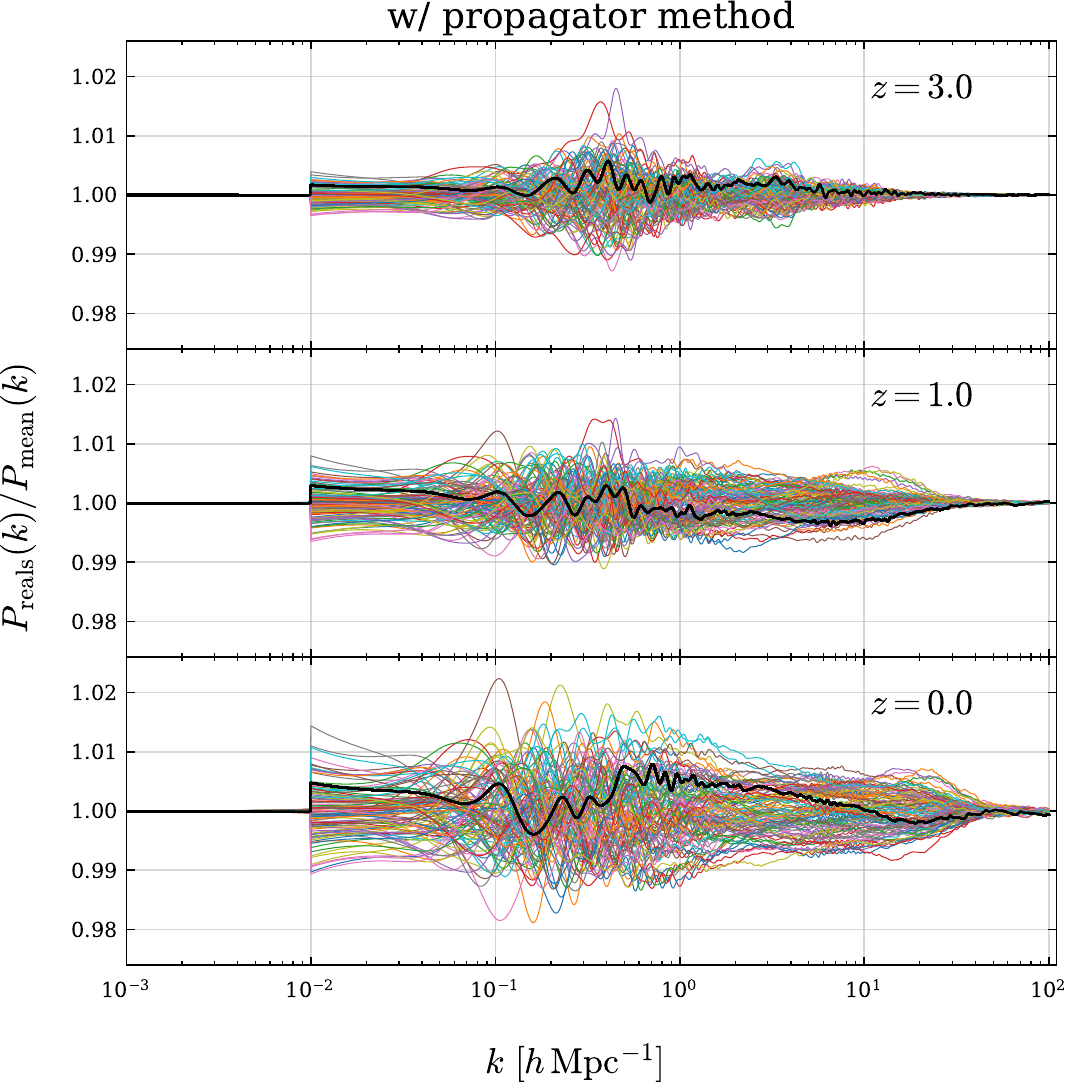}
  \caption{ Ratio of the nonlinear total matter power spectrum of each realization
  to the mean over 100 realizations of the fiducial cosmology in the low-resolution
  model \textsc{LRfid}. The left panel shows the ratios without the propagator
  method, and the right panel shows the ratios after applying the propagator
  method of \citet{Ginkaku26}. The black curves highlight the realizations
  corresponding to the fixed-phase models used in the emulator training. The impact of realization-dependent fluctuations is significantly reduced by the propagator method, but residual realization-to-realization scatter of order $2\%$ remains at the BAO scale. The apparent convergence of the curves at high $k$
  is an artificial effect caused by the approach to the constant shot noise limit
  in the \textsc{LRfid} simulations. 
  {Alt text: Without the propagator method, the realization ratios show large oscillatory scatter on broad scales. With the propagator method, the scatter is strongly reduced but residual baryon acoustic oscillation scale fluctuations remain.}
  }
  \label{fig:pk_reals_check}
\end{figure*}

Figure~\ref{fig:pk_reals_check} summarizes the impact of fluctuations between realizations in the fiducial low-resolution model \textsc{LRfid}. 
For this cosmology, we generate 100 realizations with independent Gaussian random phases in the initial conditions and, for each realization, plot the ratio of its measured nonlinear total matter power spectrum to the mean over the 100 realizations.
The left panel shows these ratios when the nonlinear power spectrum is estimated with a standard FFT-based estimator. 
The right panel shows results from our hybrid estimator~\citep{Ginkaku26}, in which large-scale modes are reconstructed with the propagator method and very small-scale modes are obtained from a direct pair-count correlation estimator. 
Even after applying this procedure, the ratios at the BAO scale exhibit residual scatter of order $2\%$ around unity.

On sufficiently large scales where nonlinear effects are negligible, the ensemble mean nonlinear power spectrum should smoothly match the linear theory power spectrum. 
In practice, we therefore replace the measured nonlinear spectrum by the linear theory prediction from the Boltzmann code for $k<0.01 \, \hMpci$ in all subsequent analyses. In the high-$k$ regime, all realizations appear to converge toward similar values. 
This behavior is not due to a genuine reduction of cosmic variance, but rather reflects the fact that the spectra in the \textsc{LRfid} setup approach the constant shot noise level
given by equation~\ref{eq:pksn}. 
Below this convergence regime, however, realization-to-realization differences remain visible even at relatively high $k$. 
From figure~\ref{fig:pk_reals_check}, we infer that the realization corresponding to the fixed-phase fiducial model used in the emulator training deviates from the 100 realization mean by approximately $\sim 1\%$ at $k \simeq 1 \, \hMpci$.

\subsubsection{White noise parametrization}
\label{subsubsec:cv_whitenoise}

The initial Gaussian random field is represented through one-dimensional white noise variables derived from its binned power spectrum, which are supplied as additional inputs to the neural network. 
This construction is intended as a reduced description of realization dependence at the power spectrum level and does not retain the full phase information of the three-dimensional field. 
In Fourier space, the complex mode amplitudes are Gaussian, and the estimated power in a spherical $k$-bin follows a $\chi^{2}$ distribution. 
In a finite periodic box, modes within the same $k$-bin can therefore be treated, to a good approximation, as independent samples of the same underlying Gaussian field.

For each simulation realization, we define the white-noise field $\delta_{\mathrm{w}}(\mathbf{x})$ corresponding to the Gaussian random field used to generate the initial conditions, normalized so that its ensemble power spectrum is unity, $P_{\mathrm{w}}(k)=1$. 
We estimate its binned power spectrum $\hat{P}_{\mathrm{w}}(k)$ from the Fourier amplitudes of $\delta_{\mathrm{w}}(\mathbf{x})$ and define the dimensionless white-noise variable as
\be 
W(k) \equiv \hat{P}_\mathrm{w}(k)\,. \label{eq:white_noise_def} 
\ee 
For Gaussian initial conditions, the normalized binned power estimator satisfies $n_{k}\,W(k) \sim \chi^{2}_{n_k}$ to a good approximation, and therefore
\be 
\langle W(k) \rangle = 1, \qquad \mathrm{Var}[W(k)] = \frac{2}{n_{k}}\, ,
\label{eq:white_noise_chi2} 
\ee
where $n_{k}$ is the effective number of degrees of freedom in the $k$-bin, set by the number of approximately independent Fourier modes contributing to that bin.

For Monte Carlo estimates of ensemble-averaged emulator outputs, we also use a simple white-noise generator based on this distribution. 
Rather than explicitly generating a full three-dimensional Gaussian field for each emulator call, we directly draw $W(k)\sim\chi^{2}_{n_k}/n_k$ independently in each $k$-bin. 
This approximation neglects correlations between different $k$-bins and retains only the binned power-level information used by the present emulator. 
When estimating ensemble averages, we draw independent samples from this generator and average the corresponding emulator outputs.

\begin{figure}[htbp]
  \centering
  \includegraphics[width=0.7\linewidth]{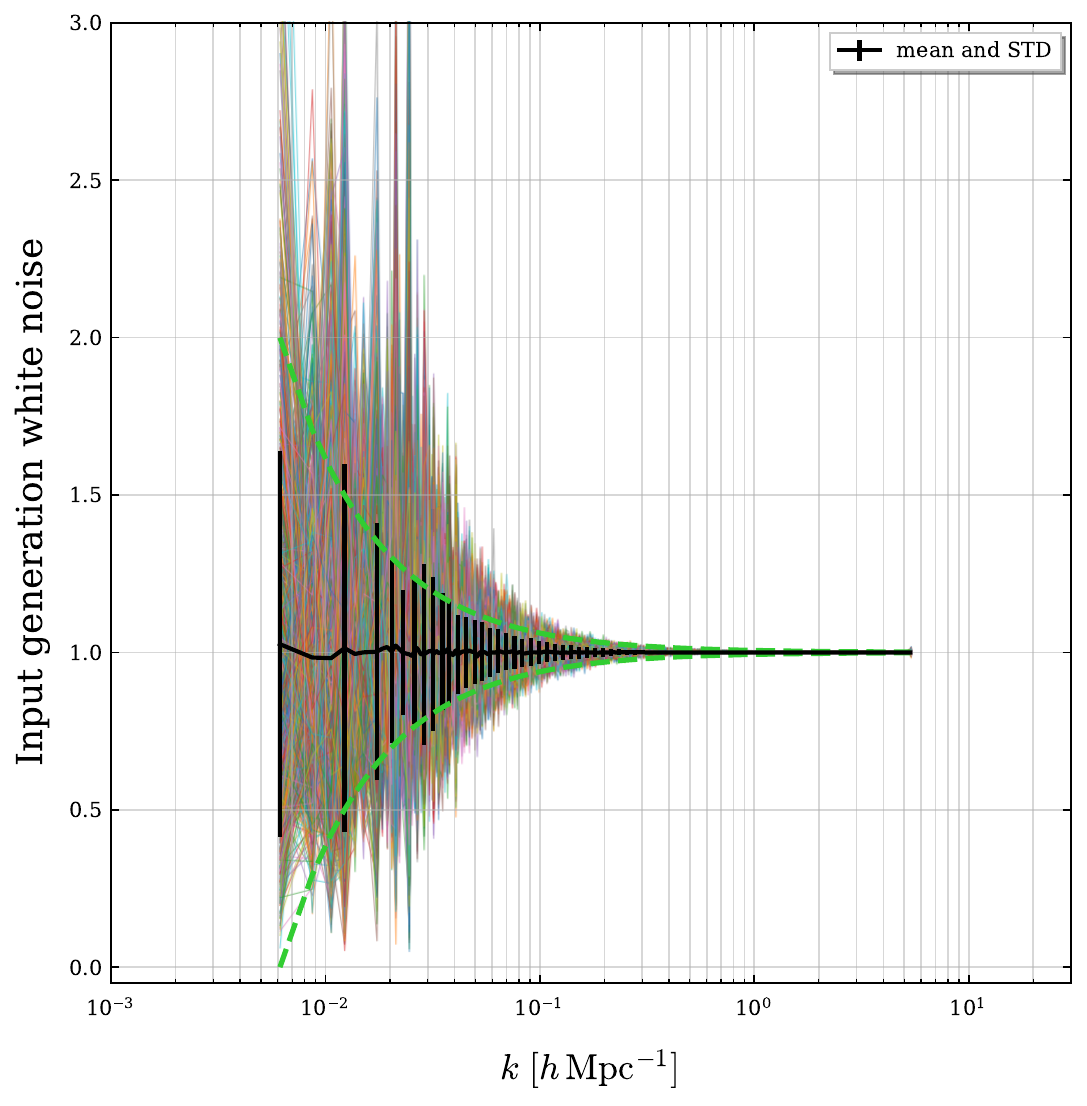}
  \caption{ White noise generator used as emulator input. For each $k$-bin, we
  generate 1,000 realizations of the white noise variable $W(k)$ drawn from a $\chi^2_{n_k}/n_k$ distribution, whose expectation value is unity. The black line with error bars shows the mean and standard deviation over the 1,000 realizations. The green dashed lines show a simple analytic envelope for the expected scatter
  based on the $\chi^{2}$ statistics with $k$-dependent degrees of freedom. 
  {Alt text: The generated white noise variables fluctuate strongly at the lowest wavenumbers and converge toward unity at higher wavenumbers as the number of contributing Fourier modes increases.}
  }
  \label{fig:white_noise}
\end{figure}

Figure~\ref{fig:white_noise} illustrates the behavior of our white noise generator based on these statistics. 
For each $k$-bin, we draw 1,000 realizations of $W(k)$ with the appropriate $k$-dependent number of $n_{k}$. 
The black line with error bars shows the mean and standard deviation over the 1,000 realizations, while the green dashed lines indicate the analytic expectation from the $\chi^{2}$ distribution.
These results confirm that the generator reproduces the correct mean and scatter implied by Gaussian initial conditions.

\subsubsection{Validation of the cosmic variance emulator}
\label{subsubsec:cv_validation}

\begin{figure*}[htbp]
  \centering
  \includegraphics[width=0.8\linewidth]{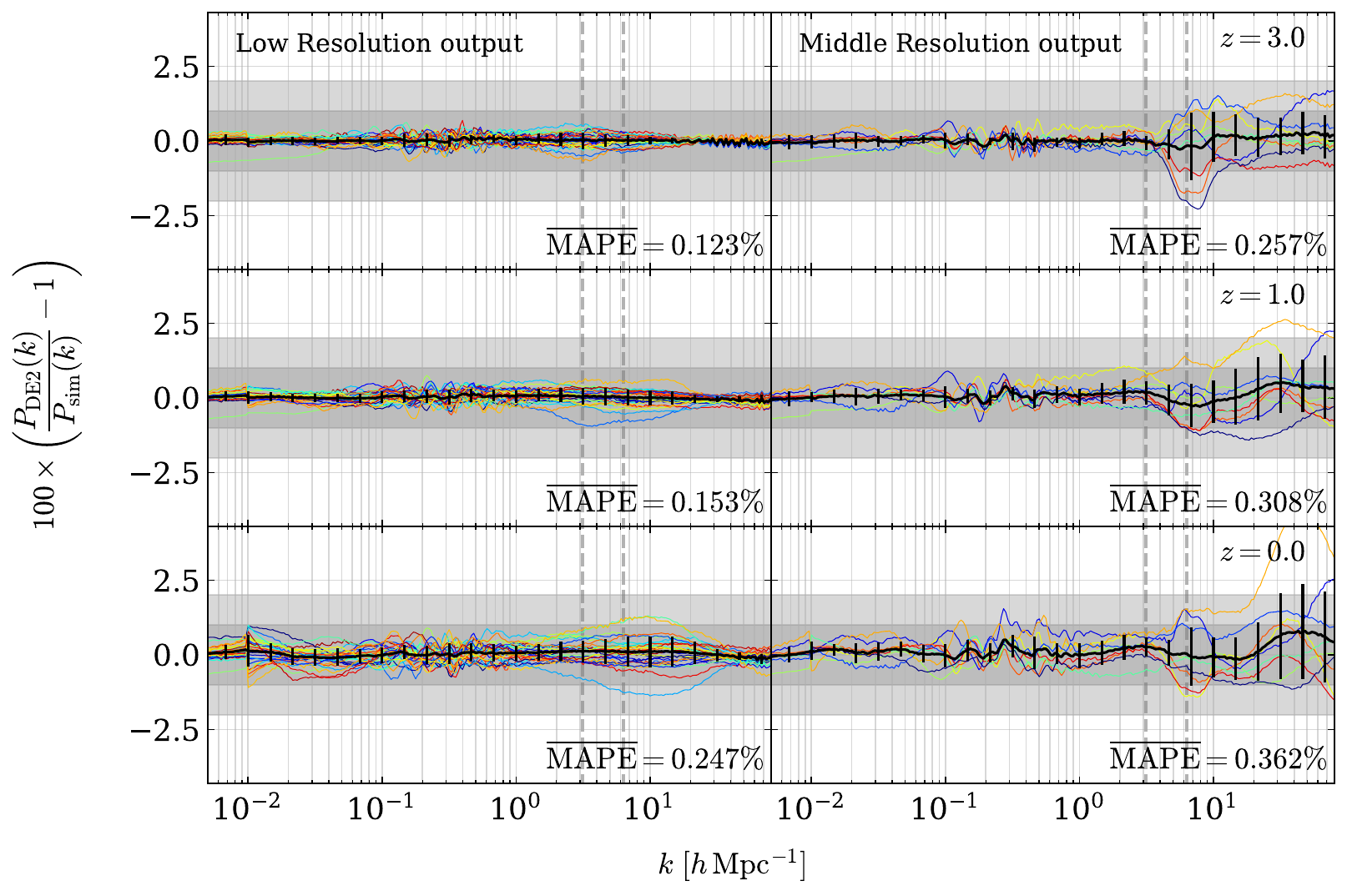}
  \caption{Conditional consistency test of the cosmic-variance-aware emulator when random-phase models are included in the training data. The left panels show results for the low-resolution training configuration (\textsc{LF}+\textsc{LR}+\textsc{LRfid}), and the
  right panels additionally include middle-resolution models (+\textsc{MF}+\textsc{MR}). For each test cosmology, we compare the mean of 10 $N$-body realizations to the mean of 10 emulator outputs evaluated at the same white noise realizations. The black line with error bars shows the mean and standard deviation of the fractional error over 10 or 30 test cosmologies. The light and dark shaded regions indicate 2\% and 1\% error ranges, respectively. Vertical dashed lines mark the $\kny$ of the low- and middle-resolution simulations from left to right. Since the power spectrum at $k>\kny$ is dominated by shot noise, we concentrate our MAPE assessment on $k<50\,\hMpci$. Note that averaging over 10 realizations is not sufficient to fully converge the ensemble mean.
  {Alt text: Residuals remain close to zero across most of the plotted range for both training configurations, with larger fluctuations near the high-wavenumber end and in the middle-resolution output panels.}
  }
  \label{fig:pk_rand_check}
\end{figure*}

\begin{figure*}[htbp]
  \centering
  \includegraphics[width=\linewidth]{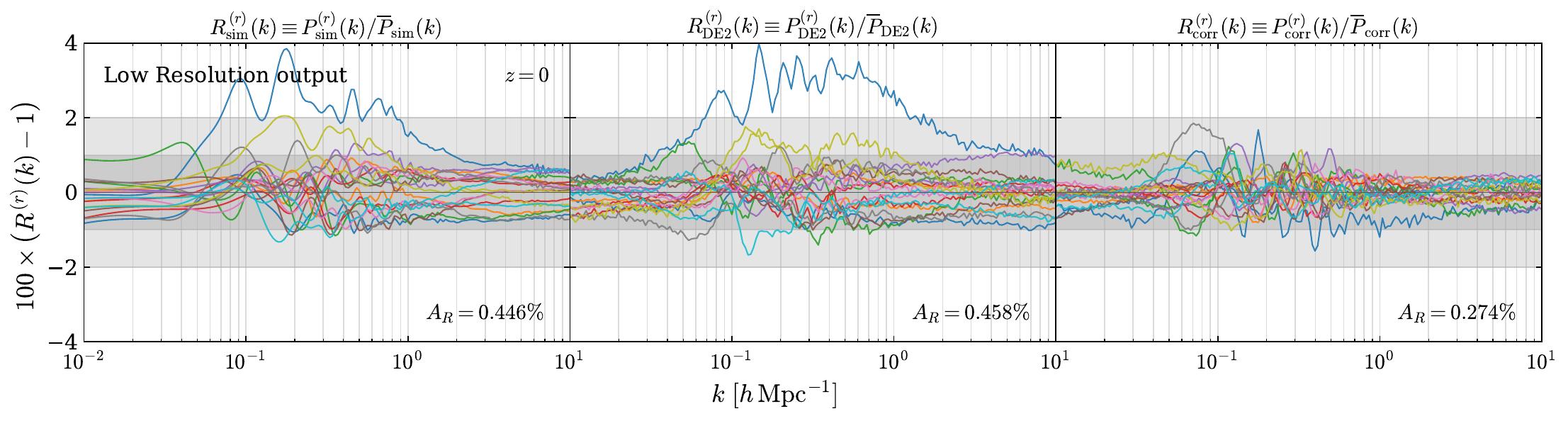}
  \includegraphics[width=\linewidth]{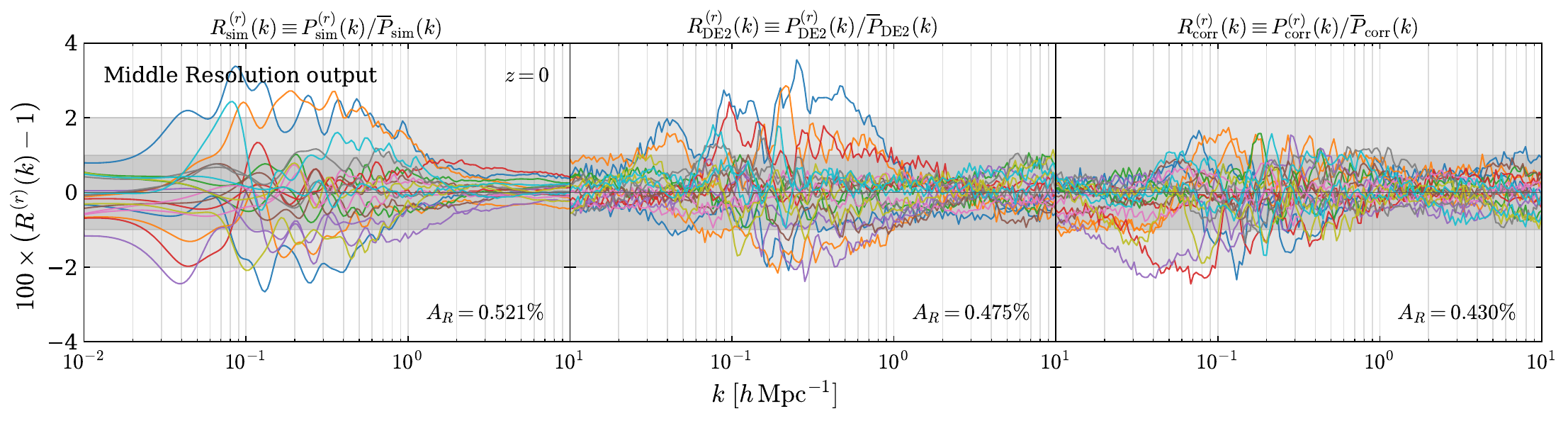}
  \caption{Realization-level response of the white-noise-conditioned emulator at $z=0$. The upper and lower rows show 20 selected curves from the low- and middle-resolution random-phase test sets, respectively, spanning different cosmological parameters and initial-condition white-noise patterns. Within each row, the panels show the simulation response, the emulator response with the same white-noise inputs, and the corrected simulation response obtained by dividing by $R_{\mathrm{DE2}}^{(r)}(k)$. Shaded bands indicate the $\pm1\%$ and $\pm2\%$ ranges, and $A_R$ denotes the mean absolute response amplitude over $0.01 \leq k \leq 10\,\hMpci$.
  {Alt text: Two sets of three panels compare realization-dependent responses in simulations, emulator predictions, and corrected simulation spectra.}
  }
  \label{fig:pk_r2r}
\end{figure*}

For the tests presented in this subsection, we do not use the white noise generator. 
Instead, we supply the emulator with the exact binned white noise vector $\mathbf{W}$ extracted from the corresponding initial conditions. 
This provides a controlled test of how much realization-dependent scatter at the power-spectrum level is captured by the truncated white-noise representation.
Combining the above ingredients, the cosmic-variance-aware emulator appends the truncated white-noise vector to the mixed-resolution input. For a white-noise vector $\mathbf{W}\equiv(W_1,\dots,W_{N_W})$ with $W_i\equiv W(k_i)$, the input vector is $\left(\boldsymbol{\theta},\, z,\, \mathbf{p}_{\mathrm{lin}},\,
N_{\mathrm{p,1D}},\, L_{\mathrm{box}},\, \mathbf{W} \right)$,
where $\mathbf{W}$ collects the white-noise variables for the first $N_W=150$ $k$ bins, corresponding to $k\simeq 6\times10^{-3}$--$1.3\,\hMpci$.

Figure~\ref{fig:pk_rand_check} summarizes the conditional performance of the emulator when the white noise variables are included as inputs. 
For each test cosmology, we generate 10 independent $N$-body realizations, compute their mean nonlinear power spectrum, and compare it with the mean of 10 emulator outputs evaluated at the corresponding exact white noise vectors $\mathbf{W}$. 
Constructing substantially larger multi-realization test sets is computationally expensive, so we adopt $N_{\mathrm{real}}=10$ as a compromise. 
Because this number is not sufficient to fully converge the ensemble mean, the comparison should be interpreted as a consistency check of the conditional emulator rather than as a strict determination of the ultimate ensemble mean accuracy. 
Within this controlled setting, the residual error scatter remains below the percent level up to the Nyquist scale for both the purely low-resolution training configuration (\textsc{LF}+\textsc{LR}+\textsc{LRfid}) and the mixed-resolution configuration (\textsc{LF}+\textsc{LR}+\textsc{LRfid}+\textsc{MF}+\textsc{MR}).
In this high-$k$ regime the power spectrum is dominated by shot noise, and the mixed-resolution training set combines simulations with different particle numbers and hence different shot noise levels. 
Together with the limited number of realizations, this means that the training data place only weak constraints on the emulator behavior in this regime. 
The most straightforward way to improve the accuracy beyond $\kny$ would be to increase the number of high-resolution simulations included in the mixed-resolution training set, which we leave to future updates of the emulator. 
In practice, the accuracy at $k>\kny$ is of limited concern for the downstream
integrals considered here, because we replace the high-$k$ plateau with a
smoothly decaying tail as described in appendix~\ref{append:extrapolation}.
We focus our quantitative error metrics, such as MAPE, on $k<50\,\hMpci$.

We next examine whether the white-noise inputs capture realization-dependent structure, rather than only reproducing the average over initial condition realizations. For each fixed cosmological parameter vector and resolution, we define the realization-averaged simulation and emulator spectra as
\be
\overline{P}_{\mathrm{sim}}(k)
\equiv
\frac{1}{N_{\mathrm{real}}}
\sum_{r=1}^{N_{\mathrm{real}}} P_{\mathrm{sim}}^{(r)}(k),
\quad
\overline{P}_{\mathrm{DE2}}(k)
\equiv
\frac{1}{N_{\mathrm{real}}}
\sum_{r=1}^{N_{\mathrm{real}}} P_{\mathrm{DE2}}^{(r)}(k),
\ee
where the average is taken over different initial condition realizations at the same cosmological parameter vector and resolution. We then define the realization responses as
\be
R_{\mathrm{sim}}^{(r)}(k)
\equiv
\frac{P_{\mathrm{sim}}^{(r)}(k)}
{\overline{P}_{\mathrm{sim}}(k)},
\qquad
R_{\mathrm{DE2}}^{(r)}(k)
\equiv
\frac{P_{\mathrm{DE2}}^{(r)}(k)}
{\overline{P}_{\mathrm{DE2}}(k)} .
\ee
If the emulator captures the realization-dependent response in the simulation, dividing the simulated spectrum by the emulator response should reduce the corresponding finite-realization fluctuation. We therefore define
\be
P_{\mathrm{corr}}^{(r)}(k)
\equiv
\frac{P_{\mathrm{sim}}^{(r)}(k)}
{R_{\mathrm{DE2}}^{(r)}(k)},
\qquad
R_{\mathrm{corr}}^{(r)}(k)
\equiv
\frac{P_{\mathrm{corr}}^{(r)}(k)}
{\overline{P}_{\mathrm{corr}}(k)} ,
\ee
where $\overline{P}_{\mathrm{corr}}(k)$ is averaged over the corrected spectra at the same cosmological parameter vector and resolution.

Figure~\ref{fig:pk_r2r} shows selected realization-level examples at $z=0$ for the low- and middle-resolution test sets. For each resolution, we select 20 individual cosmology--realization pairs from the corresponding random-phase test set to illustrate cases in which the emulator response shows a visible correspondence with the simulation response. 
These examples should therefore be regarded as illustrative diagnostics, not as a representative sample of the full test set. The quantity displayed in each panel is $100[R^{(r)}(k)-1]$. We also report $A_R \equiv 100 \times \left\langle |R^{(r)}(k)-1| \right\rangle$, where the average is taken over the displayed curves and over $0.01 \leq k \leq 10\,\hMpci$. This quantity measures the typical amplitude of the plotted realization response and is not the emulator MAPE defined in section~\ref{subsec:acc_metrics}. 
In these selected examples, the corrected responses have smaller $A_R$ than the original simulation responses, indicating that the white-noise-conditioned emulator can capture part of the realization-dependent response at the power-spectrum level in favorable cases.

Accurate realization-by-realization predictions would require a training and validation design optimized for that purpose, for example with multiple initial condition realizations per cosmology and a loss function or validation metric targeted directly at the realization response.  The present emulator is instead designed primarily for accurate and efficient ensemble-mean predictions. For this purpose, the combination of white-noise conditioning and calibrated averaging achieves the accuracy targets considered in this work.

Finally, although the binned white-noise inputs encode part of the initial
fluctuation pattern at the power-spectrum level, using this information for an inverse problem is beyond the scope of this work. Such a reconstruction would be highly degenerate, because realization-dependent fluctuations in $P(k)$ must be separated from changes caused by the cosmological parameters and by measurement or modeling uncertainties. In addition, the scale-dependent features carrying this information would need to be measured with substantially higher precision than assumed here. We therefore use the white-noise inputs only to model and average over realization dependence in the nonlinear power spectrum, not to reconstruct individual initial condition realizations.

\subsubsection{Calibrated ensemble-mean prediction}
\label{subsec:ave_cv}

\begin{figure*}[htbp]
  \centering
  \includegraphics[width=0.33\linewidth]{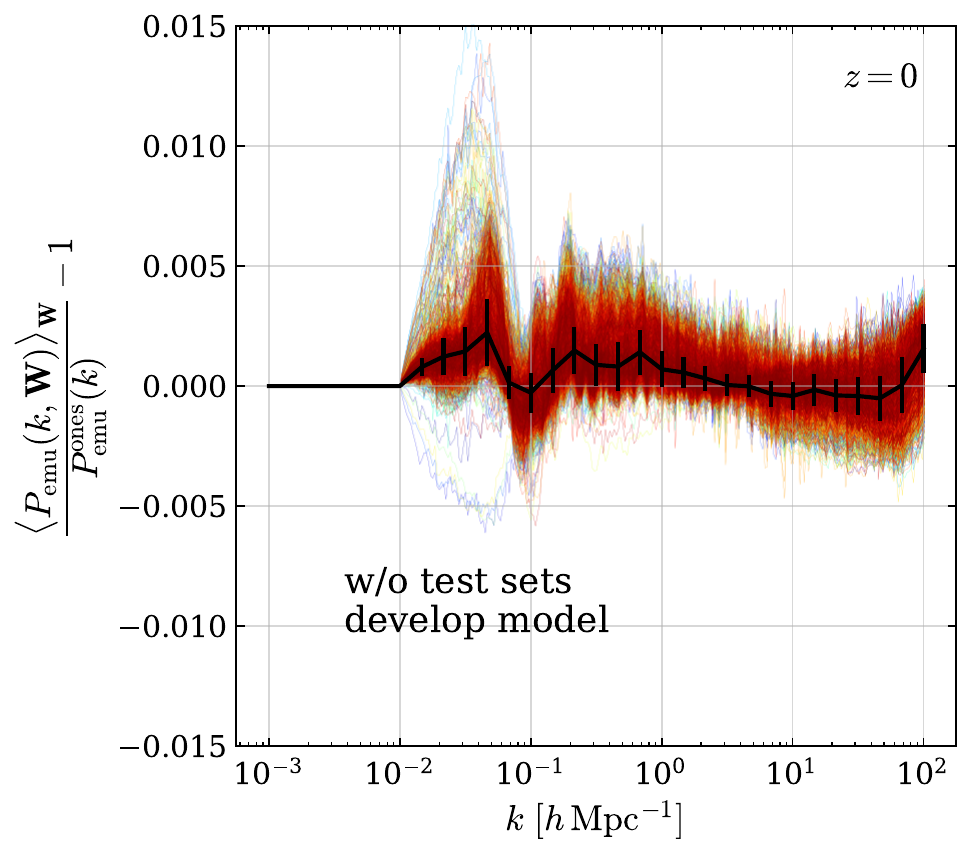}
  \includegraphics[width=0.32\linewidth]{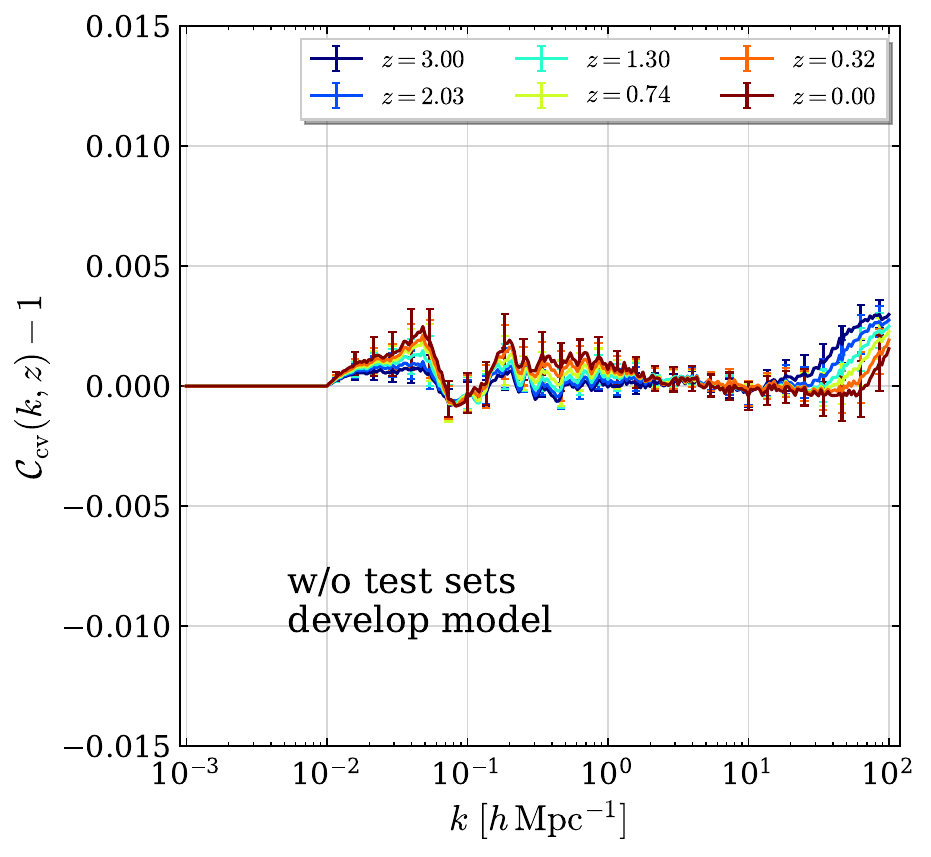}
  \includegraphics[width=0.32\linewidth]{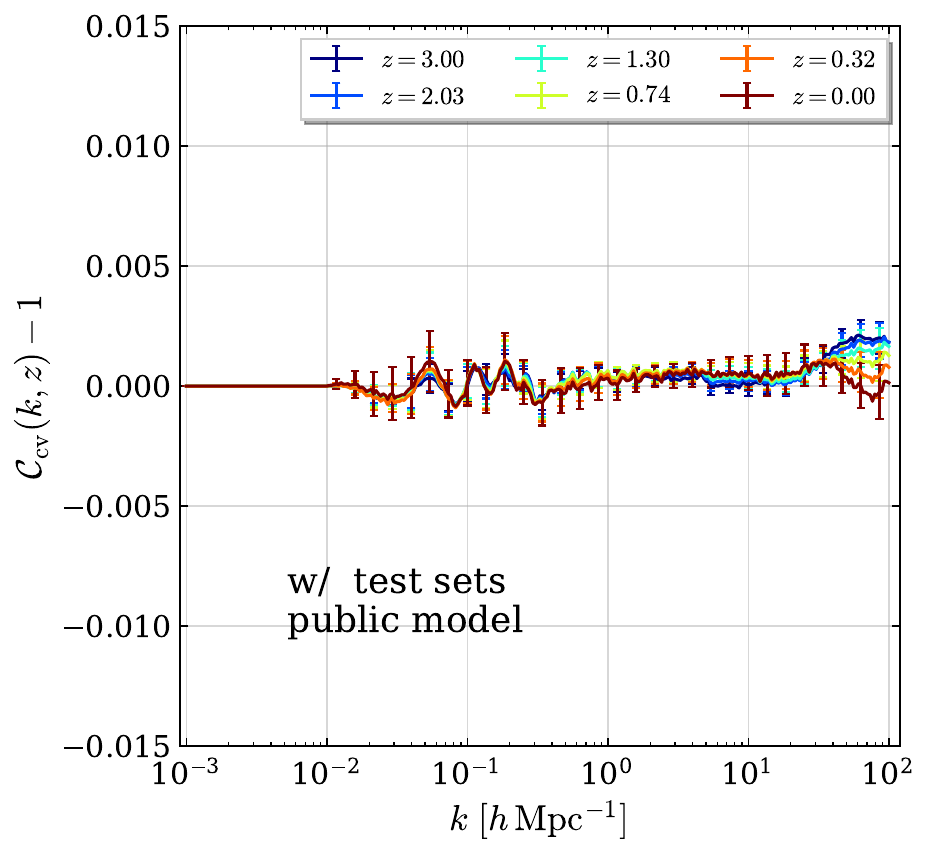}
  \caption{Comparison between the ensemble average obtained by explicitly averaging over white noise realizations and the single emulator evaluation at the all-ones white noise input.
  The left panel shows the fractional difference between the average over 500
  emulator evaluations with independent white noise inputs, $\langle P_{\mathrm{emu}}
  (k,z,\boldsymbol{\theta},\mathbf{W}) \rangle_{\mathbf{W}}$, and the emulator
  prediction evaluated once with the white noise vector set to unity, $P_{\mathrm{emu}}
  ^{\mathrm{ones}}(k,z,\boldsymbol{\theta}) \equiv P_{\mathrm{emu}}(k,z,\boldsymbol
  {\theta},\mathbf{W}\equiv\mathbf{1})$, for 5,000 randomly sampled
  cosmological models at $z=0$. The black line with error bars shows the mean and
  standard deviation of this fractional difference over the 5,000 cosmologies;
  this quantity is equal to $\mathcal{C}_{\mathrm{cv}}(k,z{=}0)-1$ as defined
  in equation~\ref{eq:cv_calibrated_factor}. Although there is some cosmology-to-cosmology variation, the overall pattern of the fractional difference is very similar across models. The
  middle and right panels show the redshift evolution of the mean calibration factor
  $\mathcal{C}_{\mathrm{cv}}(k,z)$ for the evaluation version (trained without
  the test sets) and the public-release version (retrained including the test
  sets), respectively. The vertical range is intentionally narrow, as the deviations from unity are at the $\lesssim 0.1\%$ level.
  {Alt text: The left panel shows scatter among models around a small mean correction, while the middle and right panels show smooth redshift-dependent calibration factors close to zero fractional deviation.}
  }
  \label{fig:calibration}
\end{figure*}

Once the cosmic variance emulator has been trained, the ensemble mean nonlinear power spectrum for a given cosmology can in principle be obtained by averaging the emulator outputs over many white noise realizations. 
In practice, this direct Monte Carlo average increases the evaluation cost in proportion to the number of realizations and becomes a bottleneck in parameter-inference pipelines. 
We therefore construct a faster approximation that suppresses cosmic variance while requiring only a single emulator evaluation per cosmology.

By construction, the white noise variables satisfy $\langle W(k) \rangle = 1$, so the expected input vector is the all-ones vector, $\mathbf{W}=\mathbf{1}$. 
Because the emulator mapping is nonlinear, it is not guaranteed that
\be
\langle P_{\mathrm{emu}}(k,z,\boldsymbol{\theta},\mathbf{W}) \rangle_{\mathbf{W}} 
= P_{\mathrm{emu}}(k,z,\boldsymbol{\theta},\mathbf{W}=\mathbf{1}).
\ee
To quantify this mismatch, we generate a sample of 5,000 cosmological models.
For each cosmology, we draw 500 independent white-noise realizations from the $\chi^{2}$-based generator and evaluate the emulator on each realization.
We then compute
$\langle P_{\mathrm{emu}}(k,z,\boldsymbol{\theta},\mathbf{W}) \rangle_{\mathbf{W}}$
as the mean over these 500 evaluations.
For each cosmology, we form the fractional difference between this Monte Carlo average and $P_{\mathrm{emu}}(k,z,\boldsymbol{\theta},\mathbf{W}=\mathbf{1})$.

We therefore define a calibration factor, 
\be 
\mathcal{C}_\mathrm{cv}(k,z) \equiv \left\langle
\frac{\langle P_{\mathrm{emu}}(k,z,\boldsymbol{\theta},\mathbf{W}) \rangle_{\mathbf{W}}}{P_\mathrm{emu}(k,z,\boldsymbol{\theta},\mathbf{W}= \mathbf{1})}
\right\rangle_{\boldsymbol{\theta}} \, ,
\label{eq:cv_calibrated_factor} 
\ee 
where the outer average is taken over the 5,000 cosmological models. The left panel of figure~\ref{fig:calibration} shows the fractional difference between $\langle P_{\mathrm{emu}}(k,z,\boldsymbol{\theta},\mathbf{W}) \rangle_{\mathbf{W}}$ and $P_{\mathrm{emu}}(k,z,\boldsymbol{\theta},\mathbf{W}=\mathbf{1})$ as a function of $k$ at $z=0$. The middle and right panels show the redshift evolution of the cosmology-averaged calibration factor $\mathcal{C}_\mathrm{cv}(k,z)$.
We find that the Monte Carlo average and the single-evaluation prediction at $\mathbf{W}=\mathbf{1}$ do not coincide exactly: their ratio deviates from unity at the $\sim 0.1\%$ level. Nevertheless, the pattern of this deviation is smooth. It shows only weak dependence on cosmological parameters and evolves continuously with redshift.

Using this factor, we approximate the ensemble mean nonlinear power spectrum for an arbitrary cosmology $\boldsymbol{\theta}$ as
\be
\hat{P}_\mathrm{emu}(k,z,\boldsymbol{\theta})
\equiv \mathcal{C}_\mathrm{cv}(k,z)\, P_\mathrm{emu}(k,z,\boldsymbol{\theta},\mathbf{W}= \mathbf{1})
\, . 
\label{eq:cv_calibrated} 
\ee 
Although the ratio entering the cosmology average shows weak cosmology dependence, the scatter around the mean calibration factor is at the level of a few $10^{-4}$, well below our target subpercent accuracy. We therefore adopt the cosmology-averaged calibration in this work. A cosmology-dependent calibration could further reduce this residual and may be useful for applications requiring accuracy beyond the subpercent level, but we leave this refinement for future work.

At high redshift and high $k$, the evaluation version of the emulator, which was trained without including the test sets, shows small but systematic deviations from unity in the mean calibration factor (middle panel of figure~\ref{fig:calibration}). These deviations are largely removed in the public-release version, where the emulator is retrained on the full simulation suite including the multi-realization test sets (right panel of figure~\ref{fig:calibration}). This behavior indicates that the empirical calibration is better constrained once additional fixed-cosmology realization information is included.

We use this calibration not as an independent physical model of cosmic variance, but as a calibration factor that aligns the $\mathbf{W}=\mathbf{1}$ prediction with the explicit average over white-noise realizations. Although the ratio entering the cosmology average shows weak cosmology dependence, the scatter around the mean calibration factor is at the level of a few $10^{-4}$, well below our target subpercent accuracy. We therefore adopt the cosmology-averaged calibration in this work. All validation and comparison results below use the calibrated single-call prediction in equation~(\ref{eq:cv_calibrated}) unless otherwise stated. A cosmology-dependent calibration could further reduce the residual, but we leave this refinement for future applications requiring accuracy beyond the subpercent level.

\begin{figure}[htbp]
  \centering
  \includegraphics[width=\linewidth]{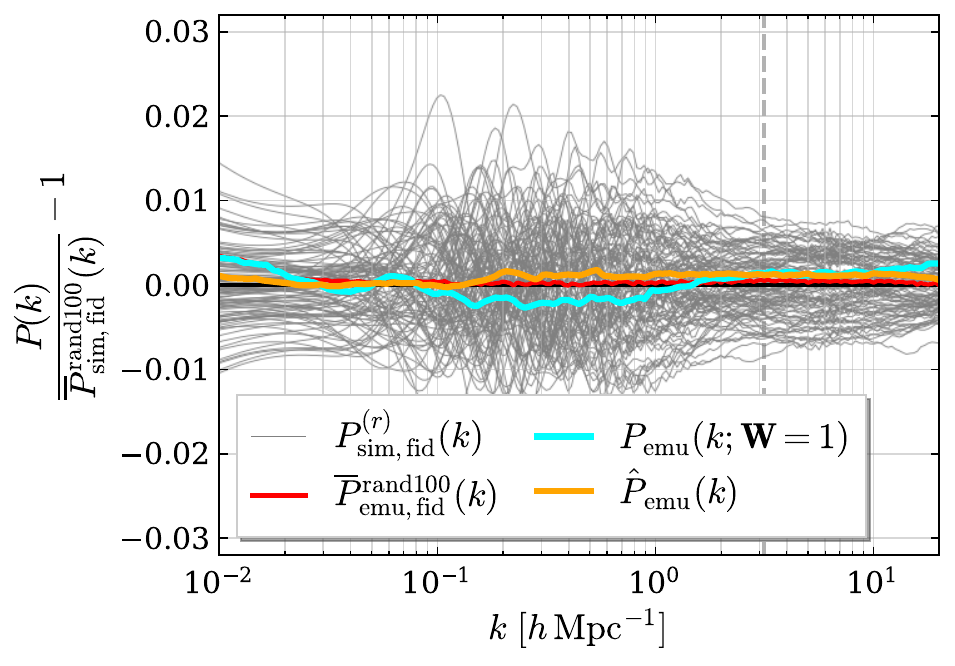}
  \caption{Validation of the calibrated fast ensemble-mean prediction for the fiducial cosmology of the low-resolution model \textsc{LRfid}. All curves are shown as ratios to the simulation mean $\overline{P}_{\mathrm{sim,fid}}^{\,\mathrm{rand100}}(k)$, minus unity. Thin gray curves show the 100 individual simulation realizations $P_{\mathrm{sim,fid}}^{(r)}(k)$. The red curve shows $\overline{P}_{\mathrm{emu,fid}}^{\,\mathrm{rand100}}(k)$, computed with the same white-noise inputs as the simulations. The cyan curve shows the single-evaluation prediction $P_{\mathrm{emu}}(k;\mathbf{W}=\mathbf{1})$, and the orange curve shows the calibrated prediction $\hat{P}_{\mathrm{emu}}(k)$ of equation~(\ref{eq:cv_calibrated}). The vertical dashed line marks the particle Nyquist wavenumber of the low-resolution model.
  {Alt text: Individual simulation realizations scatter around the fiducial mean, whereas the matched emulator average and the calibrated single-evaluation prediction remain close to zero residual over most of the plotted wavenumber range.}
  }
  \label{fig:rand_vs_fid100}
\end{figure}

It is important to verify that the calibrated fast-averaging scheme reproduces the ensemble-mean nonlinear power spectrum. We test this using the fiducial cosmology of the low-resolution \textsc{LRfid} set, for which 100 independent simulation realizations are available. We use $\overline{P}_{\mathrm{sim,fid}}^{\,\mathrm{rand100}}(k)$, the average of the 100 simulated nonlinear power spectra, as the reference ensemble mean. All emulator quantities below are evaluated at the same fiducial cosmology. We then compare three emulator-based estimates:
\begin{enumerate}
  \item[(1)] $\overline{P}_{\mathrm{emu,fid}}^{\,\mathrm{rand100}}(k)$, the average
  of 100 emulator predictions evaluated with the same white-noise inputs as the
  100 simulations;
  \item[(2)] $P_{\mathrm{emu}}(k;\mathbf{W}=\mathbf{1})$, the single-evaluation
  prediction with the white-noise input set to unity;
  \item[(3)] $\hat{P}_{\mathrm{emu}}(k)$, the calibrated single-evaluation prediction
  defined by equation~(\ref{eq:cv_calibrated}).
\end{enumerate}
Figure~\ref{fig:rand_vs_fid100} shows the ratios of these quantities to $\overline{P}_{\mathrm{sim,fid}}^{\,\mathrm{rand100}}(k)$, together with the ratios of the individual simulation realizations to the same reference mean.

The matched emulator average $\overline{P}_{\mathrm{emu,fid}}^{\,\mathrm{rand100}}(k)$ agrees with the simulation reference mean at the subpercent level over the plotted scales. The uncalibrated single-evaluation result $P_{\mathrm{emu}}(k;\mathbf{W}=\mathbf{1})$ shows a small but visible offset. After applying the calibration, $\hat{P}_{\mathrm{emu}}(k)$ closely follows the reference mean, demonstrating that the calibrated single-call prediction provides an accurate approximation to the ensemble-mean spectrum for this fiducial test.

Direct $N$-body verification of the mismatch between averaging the white noise inputs and averaging the nonlinear emulator outputs would require many realizations for the same cosmology and is therefore computationally expensive. 
Studies of the Paired-and-Fixed (PAF) method have shown that cosmic variance can be reduced efficiently, but that the covariance matrix is modified in a non-trivial, scale- and redshift-dependent manner~\citep{Pontzen2016-FP,Angulo2016-FP,Villaescusa-Navarro2018-FP,Klypin2020-FP}. 
These results, together with our calibration tests, show that any procedure intended to suppress cosmic variance must be validated explicitly. 
In our approach, the fast-averaging scheme approximates the nonlinear ensemble mean with one evaluation at $\mathbf{W} = \mathbf{1}$ and corrects the residual mismatch with the precomputed factor $\mathcal{C}_{\mathrm{cv}}(k,z)$. Over the parameter domain used for calibration, the remaining cosmology dependence of $\mathcal{C}_{\mathrm{cv}}$ is more than an order of magnitude smaller than our target subpercent accuracy.

\section{Applications}
\label{sec:apps}

This section compares the nonlinear matter power spectra predicted by \DE\ with widely used halo model fitting formulas and public emulators, and then demonstrates an application to cosmic shear observables.
Throughout this section, we use the public version of the emulator, retrained on the full simulation suite, including the training and test sets summarized in tables~\ref{table:ldataset} and~\ref{table:tdataset}, as described in section~\ref{subsec:acc_metrics}. 
The comparisons in this section are therefore application-level demonstrations, not out-of-sample validation tests.
We also apply the calibration table for cosmic variance effects (right panel of figure~\ref{fig:calibration}) throughout this section. 
Unless stated otherwise, all comparisons use the total matter power spectrum rather than the ``cb'' component.

\subsection{Comparison with fitting formulas}

We compare the predictions of \DE\ with the nonlinear fitting formulas \textsc{Halofit}~\citep{Smith03,Takahashi12} and \textsc{HMcode}~\citep{Mead2015-mn,Mead2016-vv,Mead2021-og}, evaluated using a consistent \textsc{Class} implementation.

The \textsc{Takahashi12} revision of \textsc{Halofit} recalibrated the original model~\citep{Smith03} against then state-of-the-art high-resolution $N$-body simulations covering $\LCDM$ and $\wCDM$ cosmologies, based on WMAP-era constraints and the Coyote Universe suite~\citep{Coyote1,Coyote2,Coyote3}. 
Its reported accuracy is typically about $5\%$ for $k<1\,\hMpci$ and about $10\%$ for $1<k<10\,\hMpci$. \textsc{HMcode2016} extended the halo model framework to massive neutrinos and modified gravity, with calibration mainly over the Coyote Universe Extended parameter range~\citep{Coyote_ex} in a $\wnCDM$ cosmology, and reported few-percent accuracy up to $k\lesssim 10\,\hMpci$. 
\textsc{HMcode2020} further updated the halo model parameters and introduced $\wwnCDM$ support, with calibration informed by the Mira-Titan Universe simulations and a phenomenological baryonic feedback model. 
In this work, we do not enable the baryonic feedback correction and use the dark matter only configuration. 
Unless noted otherwise, we refer to \textsc{HMcode2020} simply as \textsc{HMcode}.

\begin{figure}[htbp]
  \centering
  \includegraphics[width=\linewidth]{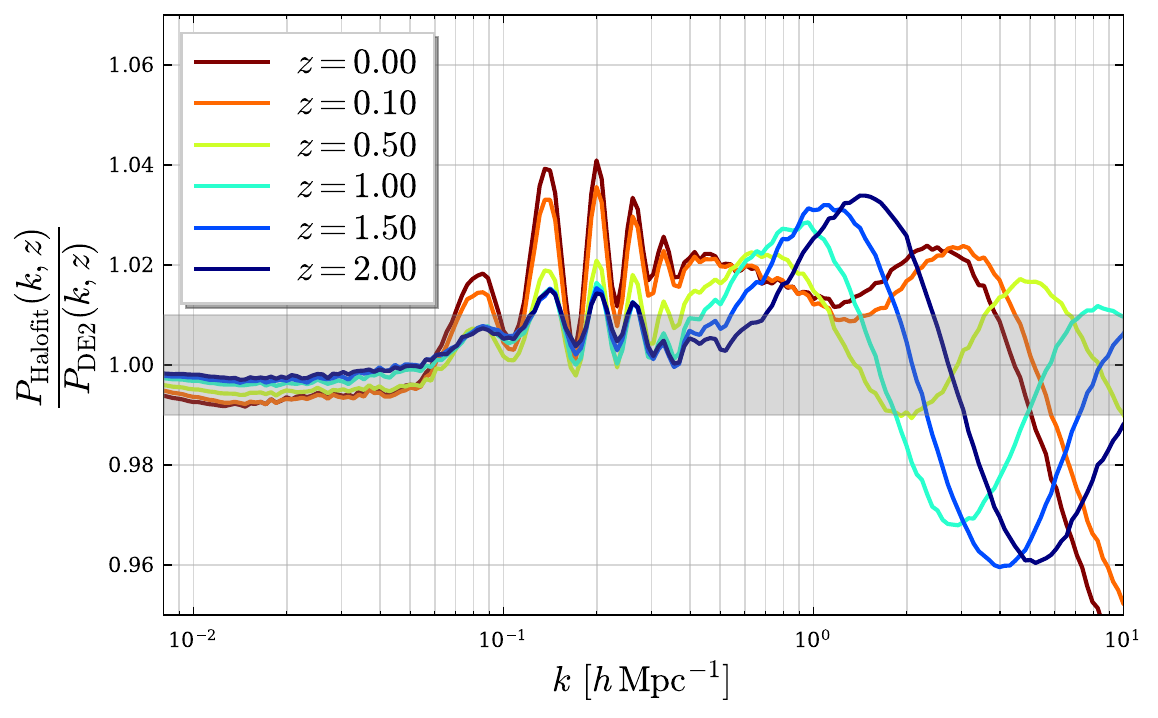}
  \includegraphics[width=\linewidth]{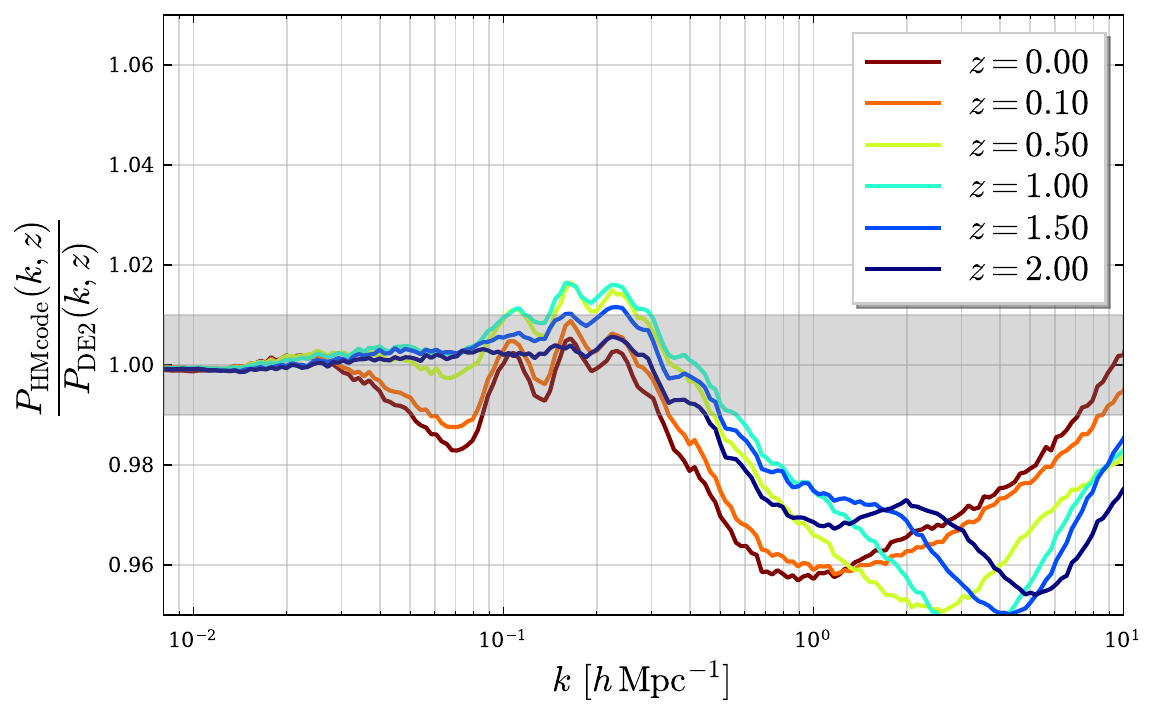}
  \caption{Comparison at the DQ2 fiducial cosmology between \DE\ and nonlinear fitting formulas. 
  Top panel: \textsc{Takahashi12} \textsc{Halofit}. Bottom panel: \textsc{HMcode} (\textsc{Class} implementation). 
  Line colors indicate redshift, and the shaded bands denote the $\pm1\%$ range. Around the BAO range ($k\sim0.1$ to $0.5\,\hMpci$), \textsc{HMcode} shows smaller oscillatory residuals than \textsc{Halofit}.
  {Alt text: Halofit shows stronger oscillatory residuals around baryon acoustic oscillation scales, while HMcode gives smoother deviations of about one percent before both fitting formulas depart more strongly at smaller scales.}
  }
  \label{fig:comp_fitting_fid}
\end{figure}

Figure~\ref{fig:comp_fitting_fid} compares the nonlinear total matter power spectrum $P_{\mathrm{tot}}(k)$ predicted by \DE\ with the fitting formulas at the DQ2 fiducial cosmology. For both \textsc{Halofit} and \textsc{HMcode}, the discrepancies are typically within about $5\%$ for $k<1\,\hMpci$ and increase to about $10\%$ at lower redshift once $k>1\,\hMpci$. Similar residual patterns were also reported by \citet{EuclidEmu2}. 
Around the BAO range, \textsc{HMcode} suppresses oscillatory residuals more effectively than \textsc{Halofit}, with deviations at the percent level from $k\sim0.1$ to $0.5\,\hMpci$, whereas several-percent differences remain at $k>0.5\,\hMpci$ relative to \DE.

\begin{figure*}[htbp]
  \centering
  \includegraphics[width=\linewidth]{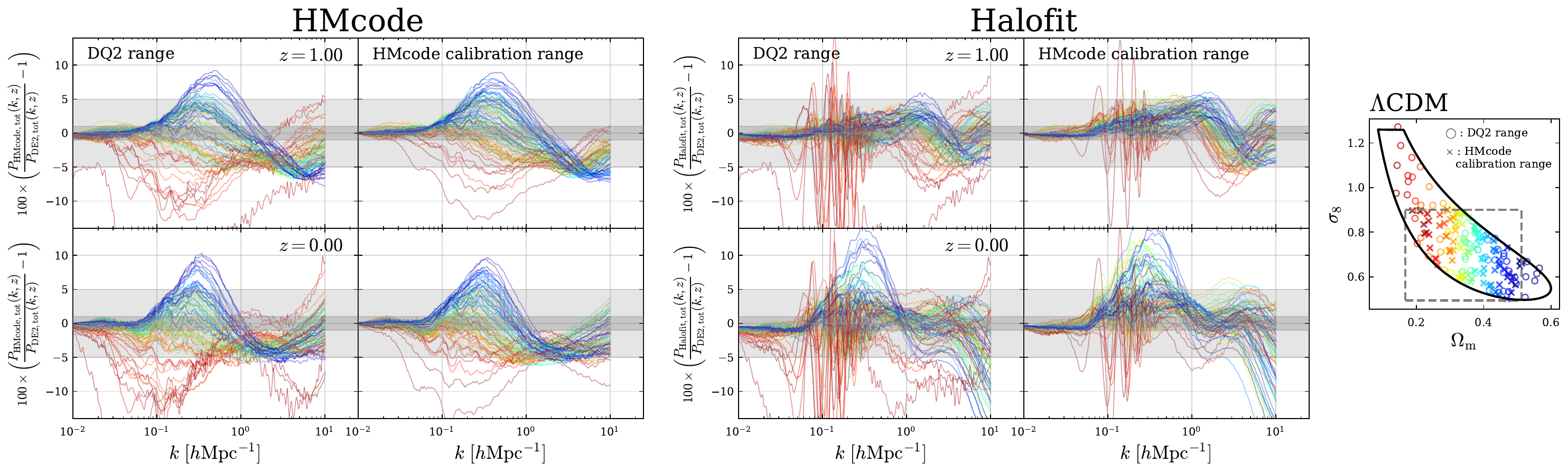}
  \includegraphics[width=\linewidth]{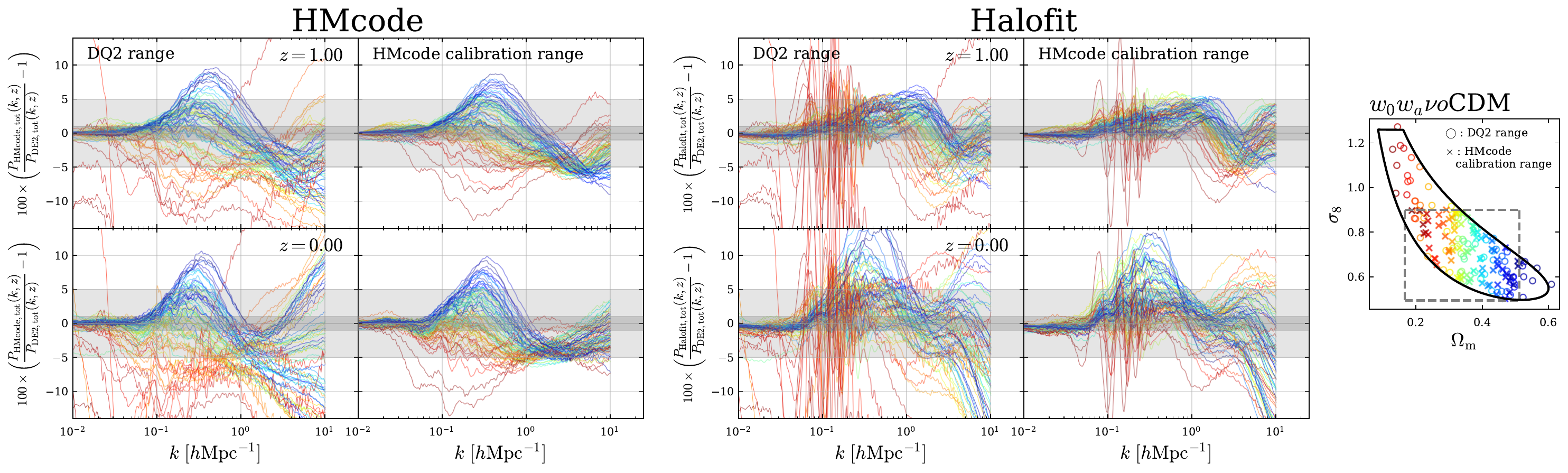}
  \caption{Comparison of \DE\ with nonlinear fitting formulas for 80 randomly sampled cosmologies. The upper block shows $\LCDM$ models, and the lower block shows extended dark energy models. In each block, the left large panel shows \textsc{HMcode}, and the middle large panel shows \textsc{Halofit}. Within each large panel, the upper row corresponds to $z=1$ and the lower row to $z=0$; the left subcolumn uses cosmologies sampled across the DQ2 range, and the right subcolumn uses cosmologies sampled within the \textsc{HMcode} calibration domain. In the lower block, the DQ2-range sample is drawn from $\wwnoCDM$, whereas the \textsc{HMcode}-domain sample is drawn from $\wnCDM$. The rightmost panels show the sampled points in the $\Om$--$\sigma_{8}$ plane. Circles mark the DQ2-range sample, and crosses mark the \textsc{HMcode}-domain sample. Marker colors encode $\Om$ and match the colors used in the ratio panels. The light and dark shaded bands indicate $\pm5\%$ and $\pm1\%$ deviations, respectively.
  {Alt text: The ratio panels show larger and more structured deviations outside the HMcode calibration domain, with particularly broad scatter for some low matter density samples and at nonlinear wavenumbers.}
  }
  \label{fig:comp_fitting_param}
\end{figure*}

To assess the robustness of the fitting formulas across cosmological parameter space, figure~\ref{fig:comp_fitting_param} presents comparisons for 80 randomly sampled cosmologies. 
The upper block shows $\LCDM$ models, while the lower block shows the extended dark energy cases. Within each block, the left subcolumns use cosmologies sampled across the DQ2 range, whereas the right subcolumns use cosmologies restricted to the \textsc{HMcode} calibration domain (see figure~\ref{fig:emus_ranges1}); in the lower block, this restricted sample is drawn from $\wnCDM$. 
Neither \textsc{Halofit} nor \textsc{HMcode} is formally calibrated over the full $\wwnoCDM$ parameter space, but both are often extrapolated beyond their nominal calibration domains in cosmological analyses~(e.g., \cite{Planck2018cosmo,Hamana2020-bh,KiDS1000:Heymans}). 
As expected, their accuracy degrades substantially outside the calibrated domains.

\subsection{Comparison with public emulators}
\label{subsec:comp_emu}

This subsection compares the predictions of \DE\ with several public nonlinear matter power spectrum emulators.
We first summarize the main features of these emulators and their training sets, and then quantify their agreement with \DE\ at the DQ2 fiducial cosmology and across the cosmological parameter space.

\subsubsection{Overview of public nonlinear $P(k)$ emulators}
\label{subsubsec:emu_overview}

\begin{figure}[htbp]
  \centering
  \includegraphics[width=0.9\linewidth]{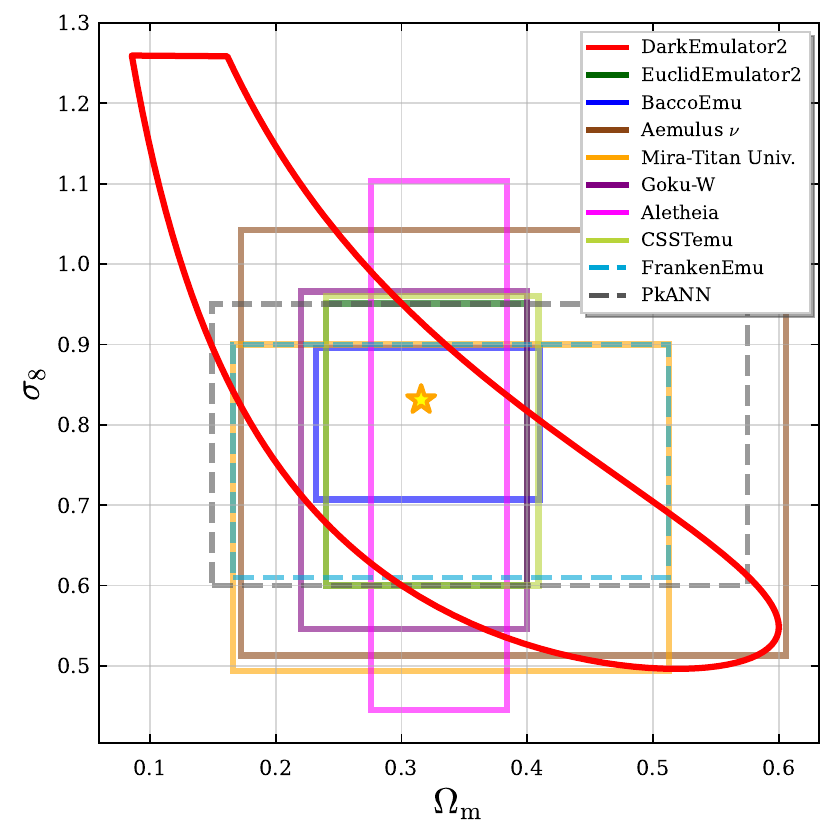}
  \caption{Parameter ranges on the $\Om$-$\sigma_{8}$ plane for this work and
  other public emulators. For several emulators that do not explicitly include
  $\Om$ or $\sigma_{8}$ as input parameters, we derive these quantities from
  their native cosmological parameters. The star symbol indicates the \textit{Planck}~2015
  best-fit cosmology and the fiducial cosmology of DQ2. 
  {Alt text: Support regions of public emulators mostly occupy narrower central parts of the broader DarkEmulator2 domain, with different shapes reflecting different native parameter choices and derived mappings.}
  }
  \label{fig:emus_ranges1}
\end{figure}

\begin{figure*}[htbp]
  \centering
  \includegraphics[width=0.8\linewidth]{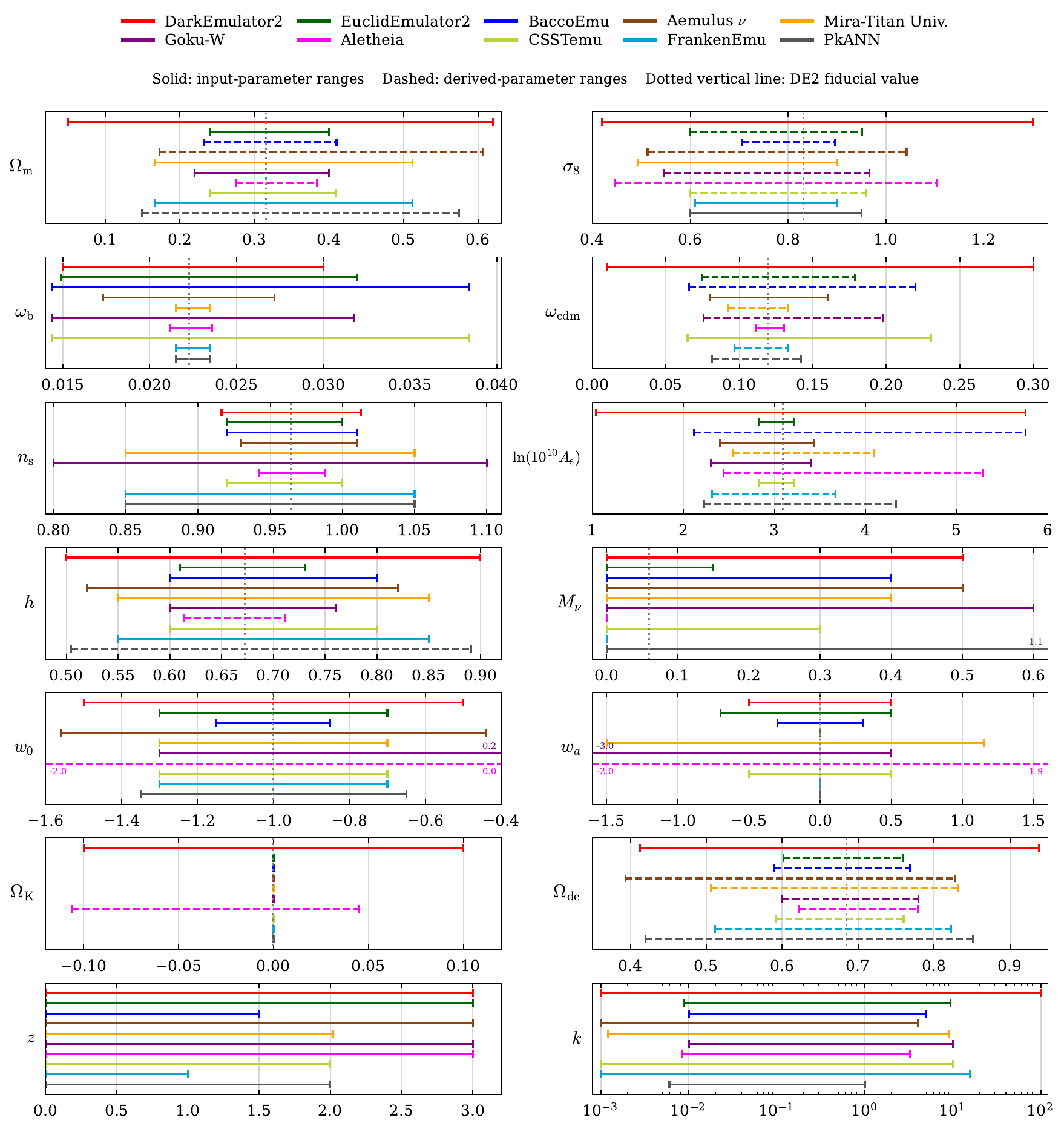}
  \caption{Parameter, wavenumber, and redshift ranges for \DE\ and other public nonlinear matter power spectrum emulators considered in this work. 
  For each parameter, horizontal segments show the support range covered by each emulator. Solid lines denote the ranges of the native input parameters used in the design of each emulator. Dashed lines indicate the ranges of derived parameters inferred from over 1,000 Monte Carlo samples of the input parameter space. Where necessary, we convert between physical densities and density parameters using $\omega_{i}\equiv \Omega_{i}h^{2}$ (equivalently, $\Omega_{i}= \omega_{i}/ h^{2}$). 
  The gray vertical dotted line marks the fiducial cosmology adopted for \DE. For parameters extending beyond the plotted range, the corresponding endpoint values are printed near the associated bar.
  Note that \DE\ can generate matter power spectra up to $k = 100\,\hMpci$ by using direct particle-pair correlations on small scales, although the accuracy degrades once shot noise becomes dominant. 
  {Alt text: Horizontal range bars show that emulator support varies strongly by parameter, wavenumber, and redshift. DarkEmulator2 covers a broad cosmological domain and extends to higher wavenumbers than most public emulators.}
  }
  \label{fig:emus_ranges2}
\end{figure*}

We consider the following public nonlinear matter power spectrum emulators: \EE\footnote{\url{https://github.com/miknab/EuclidEmulator2}}~\citep{EuclidEmu1,EuclidEmu2},
\BCE\footnote{\url{https://bitbucket.org/rangulo/baccoemu/}}~\citep{BACCO,BACCO_bias,BACCO_baryon},
\AEMU\footnote{\url{https://github.com/AemulusProject/aemulus_nu_public}}~\citep{Aemulus1,Aemulusnu},
\FEMU\footnote{\url{https://www.hep.anl.gov/cosmology/CosmicEmu/emu.html}}~(\CEMU ver.2; \cite{Coyote_ex}) and \textsc{Mira-Titan Universe}\footnote{\url{https://github.com/lanl/CosmicEmu}}
~(\CEMU ver.3.1; \cite{MiraTitan1,MiraTitan2,MiraTitan3,MiraTitan4}),
\GOKUN\footnote{\url{https://github.com/astro-YYH/GokuNEmu}}~\citep{Goku,GokuN},
\ALETHEIA\footnote{\url{https://gitlab.mpcdf.mpg.de/arielsan/aletheia}}~\citep{Aletheia},
\CSST\footnote{\url{https://github.com/czymh/csstemu}}~\citep{CSST-emu-I},
and \PA\footnote{\url{https://zuserver2.star.ucl.ac.uk/~fba/PkANN/}}~\citep{PkANN1,PkANN2}.
Our main comparison focuses on recent emulators trained on high-resolution simulation suites, \EE\ (EE2), \BCE\ (Bacco), \AEMU\ (AE$\nu$), \textsc{Mira-Titan Universe IV} (MTU~IV), \GOKUN\ (GokuN), \ALETHEIA, and \CSST\ (CSSTemu), whose training simulations have resolutions comparable to those used in the \DE\ training data sets.
In contrast, \FEMU\ and \PA\ are included only as historical references for fiducial cosmology tests, since they were developed in the early stage of cosmological emulation and adopt different resolutions and cosmological models.
Table~\ref{table:emus} summarizes the key specifications of the emulators considered in this comparison.
Figure~\ref{fig:emus_ranges1} presents their support coverage in the $\Om$-$\sigma_{8}$ plane, while figure~\ref{fig:emus_ranges2} shows the one-dimensional ranges for each cosmological parameter, wavenumber, and redshift.

\EE supports the $\wwnCDM$ cosmology. 
It employs the $N$-body code \textsc{Pkdgrav3}~\citep{potter2017}, which uses a parallel fast multipole method~(FMM; \cite{Greengard1987-ao,Dehnen2000-sb,Dehnen2002-ek}). 
A total of 254 cosmological models were generated using the Paired-and-Fixed (PAF) technique, yielding 508 high-resolution simulations with $N_{\mathrm{p}}=3000^{3}$ particles in a box of $\Lbox=1 \,\hiGpc$. 
The learning model combines principal component analysis (PCA) with polynomial chaos expansion (PCE)~\citep{Xiu2002-PCE}. 
In \EE, the emulator is trained on a nonlinear correction (NLC) factor $B(k,z)$ in equation~\ref{eq:B}, and the nonlinear matter power spectrum is reconstructed by combining this factor with the linear power spectrum. 
Moreover, the emulator adopts the $N$-body gauge~\citep{2015PhRvD..92l3517F}, consistent with that used in \DE.

\BCE\ supports the $\wwnCDM$ cosmology. 
It employs high-resolution simulations performed with the \textsc{L-Gadget3} code~\citep{gadget2,Angulo12}, using $N_{\mathrm{p}}=4320^{3}$ particles within a box of $\Lbox=1440 \,\hiMpc$. 
Three representative cosmologies were simulated, each with PAF realizations (six runs in total). 
To explore cosmological dependence beyond these simulations, the cosmology-rescaling technique~\citep{angulo10} was applied, effectively extending the parameter coverage to 800 models from which the nonlinear power spectra were derived. 
Instead of directly emulating the nonlinear power spectrum, \BCE\ learns the logarithmic ratio of $P(k,z)$ to a BAO-smoothed linear power spectrum $P^{\mathrm{smeared-BAO}}_{\mathrm{lin}}(k,z)$.
Two regression schemes are implemented in \BCE: a GP model and a NN model. 
The GP variant is used only to optimize the sampling design, while the public nonlinear $P(k)$ emulator adopts a fully connected NN. 
For the amplitude parameter, \BCE\ can accept either $\sigma_{8,\mathrm{cb}}$ or $\As$; in this work we adopt $\As$ for consistency across comparisons.
The underlying simulations reach a Nyquist wavenumber of $\kny \simeq 9.4\,\hMpci$, but this emulator is calibrated only up to $k_{\max}=\kny/2$.

\MTU\ supports the $\wwnCDM$ cosmology. 
It provides nonlinear matter power spectra across a wide dynamical range, combining high-resolution simulations performed with the \textsc{Hacc} code~\citep{Habib2016-jj,Habib2016-qq}, 16 particle-mesh (PM) realizations for lower-resolution coverage, and Time-Renormalization Group (TimeRG) perturbation theory~\citep{Pietroni:2008qy,Lesgourgues2009-cw} for large-scale modes. 
The emulator was trained on 111 cosmological models. 
The underlying simulation campaign follows a set of nested LHDs, enabling sequential extensions of the training set and systematic improvements in the emulator accuracy. 
The high-resolution simulations adopt $N_{\mathrm{p}}=3200^{3}$ particles in a box of side length $L=2100\,\mathrm{Mpc}$. The nonlinear power spectra are represented using a principal component basis, and the coefficients are modeled with GP regression.

\AEMU\ supports the $\wnCDM$ cosmology. 
Unlike the other emulators considered here, it does not include a time-varying dark energy parameter $w_{a}$. The base simulation consists of $N_{\mathrm{cb}}=1400^{3}$ CDM+baryon particles and $N_{\nu}=1400^{3}$ massive neutrino particles within a box of $L=1050\,\hiMpc$, evolved using the \textsc{Gadget-3} code. 
\AEMU\ explicitly incorporates massive neutrinos as separate particle species, thereby capturing nonlinear backreaction effects between the cold and neutrino components. 
The emulator employs a surrogate modeling approach applied to the Hybrid Effective Field Theory (HEFT) framework~\citep{Modi2020-HEFT}, which combines $N$-body simulations with perturbative modeling. 
The simulated HEFT power spectra are further refined using the Zel'dovich Control Variates (ZCV) technique~\citep{Kokron2022-mh,DeRose2023-wn} to suppress cosmic variance. 
The training is performed not on the HEFT power spectrum itself but on the logarithmic ratio between the HEFT prediction and the one-loop Lagrangian perturbation theory (LPT) result. 
The resulting model is constructed using PCA combined with PCE.

\GOKUN\ is based on the Goku simulation suite, which samples a ten-dimensional $N_{\mathrm{eff}}\alpha_{\mathrm{s}}\wwnCDM$ parameter space. 
The simulation design uses two sliced LHDs~\citep{Ba2015-jp}. 
The wide-box suite Goku-W broadly covers this ten-dimensional space, whereas the narrower-box suite Goku-N focuses on the region around the \textit{Planck}~2018 constraints. Each of Goku-W and Goku-N covers 564 cosmological models, each with a pair of low-fidelity simulations at two distinct resolutions, supplemented by 21 (Goku-W) and 15 (Goku-N) high-fidelity simulations.
The high-fidelity runs use $N_{\mathrm{p}}=3000^{3}$ particles in a box of side length $L_{\mathrm{box}}=1000\,\hiMpc$, and the low-fidelity levels, L1 and L2, each use $N_{\mathrm{p}}=750^{3}$ particles. 
L1 shares the same box size as the high-fidelity runs, whereas L2 is run in smaller volumes with $L_{\mathrm{box}}=250\,\hiMpc$ to improve the resolution on small scales. 
All simulations are performed with the \textsc{MP-Gadget} code~\citep{Bird2022-MPGadget,Ni2022-MPGadget,MP-GADGET}, which includes massive neutrinos via a linear response scheme on the particle-mesh grid.
The first emulator built on this suite, \GOKU, is a graphical multi-fidelity Gaussian-process (GMGP) model~\citep{Ji2021-dh}. 
In this work we instead use the neural network version, \GOKUN, which employs a fully connected multi-fidelity deep neural network. 
\GOKUN builds two separate multi-fidelity emulators for the low- and high-$k$ ranges, trained on the L1+HF and L2+HF spectra, respectively. 
The final prediction is obtained by smoothly combining the two outputs over the overlap region $0.025\,\hMpci < k < 1.5\,\hMpci$. 
The neural network takes the ten cosmological parameters as input and predicts the blended high-fidelity nonlinear matter power spectrum on a $(k,z)$ grid, while the low-fidelity spectra serve as auxiliary training targets, allowing the many low-resolution simulations to constrain the sparse high-resolution ones. 
In the comparisons presented in this paper, we fix $N_{\mathrm{eff}}=3.046$ and $\alpha_{\mathrm{s}}=0$ and use the usual equal-mass approximation for the neutrino sector, while the underlying \GOKUN\ training simulations assume the normal mass hierarchy.

\ALETHEIA\ employs two GP emulators trained on the \textsc{AletheiaEmu} simulation suite. 
The boost emulator $\varepsilon_{B}(k)$ is trained on 100 cosmological models sampled with a four-dimensional LHD in $(\ob,\oc,\ns,\sigma_{12})$, covering $\pm5\sigma$ around the \textit{Planck}~2018 constraints and $0.2 \le \sigma_{12} \le 1.0$. 
The parameter $\sigma_{12}$ is a derived quantity that encodes the impact of the background and clustering parameters $(\ob,\oc,\ns,h,\As,w_{0},w_{a},\ode,\ok)$ on the growth of structure. 
Each model is realized with a PAF $N$-body simulation performed with \textsc{Gadget-4}~\citep{gadget4}, using $N_{\mathrm{p}}=2048^{3}$ CDM particles in a box of $L_{\mathrm{box}}=1500\,\mathrm{Mpc}$.
The boost emulator predicts the ratio of the nonlinear matter power spectrum to a ``de-wiggled'' linear spectrum $P_{\mathrm{lin}}^{\mathrm{dw}}(k)$, in which the nonlinear BAO damping is approximately accounted for.
The derivative emulator uses the same 100 base cosmologies, supplemented with five additional lower-resolution simulations per cosmology in which the evolution parameters are varied, to model the response to changes in an integrated growth-history parameter $\tilde{x}$ and to predict the corresponding derivative. In these simulations, $\As$ and $h$ are fixed to the \textit{Planck}~2018 base-$\Lambda$CDM best-fit values, so that evolution parameters such as $w_{0}$, $w_{a}$, $\ode$, and $\ok$ affect the nonlinear power spectrum only indirectly through their impact on $\sigma_{12}$ and the growth-history parameter. 
Final predictions incorporate a small-scale resolution-correction factor $C(k,\sigma_{12})$ derived from the higher-resolution companion suite \textsc{AletheiaMass}. 
The calibrated \ALETHEIA\ prediction yields nonlinear matter power spectra over $0.006 \lesssim k \lesssim 2\,\mathrm{Mpc}^{-1}$. 
In the comparisons presented in this paper, we restrict \ALETHEIA\ to spatially flat, massless-neutrino cosmologies ($\Ok=0$, $M_{\nu}=0$) for consistency with the other emulator comparisons.

\CSST\ is based on the Kun simulation suite~\citep{CSST-emu-I} and targets a spatially flat eight-parameter $\wwnCDM$ space. The suite consists of 129 high-resolution simulations performed with a modified version of \textsc{Gadget-4}, using
$3072^{3}$ particles in a $1\,\hiGpc$ box. The matter power spectrum emulator compresses the training spectra with PCA and interpolates the resulting coefficients with Gaussian process regression. Its final model emulates the ratio of the nonlinear power spectrum to a \textsc{HMcode2020} prediction~\citep{Mead2021-og}, rather than the nonlinear spectrum itself.

Large-scale cosmic variance is reduced using fixed-amplitude initial
conditions together with two matched \textsc{FastPM} simulations per
cosmology~\citep{FastPM}, and the remaining residual fluctuations are smoothed before emulation. The public emulator provides both total and cb matter power spectra. In the comparisons below, we adopt the degenerate three-species massive-neutrino convention to match the equal-mass neutrino treatment used in \DE. The published validation reports approximately 1\% accuracy for $k\leq 10\,\hMpci$ and $z\leq 2$. The broader \CSST program also provides other statistics, including halo mass functions and correlation functions~\citep{CSST-emu-II,CSST-emu-III}, but in this work we use only the matter power spectrum module.

\begin{sidewaystable*}[htbp]
  \caption{Public power spectrum emulators. The abbreviations in this table are
  as follows: NN=feed-forward neural network: PCA=principal component analysis:
  PCE=polynomial chaos expansion: GP=Gaussian process: GMGP=graphical multi-fidelity
  Gaussian process: WSDs=weighted sequential designs: LHDs=Latin hypercube
  designs: PT=perturbation theory: W,N=Goku-W,N: HR,MR,LR=high,middle,low-resolution.}
  \label{table:emus}
  \centering
  \scalebox{0.72}{
  \begin{tabular}{lllllllll}
    \hline
    \hline
    Emulator                    & \DE                                                                      & \EE                                   & \BCE                                                                        & \MTU                                       & \AEMU                                                  & \GOKUN                                                                     & \ALETHEIA                                                                                             & \CSST                                                                  \\
    \hline
    Cosmology                   & $\wwnoCDM$                                                               & $\wwnCDM$                             & $\wwnCDM$                                                                   & $\wwnCDM$                                  & $\wnCDM$                                               & $N_{\mathrm{eff}}\alpha_{\mathrm{s}}\wwnCDM$                               & $w_{0}w_{a}o \mathrm{CDM}^{\dagger\dagger}$                                                        & $\wwnCDM$                                                        \\
                                & \scriptsize $[\Om, \oc, \ob, h]_{3}, [\sigma_{8},\As,\lnAs]_{1},$        & \scriptsize $\Om, \ob, M_{\nu}, \ns,$ & \scriptsize $[\sigma_{8,\mathrm{cb}}, \As]_{1}, \Omega_{\mathrm{cb}}, \ob,$ & \scriptsize $\om, \ob, \omega_{\nu}, \ns,$ & \scriptsize $\oc, \ob, M_{\nu}, \ns, w_{0},$           & \scriptsize $\Om, \Ob, h, \As, \ns,$                                       & \scriptsize $\oc, \ob, \ns$                                                                        & \scriptsize $\Omega_{\mathrm{cb}}, \Omega_{\mathrm{b}}, H_{0}, \ns,$ \\
                                & \scriptsize $[\Ok \Ode]_{1}\, , M_{\nu}, \ns, w_{0}, w_{a}$ $^{\dagger}$ & \scriptsize $w_{0}, w_{a}, h, \As$    & \scriptsize $\ns, w_{0}, w_{a}, h, M_{\nu}$                                 & \scriptsize $w_{0}, w_{a}, h, \sigma_{8}$  & \scriptsize $10^{9}\As, H_{0}$                         & \scriptsize $w_{0}, w_{a}, M_{\nu}, N_{\mathrm{eff}}, \alpha_{\mathrm{s}}$ & \scriptsize $\sigma_{12}(\scriptscriptstyle \oc, \ob, \ns, h, \As, w_{0}, w_{a}, \ok)$ \vspace{0.2cm} & \scriptsize $\As, w_{0}, w_{a}, M_{\nu}$ \vspace{0.2cm}              \\
    Code                        & \textsc{Ginkaku}                                                         & \textsc{Pkdgrav3}                     & \textsc{L-Gadget3}                                                          & \textsc{Hacc}                              & \textsc{Gadget-3}                                      & \textsc{MP-Gadget}                                                         & \textsc{Gadget-4} \vspace{0.2cm}                                                                    & \textsc{Gadget-4} \vspace{0.2cm}                              \\
    Resolution (best)           & $L_{\mathrm{box}}=1000 \,\hiMpc$                                         & $L_{\mathrm{box}}=1000 \, \hiMpc$     & $L_{\mathrm{box}}=1440 \, \hiMpc$                                           & $L_{\mathrm{box}}=2100 \, \mathrm{Mpc}$    & $L_{\mathrm{box}}=1050 \, \hiMpc$                      & $L_{\mathrm{box}}=1000 \, \hiMpc$                                          & $L_{\mathrm{box}}=1500 \, \mathrm{Mpc}$ \vspace{0.1cm}                                             & $L_{\mathrm{box}}=1000\,\hiMpc$ \vspace{0.1cm}                         \\
                                & $N_{\mathrm{p}}=3000^{3}$                                                & $N_{\mathrm{p}}=3000^{3}$             & $N_{\mathrm{p}}=4320^{3}$                                                   & $N_{\mathrm{p}}=3200^{3}$                  & $N_{\mathrm{p}}=1400^{3}(\mathrm{cb}) + 1400^{3}(\nu)$ & $N_{\mathrm{p}}=3000^{3}$                                                  & $N_{\mathrm{p}}=2048^{3}$ \vspace{0.2cm}                                                           & $N_{\mathrm{p}}=3072^{3}$ \vspace{0.2cm}                                \\
    $\kny \,\, [\hMpci]$ (best) & $9.42$                                                                   & $9.42$                                & $9.42$                                                                      & $4.78 \, h^{-1}$                           & $4.19$                                                 & $9.42$                                                                     & $4.28 \, h^{-1}$ \vspace{0.1cm}                                                                    & $9.65$ \vspace{0.1cm}                                                   \\
    Gauge                       & $N$-body                                                                 & $N$-body                              & --                                                                          & --                                         & --                                                     & --                                                                         & -- \vspace{0.2cm}                                                                                  & -- \vspace{0.2cm}                                                       \\
    Cosmic Variance             & Propagator \& Averaging                                                  & Paired-and-Fixed                      & Paired-and-Fixed                                                            & Combine TimeRG PT                          & Zel'dovich control variate                             & Paired-and-Fixed                                                           & Paired-and-Fixed \vspace{0.2cm}                                                                    & Fixed amplitude + $2$ \textsc{FastPM} \vspace{0.2cm}                       \\
    $N_{\mathrm{model}}$        & $1000$                                                                   & $254$                                 & $800$ ($3$ original model)                                                  & $111$                                      & $150$                                                  & $564$(W),$564$(N)                                                          & $100$ \vspace{0.2cm}                                                                               & $129$ \vspace{0.2cm}                                                    \\
    $N_{\mathrm{realization}}$  & $20,50,1000$ (HR,MR,LR)                                                  & pair                                  & pair                                                                        & $1,16,16$ (HR,LR,PT)                       & $1$                                                    & pair                                                                       & pair \vspace{0.2cm}                                                                                & $1$ HR + $2$ FastPM \vspace{0.2cm}                                                      \\
    $N_{\mathrm{simulation}}$   & $40,100,2100$ (HR,MR,LR)                                                 & $508$                                 & $6$                                                                         & $111,1776$ (HR,LR)                         & $150$                                                  & $36,2256$ (HR,LR)                                                          & $200,500,50^{\ddagger\ddagger}$ \vspace{0.2cm}                                                     & $129$ HR + $258$ FastPM  \vspace{0.2cm}                                                    \\
    $N_{\mathrm{train}}$        & $40,100,2100$ (HR,MR,LR)$^{\ddagger}$                                    & $254$                                 & $800$                                                                       & $111$ (combine HR,LR,PT)                   & $150$                                                  & $36,1128$ (HR,LR)                                                          & $100,500^{\ddagger\ddagger}$ \vspace{0.2cm}                                                        & $129$ \vspace{0.2cm}                                                    \\
    \hline
    $z_{\max}$                  & $3$                                                                      & $3$                                   & $1.5$                                                                       & $2.02$                                     & $3$                                                    & $3$                                                                        & $3$ \vspace{0.2cm}                                                                                 & $3^{\mathsection}$ \vspace{0.2cm}                                      \\
    $k$ $[\hMpci]$              & $[0.001, 100]$                                                           & $[0.01, 10]$                          & $[0.01, 5]$                                                                 & $[0.001,5] \, h^{-1}$                      & $[0.001,4]$                                            & $[0.006, 10]$                                                              & $[0.006, 2.0] \, h^{-1}$ \vspace{0.2cm}                                                            & $[0.006,10]$ \vspace{0.2cm}                                             \\
    Output                      & $P_{\mathrm{tot}}(k),P_{\mathrm{cb}}(k)$                                 & $P_{\mathrm{tot}}(k)$                 & $P_{\mathrm{tot}}(k),P_{\mathrm{cb}}(k)$                                    & $P_{\mathrm{tot}}(k),P_{\mathrm{cb}}(k)$   & $P_{\mathrm{tot}}(k)$                                  & $P_{\mathrm{tot}}(k)$                                                      & $P_{\mathrm{tot}}(k)$ \vspace{0.1cm}                                                               & $P_{\mathrm{tot}}(k),P_{\mathrm{cb}}(k)$ \vspace{0.1cm}                \\
                                & $P_{\mathrm{lin,tot}}(k),P_{\mathrm{lin,cb}}(k)$                         & $P_{\mathrm{lin,tot}}(k)$             & $P_{\mathrm{lin,tot}}(k),P_{\mathrm{lin,cb}}(k)$                            & --                                         & --                                                     & $P_{\mathrm{lin,tot}}(k)$                                                  & -- \vspace{0.2cm}                                                                                  & $P_{\mathrm{lin,tot}}(k),P_{\mathrm{lin,cb}}(k)$ \vspace{0.2cm}        \\
    \hline
    Regression                  & $\mathrm{NN}$                                                            & $\mathrm{PCA|PCE}$                    & $\mathrm{NN}$ (or $\mathrm{PCA|GP}$)                                        & $\mathrm{PCA|GP}$                          & $\mathrm{PCA|PCE}$                                     & $\mathrm{NN}$ (or $\mathrm{GMGP}$)                                         & $\mathrm{GP}$ \vspace{0.2cm}                                                                       & $\mathrm{PCA|GP}$ \vspace{0.2cm}                                       \\
    Sampling                    & WSDs                                                                     & LHDs                                  & LHDs                                                                        & nested LHDs                                & LHDs                                                   & sliced LHDs                                                                & LHDs \vspace{0.2cm}                                                                                & Sobol sequence \vspace{0.2cm}                                           \\
    \hline
    \hline
  \end{tabular}
  }

  \begin{minipage}{0.98\linewidth}
    \scriptsize $^{\dagger}$ Brackets with a subscript indicate the number of alternative
    parameter choices within each group in the cosmology model.\\
    $^{\ddagger}$ $N_{\mathrm{train}}$ in \DE\ is the number of models used for
    verification; the public-release version additionally includes 330 low-resolution test simulations
    and 110 middle-resolution test simulations.\\
    $^{\dagger\dagger}$ For \ALETHEIA, the emulator is trained in the four-dimensional
    space $(\ob,\oc,n_{s},\sigma_{12})$; the effects of $(w_{0},w_{a},\ok)$
    enter only through the evolution-mapping framework and are not explicitly
    sampled in the training set.\\
    $^{\ddagger\ddagger}$ For \ALETHEIA, the first and second entries in the
    $N_{\mathrm{simulation}}$ and $N_{\mathrm{train}}$ rows correspond to the simulations
    used to train the boost and derivative emulator, respectively; in the $N_{\mathrm{simulation}}$
    row, the third entry denotes the high-resolution \textsc{AletheiaMass} suite
    used only to define the resolution-correction factor.\\
    $^{\mathsection}$ For \CSST, the public interface supports evaluations up to
    $z=3$, whereas the published validation reports approximately 1\% accuracy for
    $z\leq 2$ and $k\leq 10\,\hMpci$.
  \end{minipage}
\end{sidewaystable*}

These emulators surveyed above differ in their simulation codes, training designs, and cosmic variance treatments, but methodologically they group into a small number of regression families. PCE, GP, and NN represent three distinct surrogate-model classes used in matter power spectrum emulation. PCE expands the target response in an orthogonal polynomial basis, GPs place a Gaussian-process prior over functions and provide both mean predictions and predictive uncertainty estimates, and NNs learn flexible nonlinear mappings through layered affine transformations and nonlinear activation functions. 
The public emulators considered in this work span these three model classes.

\subsubsection{Comparison for the fiducial cosmology}
\label{subsubsec:comp_fid_param}

\begin{figure*}[htbp]
  \centering
  \includegraphics[width=\linewidth]{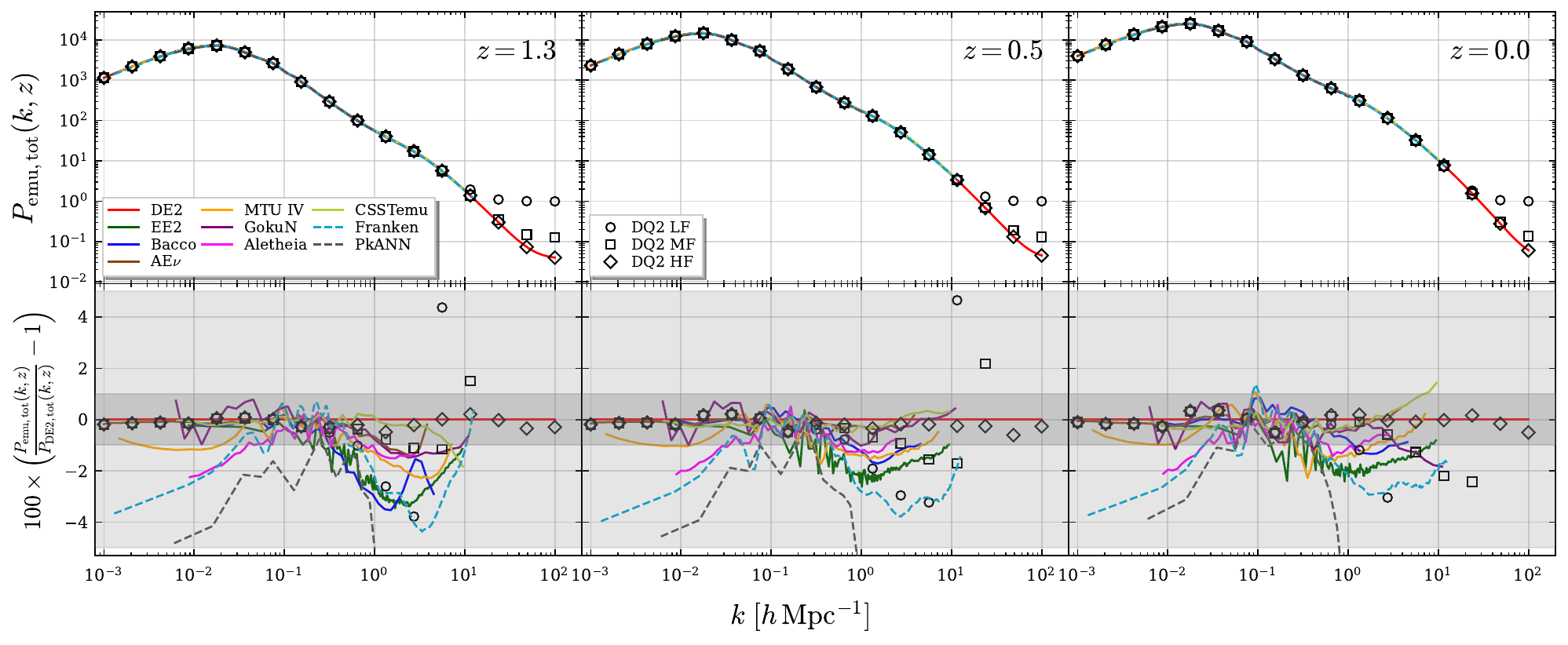}
  \caption{ Comparison of the total matter power spectra at the DQ2 fiducial cosmology.
  The top panels show $P_{\mathrm{tot}}(k)$ from each emulator, and the bottom panels show the ratios relative to \DE. From left to right, the redshift is $z=1.3$, $0.5$, and $0.0$. Light and dark shaded regions indicate $\pm 5\%$
  and $\pm 1\%$ deviations, respectively. 
  Symbols show $N$-body simulation results for the \textsc{LF}, \textsc{MF}, and \textsc{HF} models with fixed-phase initial conditions. \ALETHEIA\ and \FEMU\ do not support massive neutrinos; in this comparison, the massive neutrino density corresponding to $M_{\nu}= 0.06\,\mathrm{eV}$ is absorbed into $\oc$, and the amplitude parameter is rescaled accordingly. 
  {Alt text: Modern public emulators cluster near the DarkEmulator2 prediction on intermediate scales, while larger deviations appear at the largest and smallest wavenumbers and for older reference emulators.}
  }
  \label{fig:comp_fid_emu}
\end{figure*}

We first examine the DQ2 fiducial cosmology. 
The fiducial comparison provides a useful reference point for interpreting the trends seen in the random-parameter tests.
Figure~\ref{fig:comp_fid_emu} compares the nonlinear total matter power spectra from the public emulators and $N$-body simulations with those from \DE\ at the DQ2 fiducial cosmology. 
The high-resolution predictions of \DE, obtained from mixed-resolution training that accounts for cosmic variance, agree well with the corresponding high-resolution $N$-body results. 
The simulation symbols represent fixed-phase realizations of the initial density field, and their small scatter around the \DE\ curve primarily reflects residual cosmic variance. At lower and intermediate resolutions, the $N$-body results enter the shot-noise-dominated regime at large $k$.

Among the emulators, \PA\ and \FEMU\ show larger deviations than the others at both low and high $k$. These early generation emulators were pioneering efforts,
but they were developed before modern machine learning and computational
frameworks for high-precision nonlinear modeling had become widely available. 
For instance, \PA\ employs a neural network with a single hidden layer of 70
neurons with sigmoid activation, which limits its expressive capacity. 
By contrast, \EE, \BCE, \MTU, \AEMU\, \GOKUN and \ALETHEIA show close agreement with \DE\ on intermediate scales.

On large scales ($k \lesssim 0.01\,\hMpci$), where linear theory should dominate, the residual differences arise primarily from cosmic variance and from the different treatments of large-scale modes and linear power spectra in each pipeline. 
Because these scales are governed by the Boltzmann solver rather than by nonlinear clustering, \DE\ replaces the nonlinear prediction with the linear power spectrum below $k = 0.01\,\hMpci$. 
\AEMU\ and \DE\ are in excellent agreement down to $k \simeq 10^{-3}\,\hMpci$, while \MTU\ exhibits slightly larger deviations but still remains within $1\%$. Since \EE\ and \BCE\ are trained only for $k \ge 0.01\,\hMpci$, direct comparisons at larger scales are not meaningful.

At nonlinear scales ($k > 0.5\,\hMpci$), the emulators show more diverse behavior. 
Near the peak, \DE predicts a power spectrum that is about $1$--$2\%$ higher than the other emulators, whereas most of the others show a gradual ``spoon-like'' upturn toward their respective $k_{\max}$. 
Because the fiducial cosmology is included in the training sets of all emulators, the remaining differences are most likely attributable to how each underlying $N$-body data set models small-scale clustering. 
In particular, \DE\ uses a particle-mesh grid that is eight times finer than the mean inter-particle spacing in the TreePM calculation, and it incorporates pair-count corrections at small separations; these choices likely produce a slightly higher small-scale power amplitude. 
We have confirmed that when identical power spectrum estimators and internal parameters are used, $N$-body codes such as \textsc{Ginkaku}, \textsc{Pkdgrav3}, and \textsc{Gadget-2}-based implementations yield mutually consistent power spectra (see Appendix of \cite{Ginkaku26}).

\subsubsection{Nonlinear power spectrum comparison for arbitrary cosmologies}
\label{subsubsec:comp_nl_arb_param}

Public emulators support different ranges in cosmological parameter space. 
We therefore perform all nonlinear spectrum comparisons within the overlap between \DE\ and each public emulator. 
We emphasize that the support ranges of the public emulators considered here lie mostly within the central region of the \DE\ coverage, rather than near the boundaries where the accuracy of \DE\ gradually degrades.

\begin{figure*}[htbp]
  \centering
  \includegraphics[width=0.8\linewidth]{
    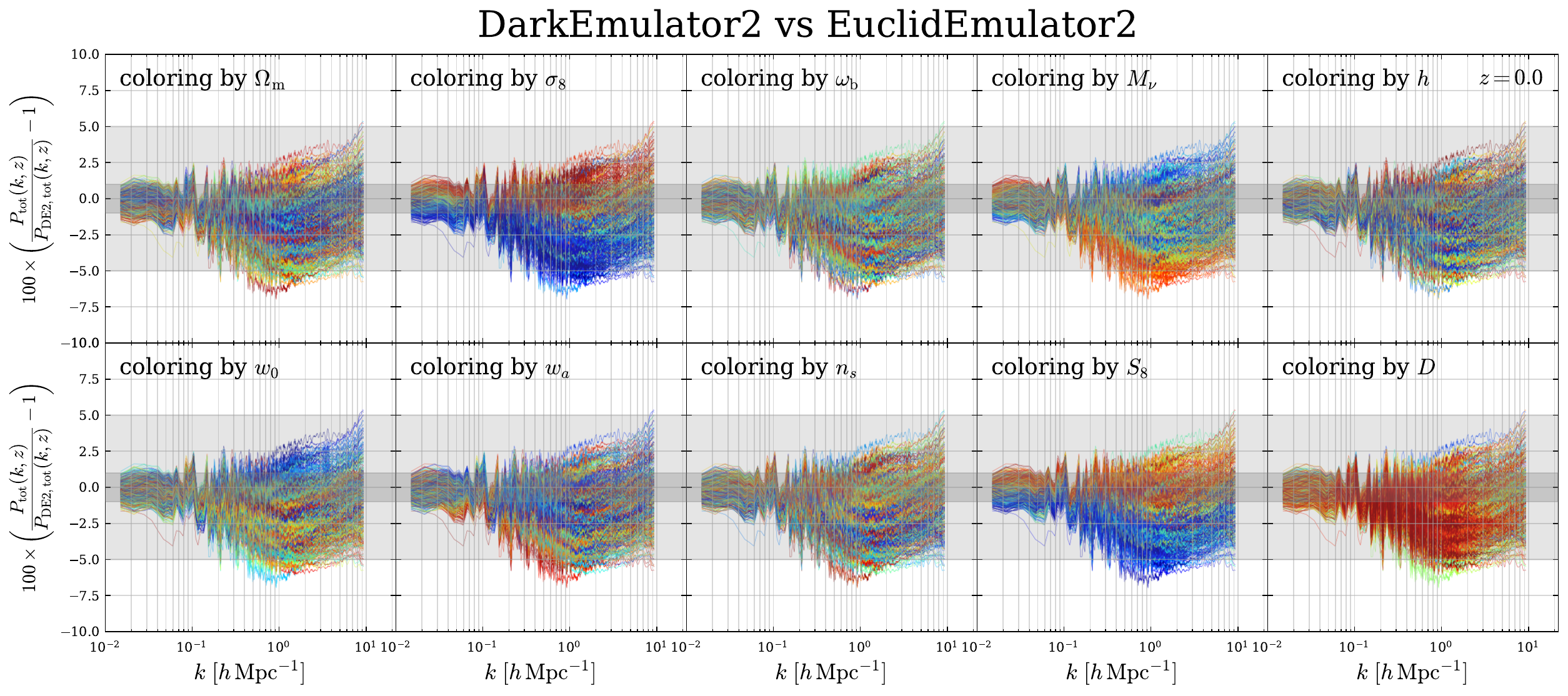
  }
  \includegraphics[width=0.8\linewidth]{
    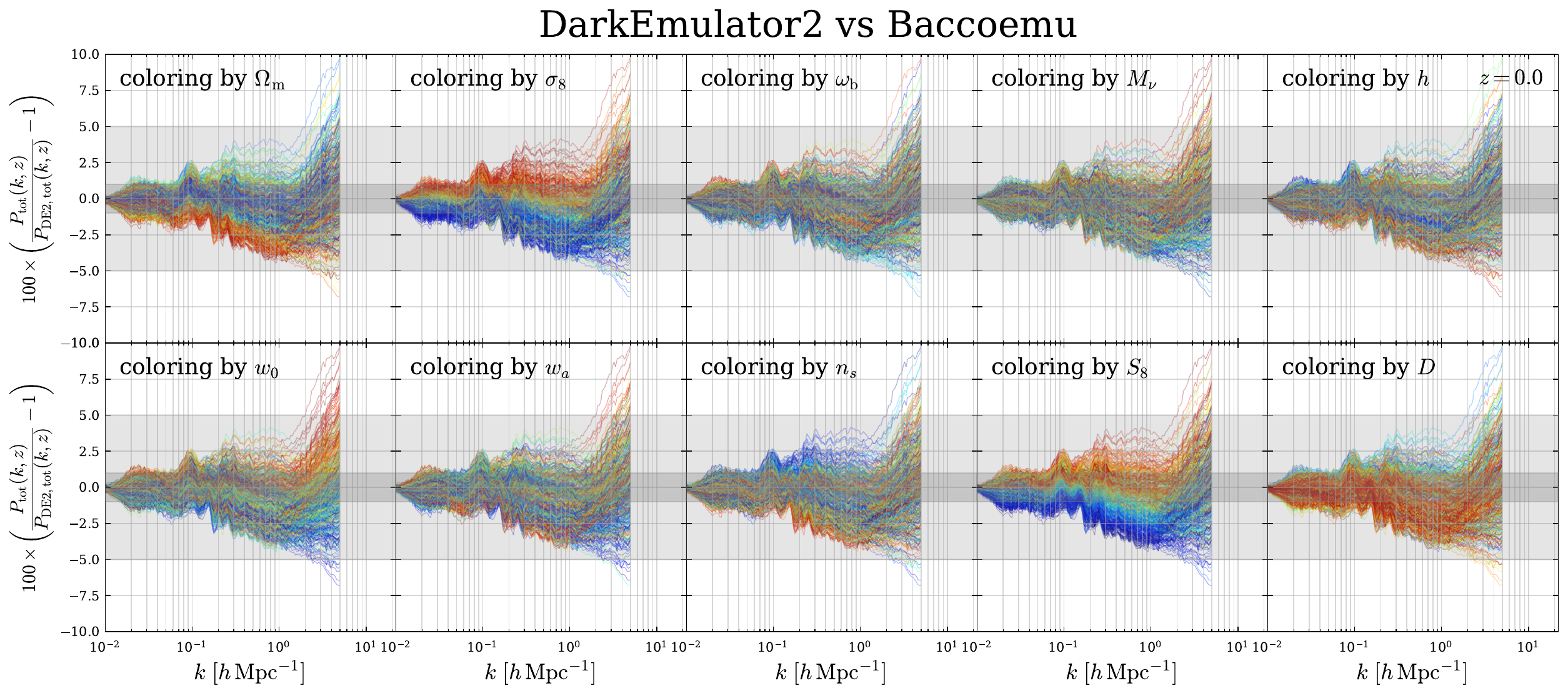
  }
  \caption{Ratios of the nonlinear total matter power spectra predicted by \EE\ (top)
  and \BCE\ (bottom) to the \DE\ prediction for 500 cosmologies randomly
  sampled from the parameter volume jointly supported by \DE\ and the
  corresponding emulator in the $\wwnCDM$ cosmology at $z=0$. Each panel shows
  the same set of ratio curves, but the curves are re-colored in each panel
  according to the value of the parameter indicated on that panel. Red-to-blue
  gradients correspond to increasing $\Om$, whereas all other parameters are
  color-coded from blue to red. $D$ denotes the distance index defined in equation~\ref{eq:p_dist}.
  Light and dark shaded regions represent $5\%$ and $1\%$ errors, respectively.
  {Alt text: EuclidEmulator2 and BaccoEmu show parameter-dependent deviations from DarkEmulator2, with the spread increasing toward nonlinear scales and visible trends with parameters related to fluctuation amplitude.}
  \label{fig:comp_nlpk_emu1}
  }
\end{figure*}

\begin{figure*}[htbp]
  \centering
  \includegraphics[width=0.8\linewidth]{
    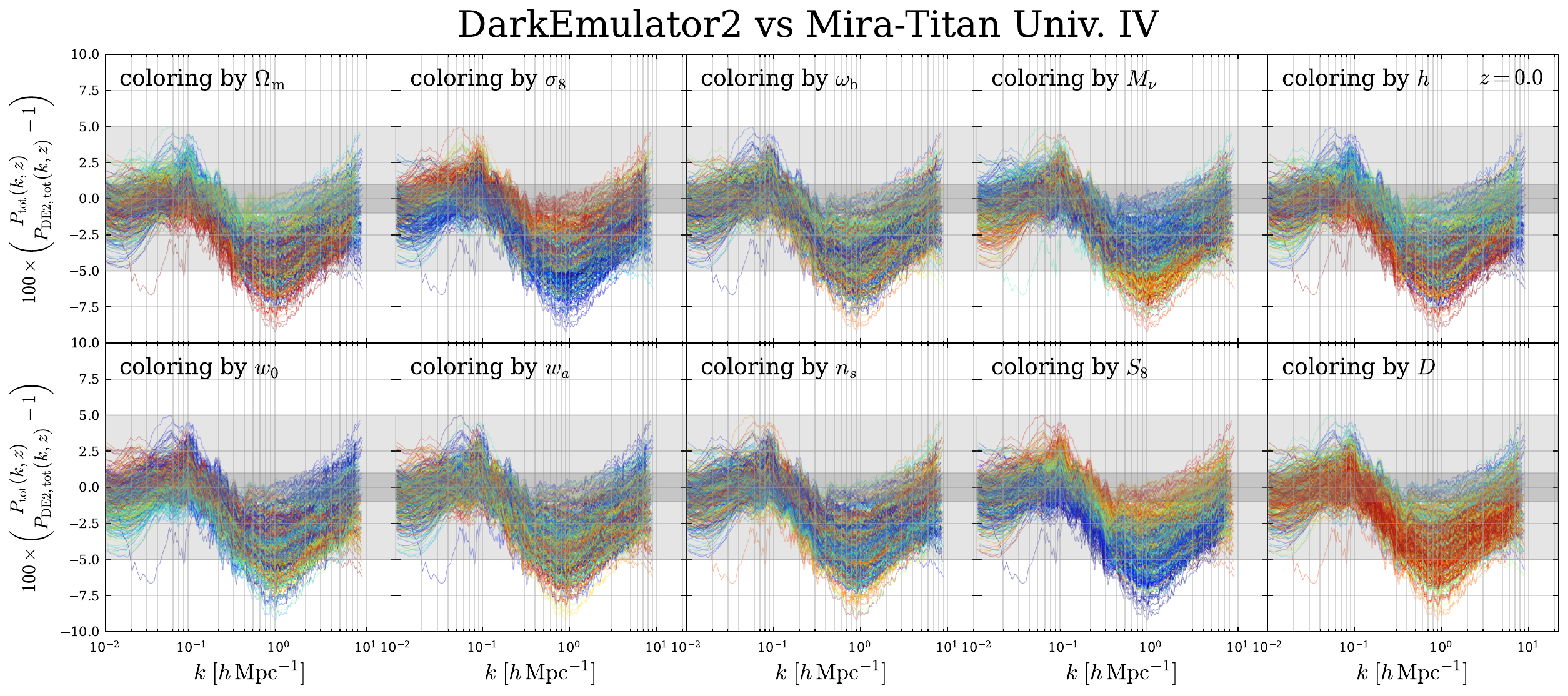
  }
  \includegraphics[width=0.8\linewidth]{
    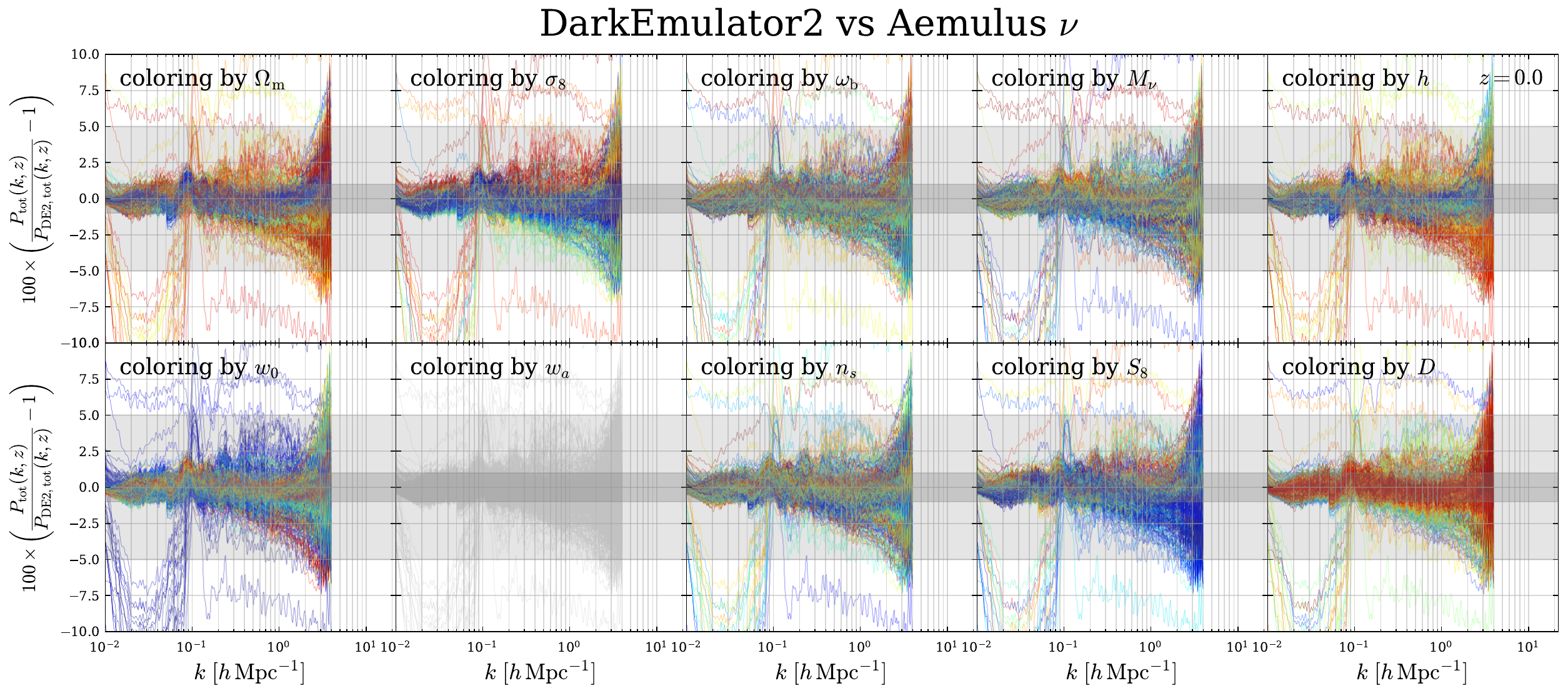
  }
  \caption{Same as figure~\ref{fig:comp_nlpk_emu1}, but for comparisons with \MTU\ (top)
  and \AEMU\ (bottom). Only \AEMU\ employs the $w_{0}\nu$CDM cosmology. Panels
  that are gray are parameters that are not in this comparison. 
  {Alt text: Mira-Titan shows broad scale-dependent suppression relative to DarkEmulator2, while Aemulus nu is mostly closer except for excursions at low wavenumber in a subset of edge cosmologies.}
  }
  \label{fig:comp_nlpk_emu2}
\end{figure*}

\begin{figure*}[htbp]
  \centering
  \includegraphics[width=0.8\linewidth]{
    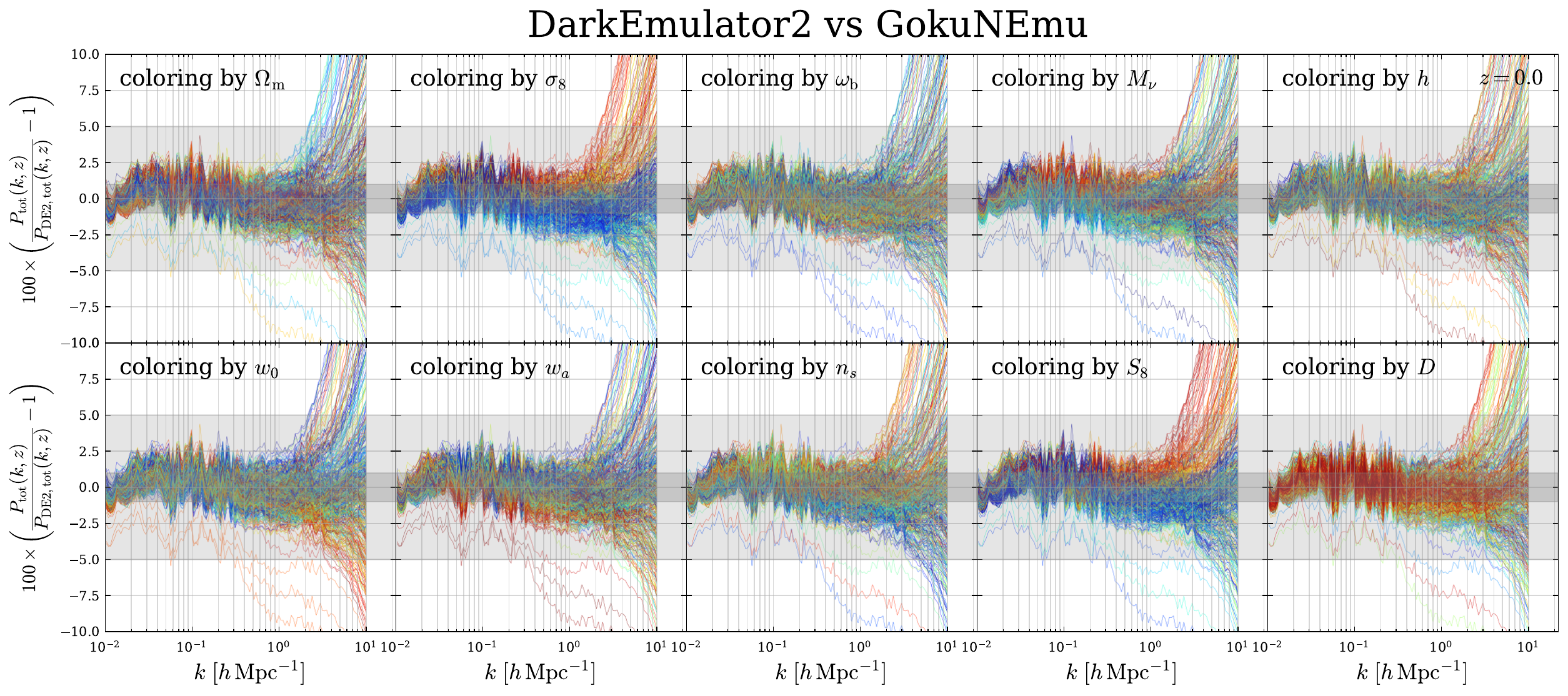
  }
  \includegraphics[width=0.8\linewidth]{
    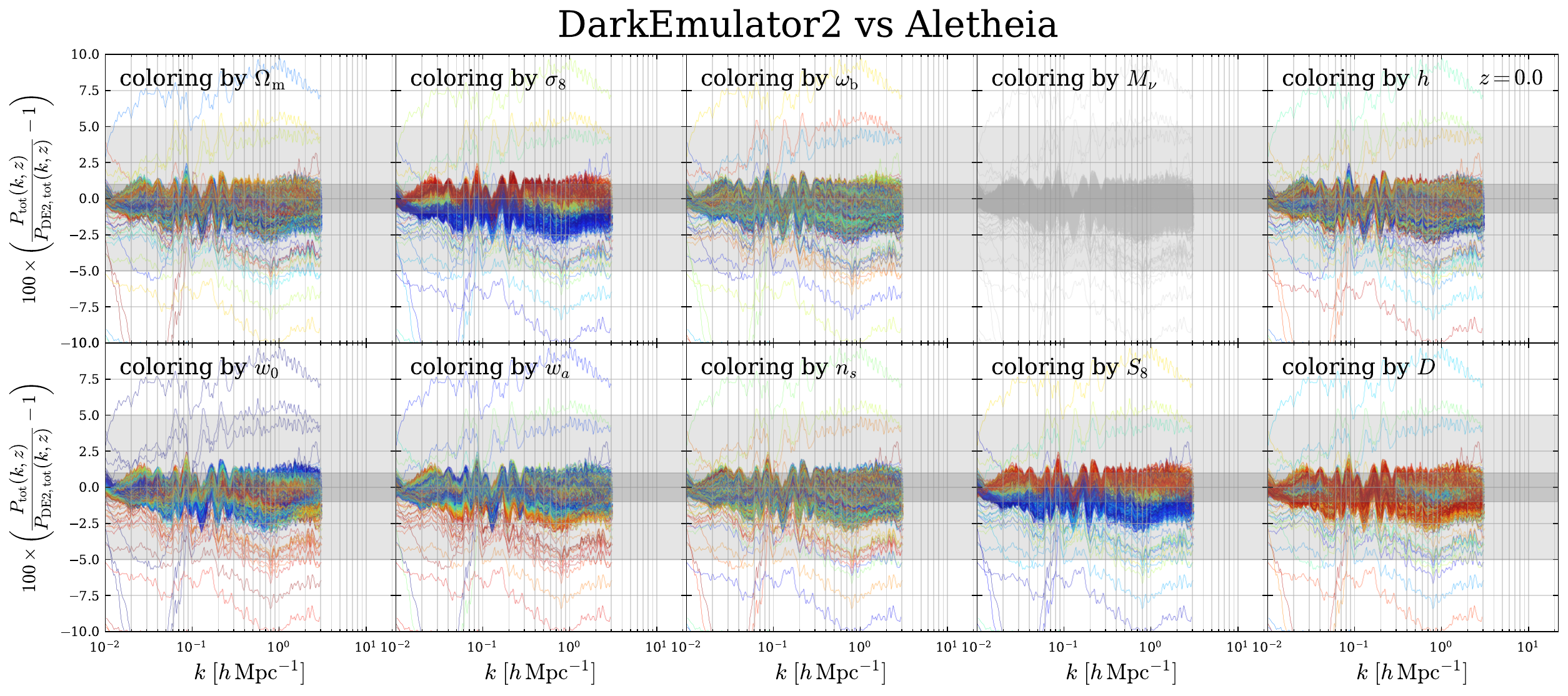
  }
  \includegraphics[width=0.8\linewidth]{
    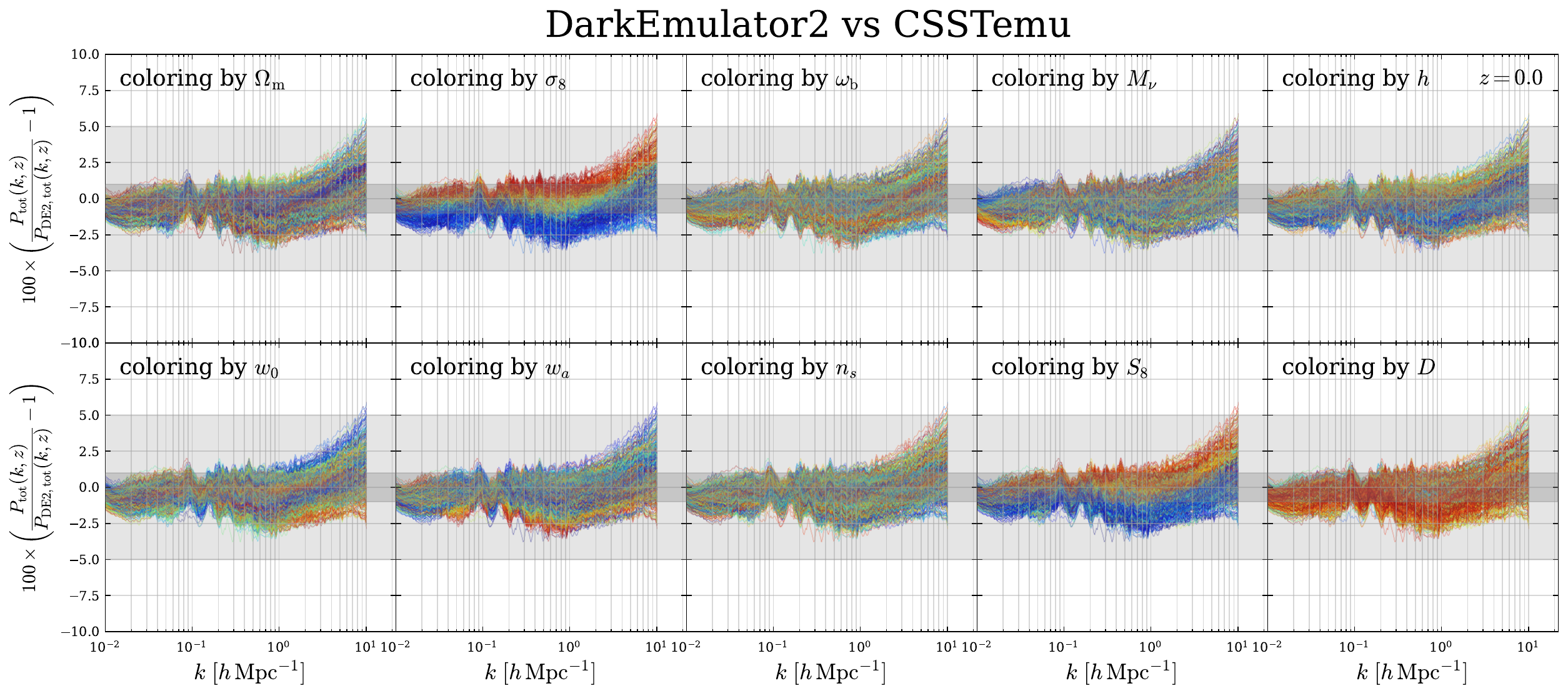
  }
  \caption{Same as figure~\ref{fig:comp_nlpk_emu1}, but for comparisons with \GOKUN\ (top), \ALETHEIA\ (middle), and \CSST\ (bottom). For \GOKUN, the additional parameters $N_{\rm eff}$
  and $\alpha_{\rm s}$ are held fixed to their default values in this comparison.
  \ALETHEIA\ does not support massive neutrinos, and we therefore restrict the
  comparison to $M_{\nu}=0$ and the curvature is also held fixed to $\Ok=0$.
  Panels that are gray are parameters that are not in this comparison.
  {Alt text: GokuNEmu remains close to DarkEmulator2 on large and mildly nonlinear scales but shows increasing scatter at high wavenumber. Aletheia displays a broader percent-level offset. CsstEmulator agrees well on large and mildly nonlinear scales, with increasing differences toward the highest wavenumbers.}
  }
  \label{fig:comp_nlpk_emu3}
\end{figure*}

Figures~\ref{fig:comp_nlpk_emu1}--\ref{fig:comp_nlpk_emu3} present ratios of emulator predictions to \DE\ for 500 randomly sampled cosmologies at $z=0$. 
Each panel displays the same set of ratio curves, color-coded according to the value of a selected cosmological parameter. 
A systematic color gradient suggests a parameter-dependent discrepancy between \DE\ and the corresponding emulator. 
By inspecting such trends across multiple emulators, one can identify which emulator exhibits parameter-dependent deviations and along which dimensions of parameter space. 
This assessment is qualitative and comparative in nature, because the sampled parameters can be correlated. Because the sampled cosmologies exhibit parameter correlations within the overlap volume, a color gradient in a given panel should be interpreted as a projected trend rather than a strictly one-parameter dependence. 
Note that panels corresponding to parameters not supported by a given emulator are shown in gray scale; we do not interpret color trends in those panels.

Relative to \DE, \EE\ exhibits a coherent scale dependence in figure~\ref{fig:comp_nlpk_emu1}.
The ratio dips by a few percent around $k \sim 1\,\hMpci$ for a subset of cosmologies, and the spread increases toward nonlinear scales. 
Clear color gradients are visible in the $\sigma_{8}$ and $S_{8}$ panels, indicating that the discrepancy depends systematically on the fluctuation amplitude and closely related parameter combinations over intermediate-to-small scales. The $w_{0}$ panel also shows an approximately scale-independent offset. Cosmologies with more negative $w_{0}$ tend to lie systematically above those with less negative $w_{0}$. 
A weaker but still systematic dependence on $M_{\nu}$ is also present.
This trend already appears at the linear level (section~\ref{subsubsec:comp_lin_arb_param}) and can therefore propagate into the nonlinear comparison, because \EE\ constructs $P_{\mathrm{nl}}$ by applying a nonlinear correction to its linear spectrum.

For \BCE, the agreement with \DE\ is tight on large scales (approximately $k\lesssim 0.1\,\hMpci$) in figure~\ref{fig:comp_nlpk_emu1}. 
Toward smaller scales, the ratio shows clear color gradients with $\Om$ and with the fluctuation-amplitude parameters ($\sigma_{8}$ and $S_{8}$), and a weaker but noticeable dependence on $w_{0}$. 
At the highest wavenumbers shown, the spread grows rapidly; in this regime, the comparison also becomes increasingly sensitive to emulator-specific small-scale modeling choices. 
Overall, this scale- and parameter-dependent pattern appears compatible with the cosmology-rescaling strategy adopted in \BCE, in which predictions over a broad parameter space are obtained by mapping a small set of high-resolution simulations to nearby cosmologies. 
In such a framework, residual mismatches would naturally be most apparent on small scales, where nonlinear growth is highly sensitive to parameters such as $\Om$ and $\sigma_{8}$.

Relative to \DE, \MTU shows a coherent scale-dependent deviation in figure~\ref{fig:comp_nlpk_emu2}.
The ratio stays close to 2\% on large scales but exhibits a broad suppression over intermediate-to-nonlinear scales, with departures reaching the several-percent level around the trough, followed by a partial recovery toward higher $k$. 
The dispersion across cosmologies is already non-negligible on large scales and increases further toward nonlinear scales; compared with the other emulator comparisons, the larger large-scale spread may reflect differences in how cosmic variance is mitigated in the underlying predictions. 
Clear color gradients are visible in the $\sigma_{8}$ and $S_{8}$ panels, indicating a systematic dependence on the fluctuation amplitude over intermediate-to-small scales, while weaker but coherent gradients are also present in the $\Om$ and $h$ panels. 
By contrast, trends with $w_{0}$ are not clearly identifiable in this comparison. The overall $k$-dependent pattern is also qualitatively similar to the low-redshift behavior of the \textsc{HMcode} fiducial model (figure~\ref{fig:comp_fitting_fid}), consistent with the fact that \MTU\ is derived from the \textsc{Mira-Titan Universe} simulation suite that underpins widely used nonlinear calibrations.

\begin{figure*}[htbp]
  \centering
  \includegraphics[width=0.8\linewidth]{
    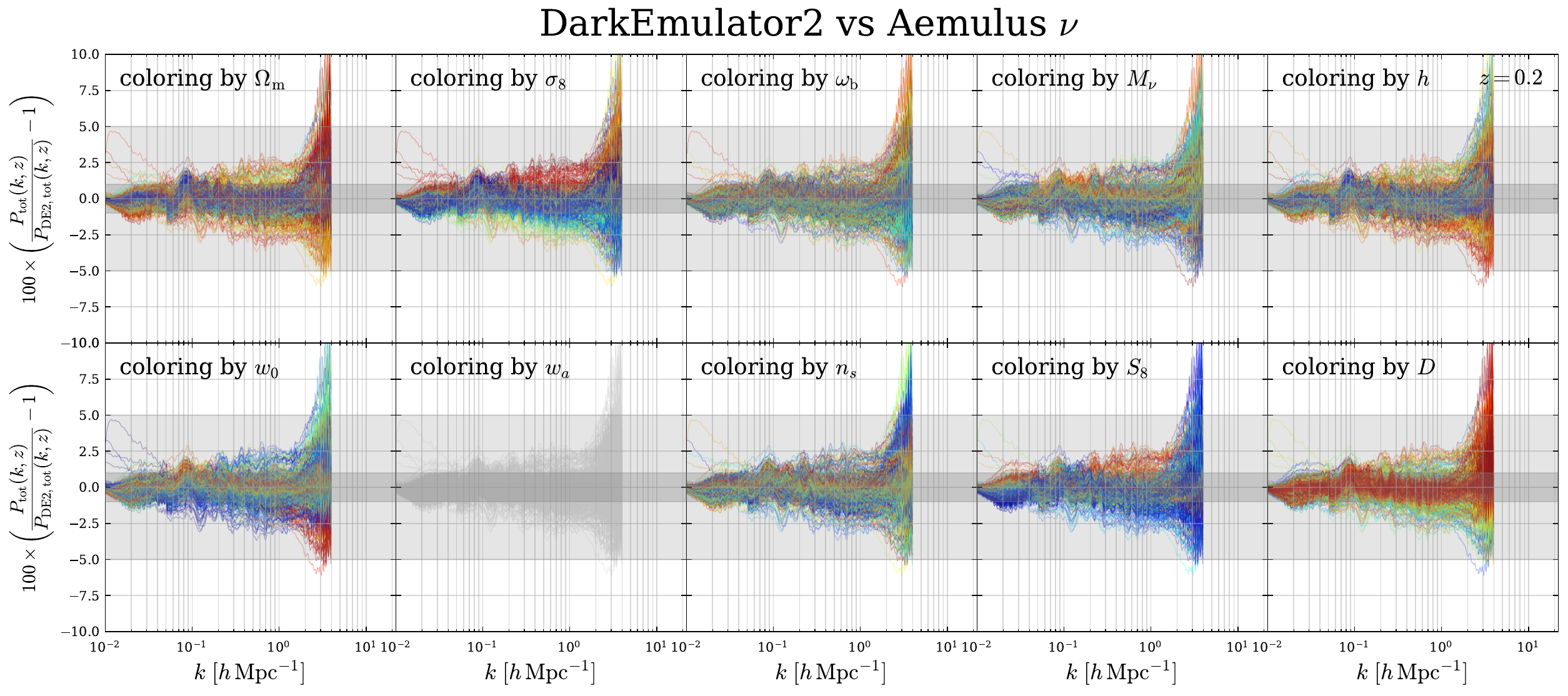
  }
  \includegraphics[width=0.8\linewidth]{
    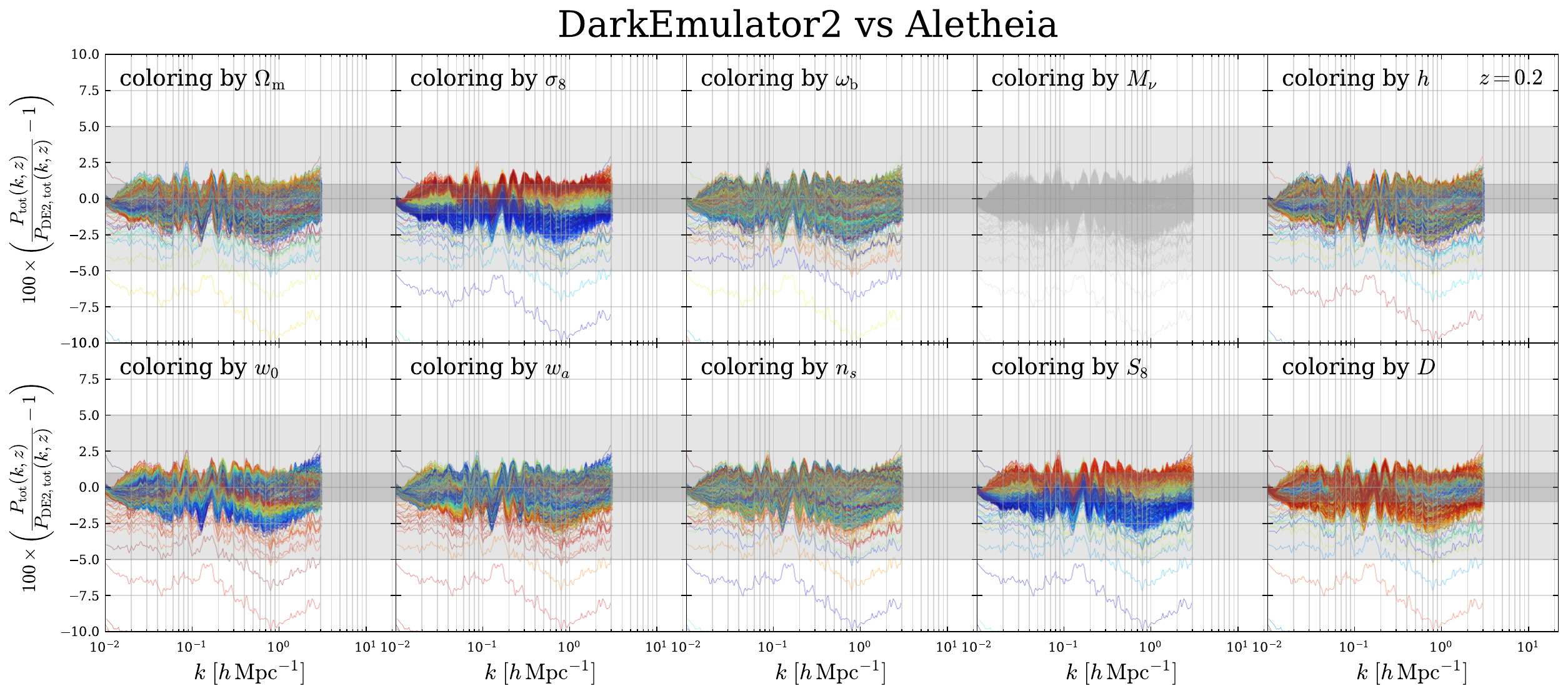
  }
  \caption{Same as Figures~\ref{fig:comp_nlpk_emu2} and ~\ref{fig:comp_nlpk_emu3},
  but showing the comparison between \DE\ and \AEMU\ or \ALETHEIA\ at $z=0.2$.
  {Alt text: At redshift 0.2, the strongest excursions at low wavenumber seen for Aemulus nu are reduced, whereas some Aletheia outliers remain visible across the plotted range.}
  }
  \label{fig:comp_nlpk_emu_z02}
\end{figure*}

In figure~\ref{fig:comp_nlpk_emu2}, \AEMU\ agrees well with \DE\ for most cosmologies, but a small subset shows pronounced excursions at the lowest wavenumbers at $z=0$, most prominently when $\Om$ or $w_{0}$ lies near the lower edge of the supported range (e.g., $w_{0}\simeq -1.5$). 
Clear gradients are not dominant, but a weak and systematic dependence on $h$ is also visible in the same figure. Appendix~\ref{append:lin_pk_emu} shows that dark energy perturbations (clustering) can become non-negligible already at the linear level for $w_{0}\lesssim -1.4$ at $z<0.05$. 
In this regime, differences in gauge conventions can become relevant at very large scales; in particular,
\DE\ adopts the $N$-body gauge~\citep{2015PhRvD..92l3517F}, whereas \AEMU\ is formulated in the synchronous gauge. Within the set of emulators considered here, only \DE, \AEMU, and \ALETHEIA\ cover such extreme $w_{0}$ values. 
The observed low-$k$ excursions at $z\simeq 0$ may therefore reflect, at least in part, the fact that the two emulators are not expected to coincide exactly at the linear level in this corner of parameter space, even before considering nonlinear modeling details. 
Away from these edge cases, \AEMU\ does not exhibit a clear, coherent offset over intermediate scales. The comparisons remain broadly consistent with unity, and the remaining dispersion mainly appears at the highest wavenumbers, where the inferred nonlinear spectrum becomes increasingly sensitive to emulator-specific small-scale prescriptions and measurement pipelines. 
Consistent with this interpretation, the low-$k$ excursions are substantially reduced at $z=0.2$ (figure~\ref{fig:comp_nlpk_emu_z02}), indicating that the discrepancy is confined to very low redshift.

\GOKUN\ remains close to the \DE\ prediction on large scales and mildly nonlinear scales, while the dispersion grows rapidly toward the highest wavenumbers shown in figure~\ref{fig:comp_nlpk_emu3}. 
Clear color gradients are visible in the $\sigma_{8}$ and $S_{8}$ panels, indicating a systematic dependence of the ratio on the fluctuation amplitude over intermediate-to-small scales. 
At the highest $k$, a small fraction of cosmologies exhibit large excursions, making the comparison increasingly dominated by small-scale behavior.

\ALETHEIA\ exhibits an approximately percent-level offset relative to \DE\ over a broad range of wavenumbers, with only a weak scale dependence in figure~\ref{fig:comp_nlpk_emu3}.
The $\sigma_{8}$ and $S_{8}$ panels show clear color gradients, demonstrating that the discrepancy depends systematically on the fluctuation amplitude (and closely related parameter combinations). 
We also observe weaker but coherent trends in several other panels; the overall spread remains moderate on large scales and increases toward smaller scales. 
Because the public \ALETHEIA\ release assumes massless neutrinos, we restrict the comparison to $M_{\nu}=0$, and we also fix the curvature to $\Ok=0$ in this test. A small subset of cosmologies shows particularly large deviations at $z=0$; however, unlike the low-$k$ excursions seen for \AEMU, these outliers are not substantially reduced when repeating the comparison at $z=0.2$ (see figure~\ref{fig:comp_nlpk_emu_z02}). 
This persistence suggests that the residual mismatch is unlikely to be driven solely by the very-low-redshift gauge-related effects discussed for \AEMU, and may instead reflect differences in modeling assumptions, calibration data, or implementation details in the public \ALETHEIA\ release.

For \CSST, the agreement with \DE\ is close on large and mildly nonlinear scales in figure~\ref{fig:comp_nlpk_emu3}. Toward the highest wavenumbers covered by \CSST, the ratio gradually departs from unity for a subset of cosmologies and reaches a few percent level. Clear projected trends are visible with $\sigma_{8}$ and $S_{8}$, while weaker trends are also present in several other panels. Since the corresponding linear spectra agree closely with those of \DE\ (figure~\ref{fig:comp_linpk_emu}), these high-$k$ differences are more likely associated with the nonlinear-emulation target, variance-reduction procedure, smoothing, or power spectrum measurement pipeline than with the underlying linear matter power spectrum.

On large (approximately linear) scales, the ratios are typically close to unity for several emulators, but some comparisons display a non-negligible dispersion or offset already at low $k$, consistent with differences in how cosmic variance is mitigated and/or how the large-scale limit is enforced. 
Toward smaller scales, the discrepancies generally become larger and more structured in both scale and parameter dependence. 
In particular, the spread grows rapidly at the highest wavenumbers shown, where the comparison becomes increasingly sensitive to the details of the power spectrum estimator and small-scale modeling and/or extrapolation. 
In our setup, \DE\ combines an FFT-based estimator with a pair-counting-based extension on small scales; differences in the underlying measurement pipelines relative to purely FFT-based approaches are therefore a plausible contributor to the increased high-$k$ dispersion.

Across the comparisons, the most clearly identifiable parameter-dependent trends
are associated with $\Om$, $\sigma_{8}$, and the derived combination $S_{8}\equiv
\sigma_{8}\sqrt{\Om/0.3}$. Because $S_{8}$ is strongly correlated with
$\sigma_{8}$ by construction, color gradients in the $S_{8}$ panels often reflect
an underlying dependence on $\sigma_{8}$ rather than an independent sensitivity
to $\Om$. At the same time, the direction and strength of these gradients are not
uniform across all emulators, which cautions against attributing any single
trend uniquely to \DE.

The comparisons also highlight that very-low-redshift behavior can depend on cosmological model assumptions and gauge conventions. 
For extreme dark energy models (e.g., $w_{0} \lesssim -1.4$) at $z \simeq 0$, we find cases where low-$k$ excursions are substantially reduced at $z=0.2$, consistent with the expectation that differences related to dark energy perturbations and/or gauge conventions are confined to very low redshift. 
However, other outliers persist when moving from $z=0$ to $z=0.2$, suggesting that additional ingredients, such as modeling assumptions, calibration data, or implementation details, can also drive residual mismatches even when the background cosmology is matched.

Overall, these results motivate (i) direct validation against $N$-body simulations within each emulator's native support range, and (ii) reporting of scale-dependent and parameter-dependent systematics, rather than relying solely on advertised parameter coverage as a proxy for cross-emulator consistency. 
While this section focuses on comparisons between \DE\ and public emulators, cross-comparisons among the public emulators themselves can also be performed in a similar manner, allowing one to assess relative accuracy, parameter dependence, and redshift evolution. 
Because a full presentation of these results would be beyond the scope of this paper, additional comparisons across redshifts and between public emulators are provided in the \textsc{Dark Quest II} supplement.~\footnote{\url{https://darkquestcosmology.github.io/dark_emulator2_supplement/}}

\subsubsection{Linear power spectrum comparison for arbitrary cosmologies}
\label{subsubsec:comp_lin_arb_param}

\begin{figure*}[htbp]
  \centering
  \includegraphics[width=0.7\linewidth]{
    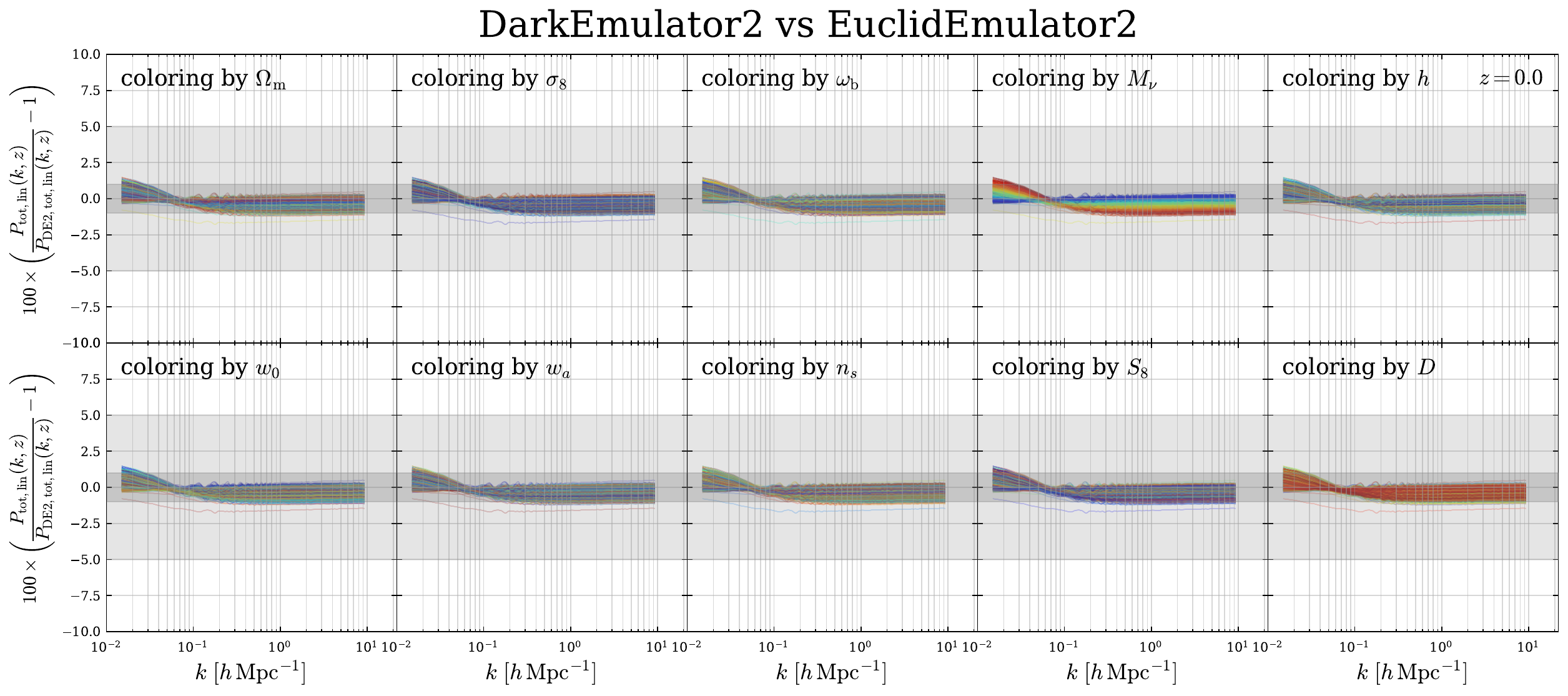
  }
  \includegraphics[width=0.7\linewidth]{
    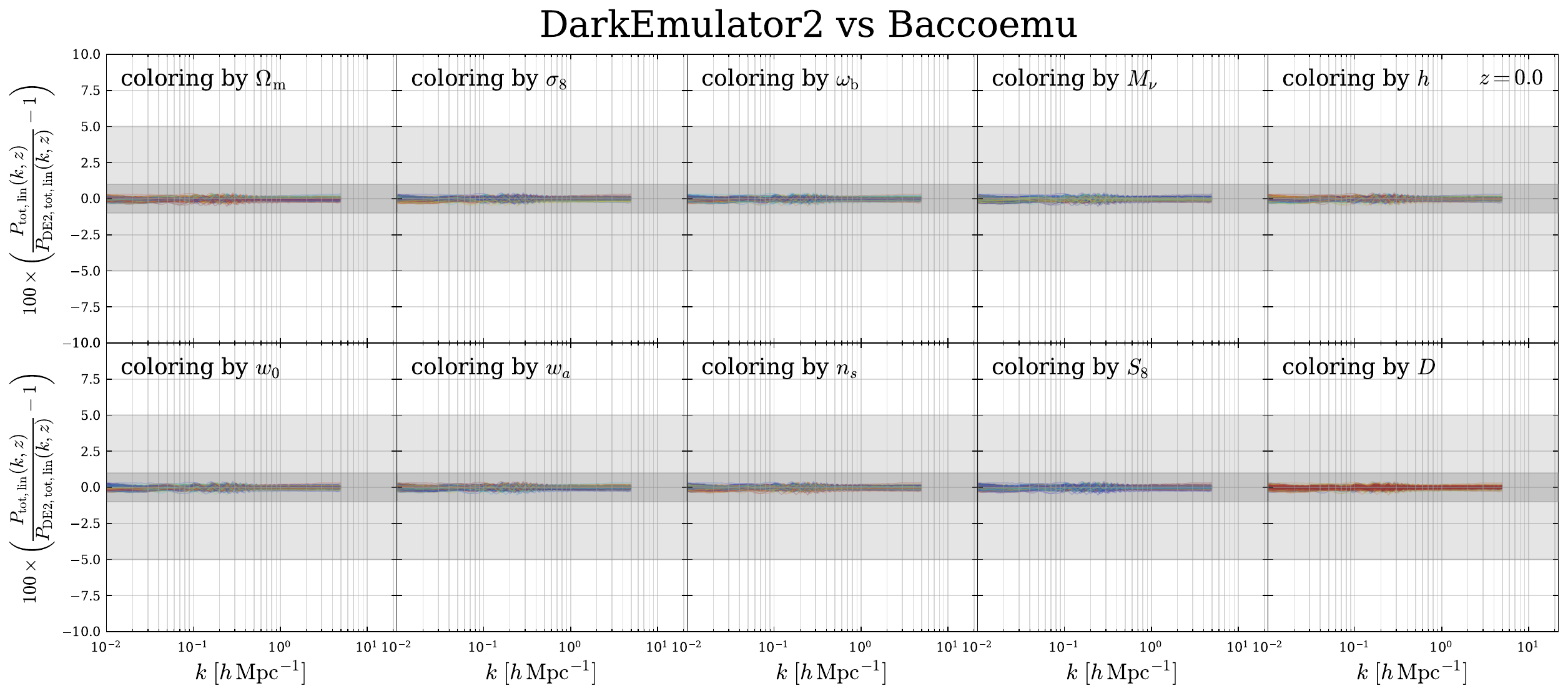
  }
  \includegraphics[width=0.7\linewidth]{
    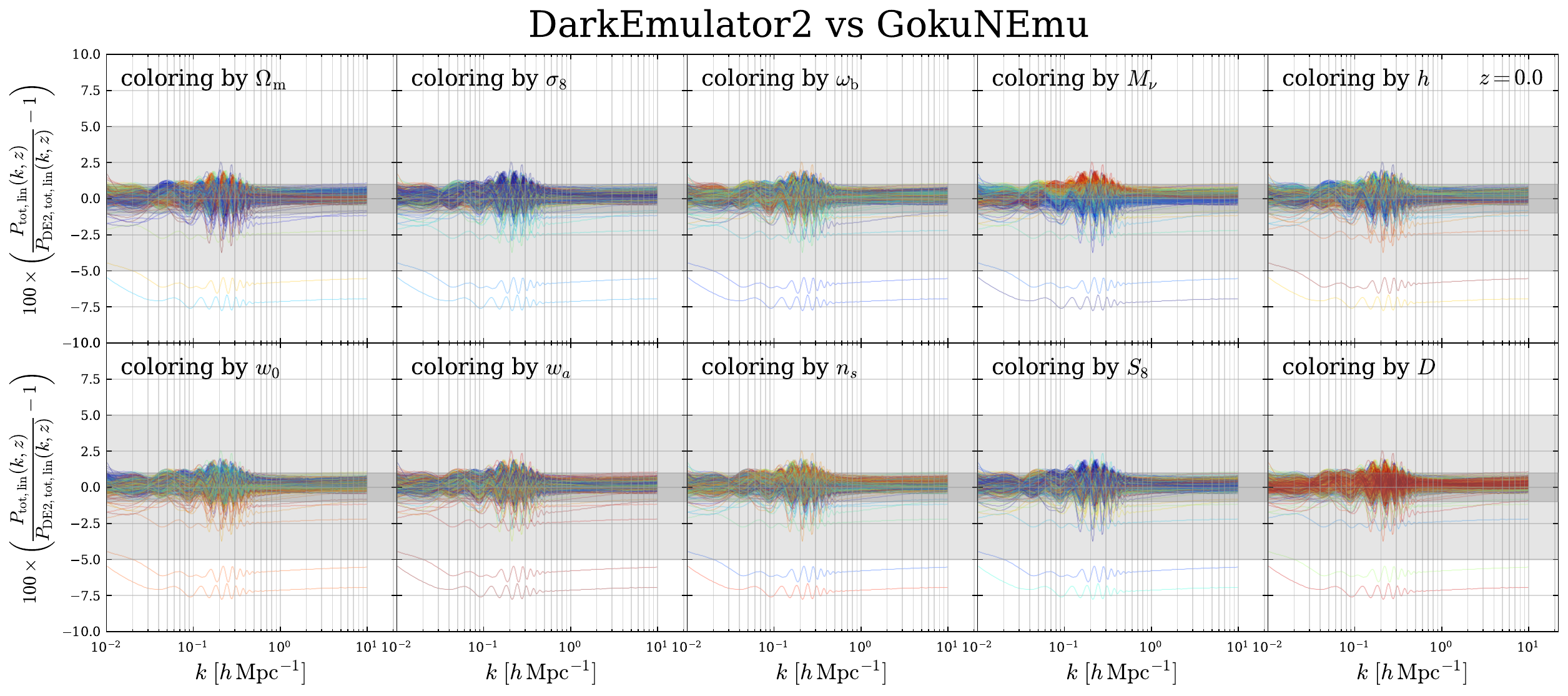
  }
   \includegraphics[width=0.7\linewidth]{
    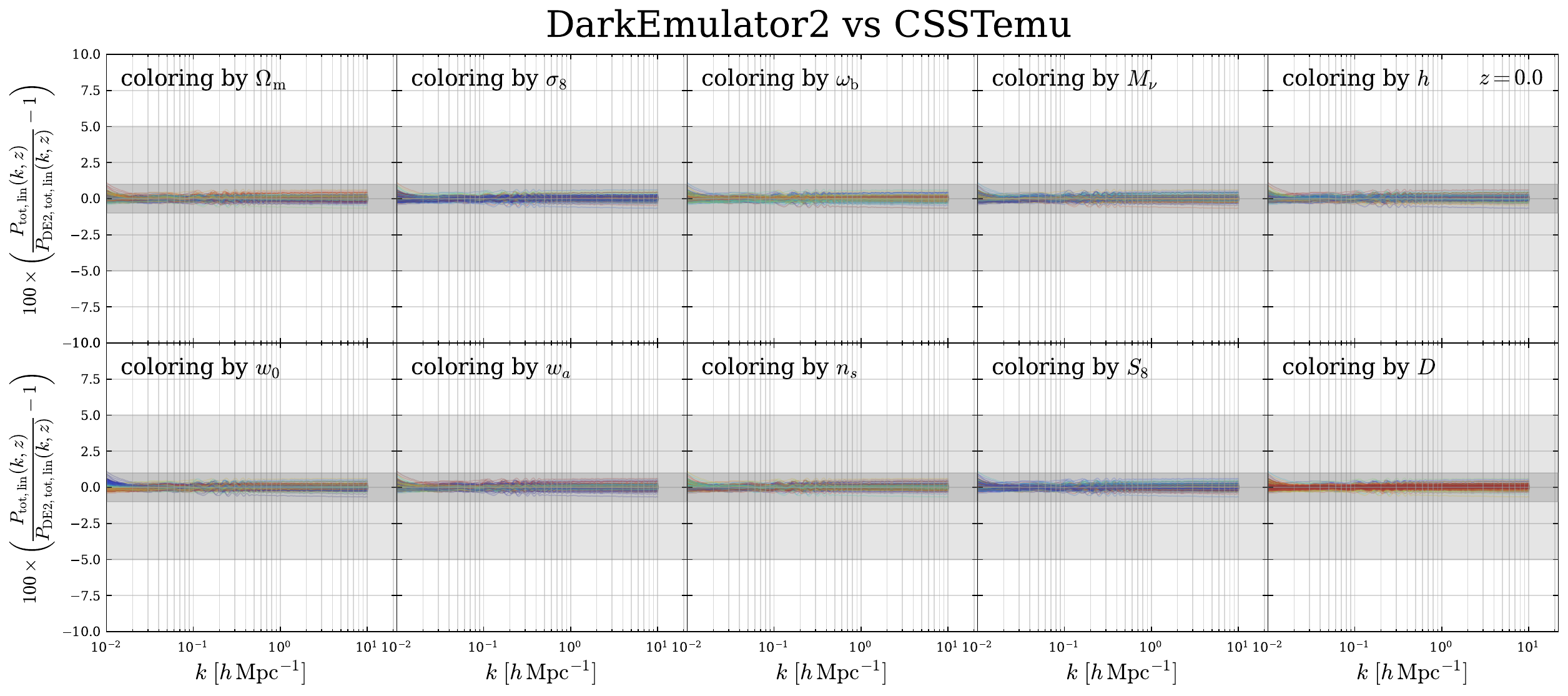
  }
  \caption{Same as figure~\ref{fig:comp_nlpk_emu1}, but for the linear total matter
  power spectrum $P_{\mathrm{tot,lin}}(k)$ at $z=0$. In addition to nonlinear
  spectra, \EE, \BCE, \GOKUN, and \CSST provide $P_{\mathrm{tot,lin}}(k)$ through their
  public interfaces. For \BCE, \GOKUN, and \CSST, the linear spectra are produced by their internal emulators, whereas \EE\ returns a \textsc{Class}-computed $P_{\mathrm{tot,lin}}(k)$ that is used together with its nonlinear correction factor.
  {Alt text: BaccoEmu and CsstEmulator agree closely with DarkEmulator2 at the linear level, while EuclidEmulator2 and GokuNEmu show more visible parameter-dependent residuals, especially trends associated with massive neutrinos.}
  }
  \label{fig:comp_linpk_emu}
\end{figure*}

In this section, we compare the linear total matter power spectrum $P_{\mathrm{tot,lin}}(k)$ at $z=0$ in the same format as the nonlinear comparison. 
\EE provides an emulator for the nonlinear correction factor $B(k,z)$ while the associated linear spectrum is computed with CLASS.
Accordingly, the ``linear'' curves shown for \EE in figure~\ref{fig:comp_linpk_emu} reflect the \textsc{Class} prediction used in our evaluation pipeline, rather than a separately emulated linear spectrum.

The linear comparisons help disentangle whether discrepancies seen in the nonlinear
regime arise from differences in the underlying linear spectra or from the nonlinear
prescriptions of each emulator. Figure~\ref{fig:comp_linpk_emu} shows that the
linear predictions of \DE\ and \BCE\ agree extremely well over the scales
shown, with differences remaining at the subpercent level and with no prominent
parameter-dependent trends. This tight agreement supports the interpretation
that the residual differences observed in the nonlinear comparison with \BCE\ primarily
originate from the nonlinear calibration/rescaling step rather than from the
linear spectrum itself.

For \EE, we find a mild but coherent dependence on the total neutrino mass $M_{\nu}$:
increasing $M_{\nu}$ suppresses power at $k \gtrsim 0.07\,\hMpci$ and
correspondingly enhances it at $k \lesssim 0.07\,\hMpci$. Since \EE\ constructs
the nonlinear spectrum by multiplying its nonlinear correction factor by a \textsc{Class}-computed
linear spectrum, any convention-level difference in how massive neutrinos are implemented
or parameterized at the interface level can propagate directly into the nonlinear
comparison in section~\ref{subsubsec:comp_nl_arb_param}. Indeed, a public issue
report has suggested that the current interface may effectively treat massive neutrinos
with a reduced number of massive species in some configurations; while we do
not attempt to resolve this here, figure~\ref{fig:comp_linpk_emu} indicates that
the dominant systematic trend in the \EE\ linear comparison is aligned with $M_{\nu}$.

Compared to \DE, \GOKUN\ shows larger residuals in the linear spectrum, including
visibly stronger deviations around BAO scales and a broader dispersion across cosmologies
than seen for \BCE\ and \EE. A systematic trend with $M_{\nu}$ is also present;
one plausible contributor is that \GOKUN\ adopts a normal-hierarchy neutrino model,
whereas \DE\ uses the common degenerate mass approximation. However, the
overall level and structure of the residuals suggest that additional
differences beyond neutrino mass splitting, such as details of the linear spectrum
emulation, parameter mappings, or implementation choices, are also likely to
contribute.

For \CSST, the linear spectrum agrees closely with \DE\ over the plotted range. The residuals remain small and show no prominent parameter-dependent structure comparable to the trends seen for \EE\ or \GOKUN. This supports the
interpretation that the nonlinear differences between \CSST\ and \DE\ in figure~\ref{fig:comp_nlpk_emu3} are dominated by the nonlinear-emulation stage and by simulation or measurement choices, rather than by differences in the underlying linear matter power spectrum.

\subsection{Cosmic shear power spectrum}
\label{subsec:cl}

\begin{figure*}[htbp]
  \centering
  \includegraphics[width=\linewidth]{
    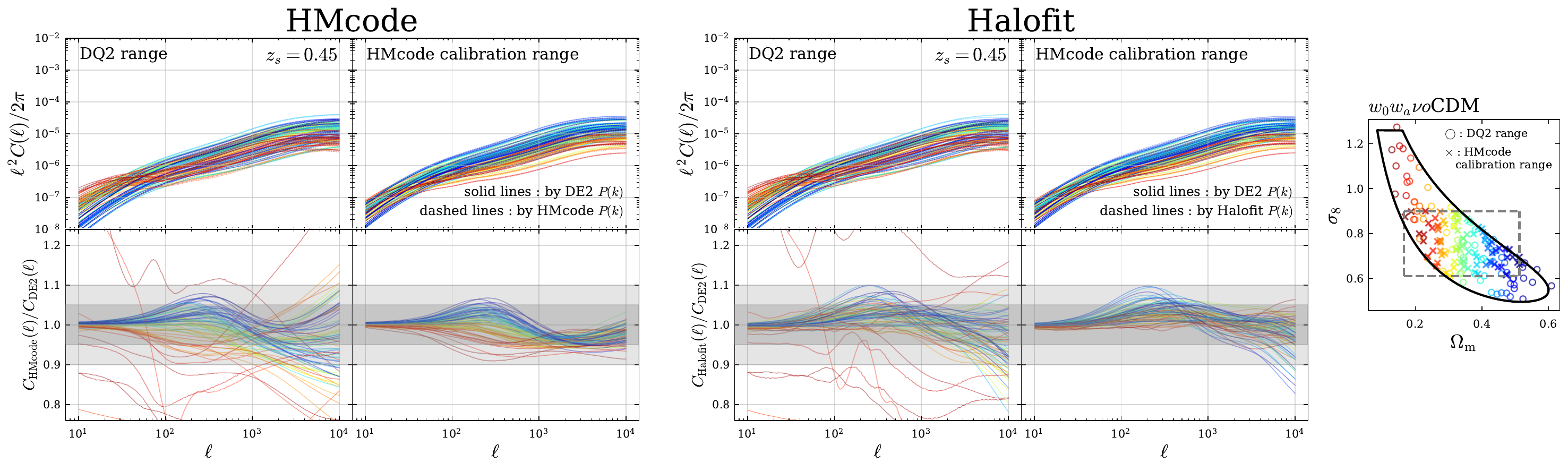
  }
  \includegraphics[width=\linewidth]{
    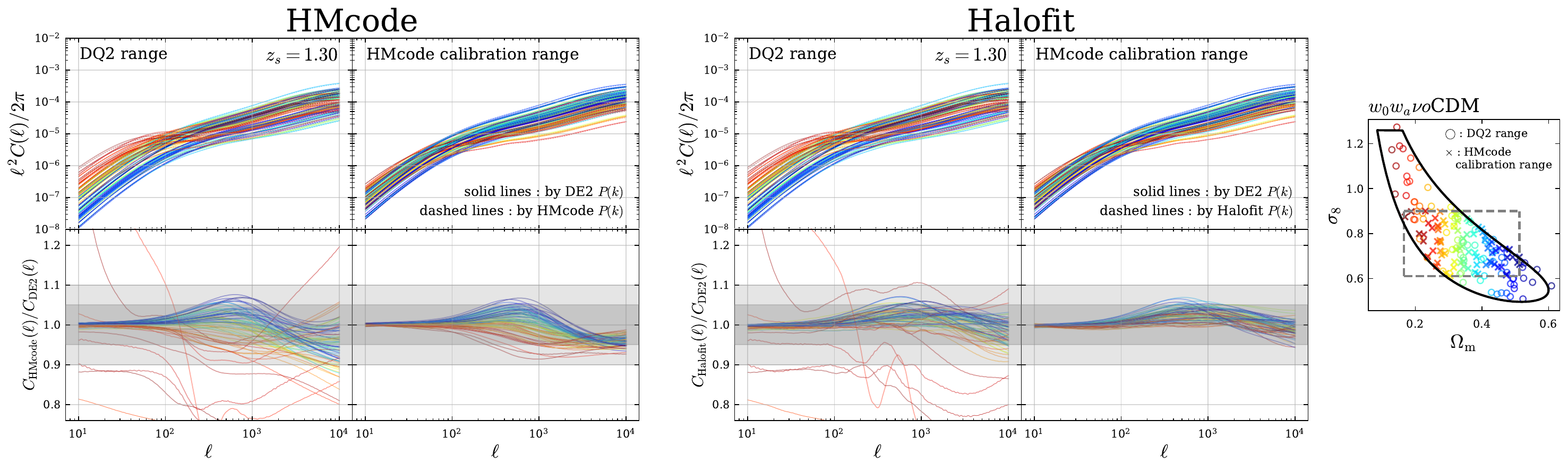
  }
  \caption{Comparison of $C(\ell)$ predicted from nonlinear $P(k)$ by \DE\ with
  \textsc{HMcode} and \textsc{Halofit} in $\wwnoCDM$ for 80 randomly sampled
  cosmologies. The top panels show $z_{s}=0.45$, and the bottom panels show
  $z_{s}=1.3$. For each $z_{s}$, the left block compares with \textsc{HMcode},
  and the right block compares with \textsc{Halofit}. Within each block, the left
  panel uses samples within the \DE\ support (DQ2 range), and the right panel
  uses samples within the \textsc{HMcode} calibration domain. The sampled
  cosmologies are shown in the rightmost $\Om$-$\sigma_{8}$ panel, where circle
  markers denote the DQ2-range samples and cross markers denote the \textsc{HMcode}-range
  samples. The curve colors in the main panels match the marker colors in this
  $\Om$--$\sigma_{8}$ panel. Shaded bands indicate $\pm5\%$ and $\pm10\%$.
  {Alt text: Ratios show broader deviations for the lower source redshift, with differences between fitting formulas growing toward higher multipoles and becoming more structured outside the calibration domain.}
  }
  \label{fig:apps_cl_fit}
\end{figure*}

\begin{figure*}[htbp]
  \centering
  \includegraphics[width=0.49\linewidth]{
    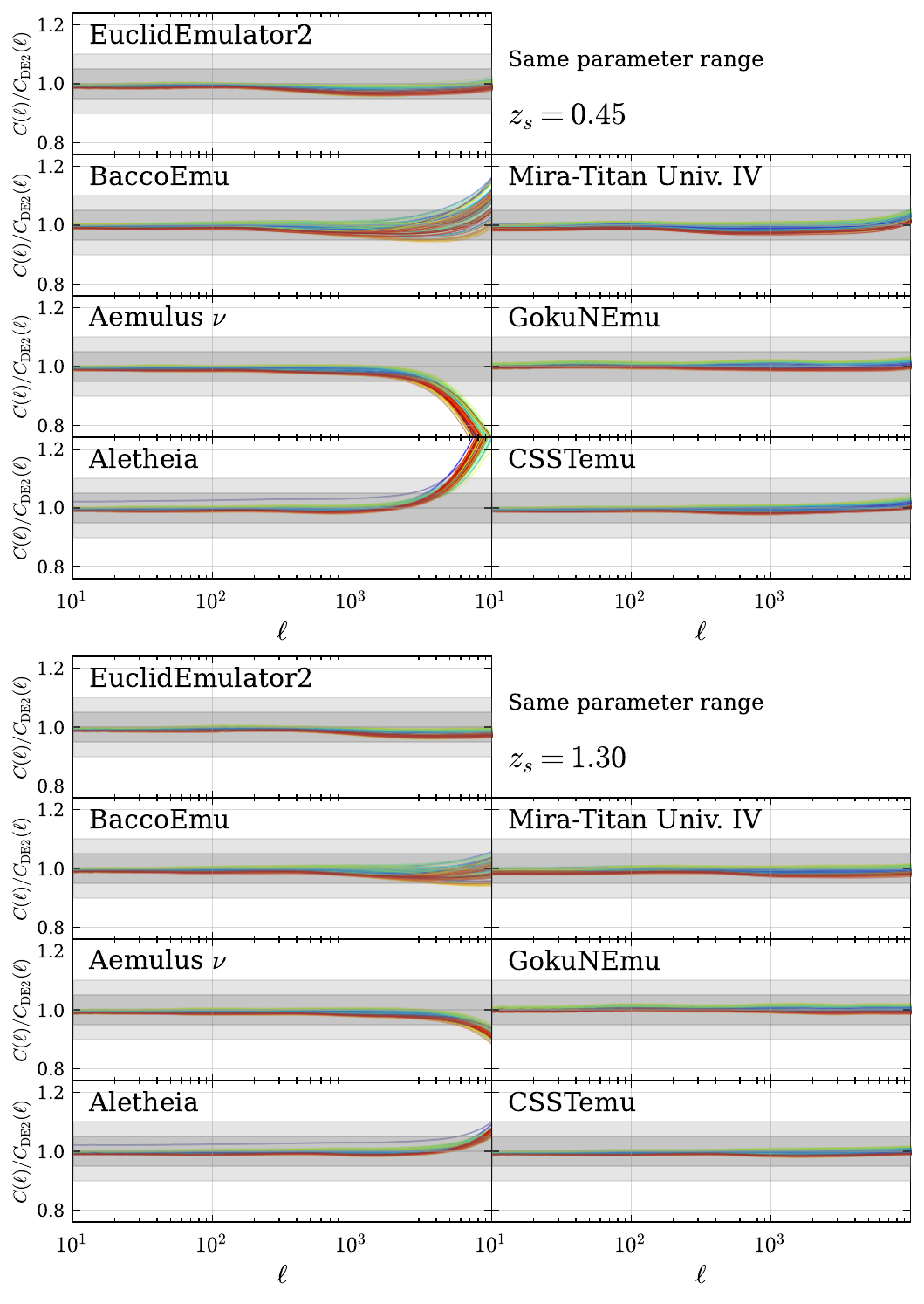
  }
  \includegraphics[width=0.49\linewidth]{
    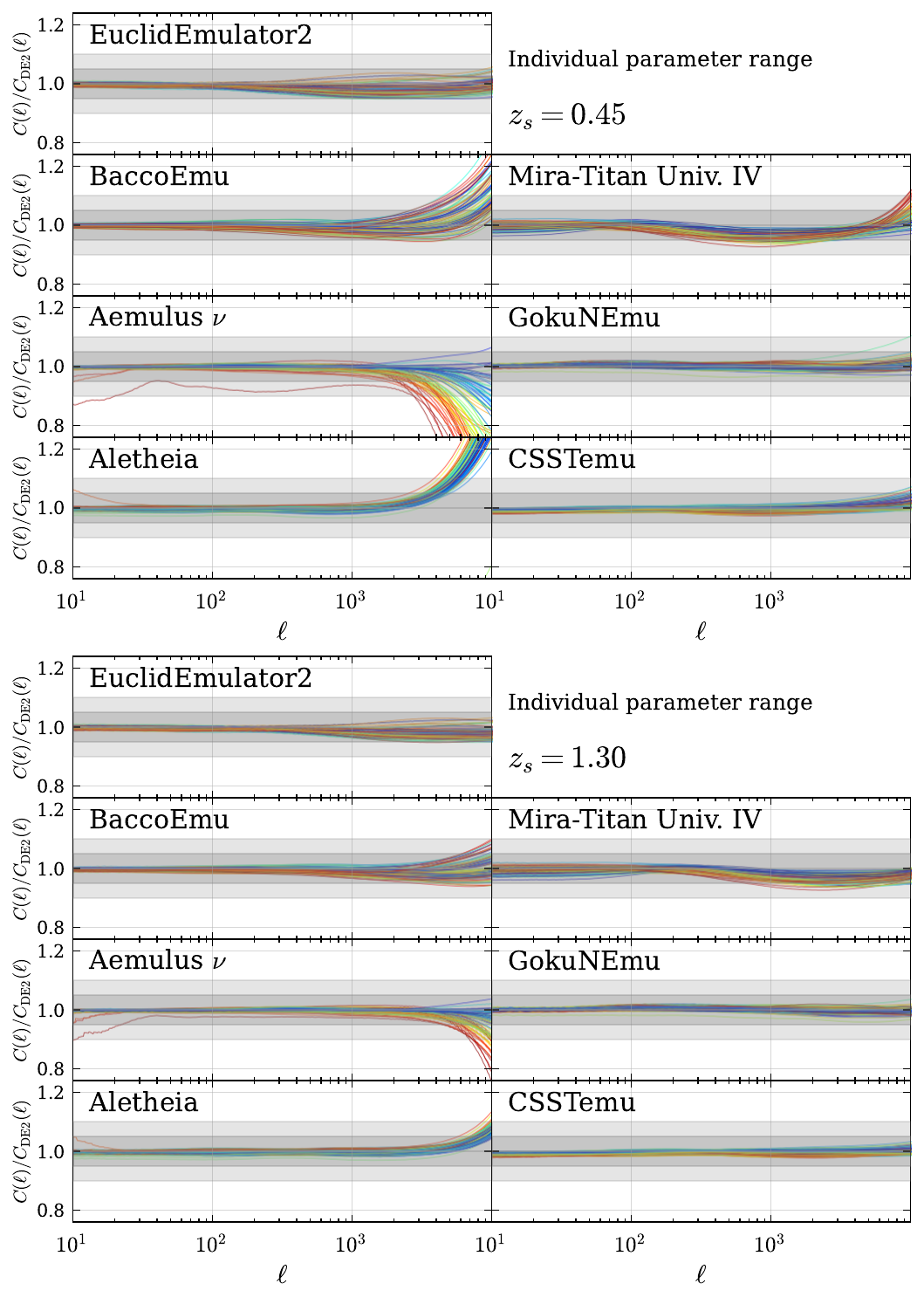
  }
  \caption{Comparison of $C(\ell)$ derived from nonlinear $P(k)$ for \DE\ and
  public emulators. We consider eight emulators in total, namely \DE\ and the
  seven public emulators shown. The left panels use 80 cosmologies sampled from the
  common overlap of all emulator parameter ranges with $M_{\nu}=0$, $w_{a}=0$
  and $\Ok=0$. The right panels use 80 cosmologies sampled from the overlap between
  \DE\ and each emulator, which probes closer to the support boundaries. The curves
  are color coded by $\Om$, with red and blue indicating lower and higher
  $\Om$, respectively. The top panels show $z_{s}=0.45$, and the bottom panels
  show $z_{s}=1.3$. Shaded bands indicate $\pm5\%$ and $\pm10\%$.
  {Alt text: Most convergence spectra derived from emulators remain close over intermediate multipoles, while the scatter increases toward higher multipoles and is larger in the samples from pairwise overlaps.}
  }
  \label{fig:apps_cl_emus}
\end{figure*}

As an example linking our nonlinear matter power spectrum to observables, we compare
weak-lensing power spectra predicted by our emulator with analytic fitting formulas
and other public emulators. For scalar gravitational lensing, the E-mode cosmic
shear power spectrum equals the convergence power spectrum, so throughout this
section we report $C(\ell)\equiv C_{\ell}^{\kappa}$ (e.g., \cite{Kilbinger2015-up}).
We then evaluate the convergence angular power spectrum under the extended
Limber approximation~\citep{LoVerde2008-oz}, given by
\be 
C(\ell) = \int_0^{\chi_s}
\! \mathrm{d}\chi\; \frac{g^{2}(\chi)}{f_{K}^{2}(\chi)}\; P_\mathrm{tot}\left(k=\frac{\ell+1/2}{f_{K}(\chi)};\;
z(\chi)\right), 
\label{eq:shear_cl_gl} 
\ee 
where $\chi=\chi(z)$ is the
comoving distance, $\chi_{s}=\chi(z_{s})$ is the source distance, and
$f_{K}(\chi)$ is the comoving angular diameter distance, 
\be 
&&f_K(\chi) =\left\{
\begin{array}{lll}
  \dfrac{\sin\left(\sqrt{K} \chi\right)}{\sqrt{K}}    & \left(K > 0 \right) \\
  \chi                                                & \left(K = 0 \right) \\
  \dfrac{\sinh\left(\sqrt{-K} \chi\right)}{\sqrt{-K}} & \left(K < 0 \right)
\end{array}
\right. \nonumber \\
\\[0.8em]
\nonumber &&\quad\qquad\qquad = \left\{
\begin{array}{lll}
  \dfrac{c \sin\left(\sqrt{-\Omega_{\mathrm{k},0}} \, \chi H_0/c \right)}{H_0 \sqrt{-\Omega_{\mathrm{k},0}}} & \left(\Omega_{\mathrm{k},0} < 0 \right) \\
  \chi                                                                                                       & \left(\Omega_{\mathrm{k},0} = 0 \right) \\
  \dfrac{c \sinh\left(\sqrt{\Omega_{\mathrm{k},0}} \, \chi H_0/c \right)}{H_0 \sqrt{\Omega_{\mathrm{k},0}}}  & \left(\Omega_{\mathrm{k},0} > 0 \right)
\end{array}
\right. , 
\ee 
where $\Omega_{\mathrm{k},0}= -K c^{2}/(a_{0}^{2}H_{0}^{2})$ and $a_{0}=1$ denote the present-day curvature density parameter and scale factor, respectively. 
These correspond to closed, flat, and open universes from top to bottom. 
The function $g(\chi)$ is the lensing kernel that describes gravitational lensing along the line of sight from the source to the observer. 
\be 
g(\chi) =
\frac{3\,\Omega_{\mathrm{m0}}H_{0}^{2}}{2\,a(\chi)\,c^{2}} \int_{\chi}^{\infty}
d\chi'\; \frac{f_{K}(\chi' - \chi)\,f_{K}(\chi)}{f_{K}(\chi')}\, n(\chi') .
\label{eq:shear_gl} 
\ee 
In general the sources are distributed in redshift
according to $n(\chi)$. For a simplified comparison in this section, however, we
assume a Dirac-$\delta$ source at $\chi_{s}$. Under this assumption the kernel
reduces to the weight 
\be 
W(\chi) = \frac{3\,\Omega_{\mathrm{m0}}H_{0}^{2}}{2\,a(\chi)\,c^{2}}\,
\frac{f_{K}(\chi_{s}- \chi)\, f_{K}(\chi)}{f_{K}(\chi_{s})} .
\label{eq:shear_wf} 
\ee 
Using $W(\chi )$, equation~\ref{eq:shear_cl_gl} becomes
\be 
C(\ell) = \int_{0}^{\chi_s} d\chi\; \frac{W^{2}(\chi)}{f_{K}^{2}(\chi)}\; P_{\mathrm{tot}}\!\left(
k=\frac{\ell + 1/2}{f_{K}(\chi)};\; z(\chi) \right). \label{eq:shear_cl_wf}
\ee 
Although we adopt the Limber approximation and a Dirac-$\delta$ source
here for simplicity, parameter inference should use the observed source redshift
distribution $n(z)$ (with photo-$z$ calibration and tomographic binning) and,
on the largest angular scales, consider corrections beyond the Limber approximation.

We tabulate $P_{\mathrm{tot}}(k,z)$ on 600 logarithmically spaced $k$ nodes over $10^{-2}<k<10^{3}\hMpci$ and 50 redshift bins from $z=0$ to $z_{s}$. Where a model's native coverage ends at $k_{\max}$, we extend the spectrum to $10^{3}\hMpci$ using the high-$k$ prescription described in appendix~\ref{append:extrapolation}.

Figure~\ref{fig:apps_cl_fit} compares equation~\ref{eq:shear_cl_wf} evaluated
with \DE-based $P_{\mathrm{tot}}(k)$ to \textsc{Halofit}~\citep{Takahashi12}
and \textsc{HMcode}~\citep{Mead2021-og} for $\wwnoCDM$, which includes spatial
curvature and thus goes beyond the original calibration domain of these fitting
prescriptions. We show two representative source redshifts, $z_{s}=0.45$ and $z
_{s}=1.3$. For surveys such as HSC, the multipole range $100<\ell<1000$ is the
most relevant. In all cases the discrepancies are larger at $z_{s}=0.45$ than
at $z_{s}=1.3$. For a fixed multipole $\ell$, the mapping
$k=(\ell+1/2)/f_{K}(\chi)$ shifts the dominant contribution of the line-of-sight
integral toward larger $k$ for the lower-redshift structures weighted more
strongly by the $z_{s}=0.45$ kernel. At $z_{s}=0.45$, the \textsc{Halofit}-based
$C(\ell)$ can differ from \DE\ by up to $\sim10\%$ at high $\Om$ within $100<\ell
<1000$, and it often exhibits a mild upward tilt relative to \DE.

Figure~\ref{fig:apps_cl_emus} shows ratios of $C(\ell)$ computed from the nonlinear
$P(k)$ of public emulators to the \DE\ prediction. In the left panels, we sample
80 cosmologies from the intersection of the parameter ranges of all emulators included
in the comparison. This restricts the test to a narrow region with $M_{\nu}=0$,
$w_{a}=0$ and $\Ok=0$ (see figure~\ref{fig:emus_ranges1} and
\ref{fig:emus_ranges2}). In the right panels, we sample 80 cosmologies from the
overlap between \DE\ and each emulator within the cosmological model supported
by that emulator. Over $100<\ell<1000$, the ratios remain close to unity for all
emulators in both sampling schemes, and most curves lie within the $\pm5\%$
band. Toward the largest multipoles shown, the scatter increases, most prominently for \BCE, \AEMU, and \ALETHEIA. \MTU\ shows a mild dependence on $\Om$, with larger offsets at lower $\Om$ in parts of the multipole range.
\EE, \GOKUN, and \CSST\ remain close to the \DE\ prediction over the main multipole range, with \GOKUN\ showing the smallest scatter across the curves shown. The \CSST-based curves show no large systematic offset in this projection, consistent with the close agreement seen in the corresponding matter power spectrum comparison on large and mildly nonlinear scales.
At large $\ell$, the Limber mapping in equation~\ref{eq:shear_cl_wf} links the signal to higher $k$, so differences in the high-$k$ behavior of $P(k)$ can translate into larger scatter in $C(\ell)$, consistent with figures~\ref{fig:comp_nlpk_emu1}--\ref{fig:comp_nlpk_emu3}.

\subsection{Demonstration for extended cosmologies}
\label{subsec:demo_cosmo}

\begin{figure*}[htbp]
  \centering
  \includegraphics[width=0.85\linewidth]{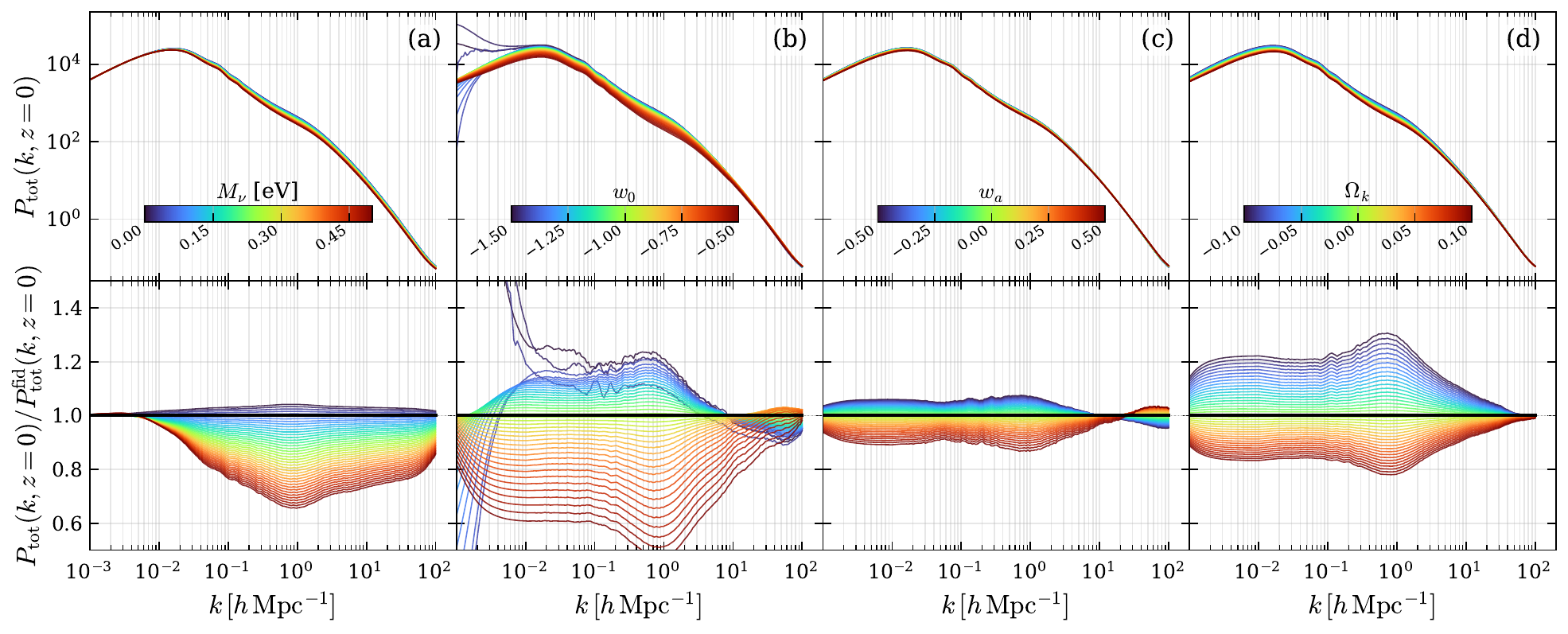}
  \caption{Response of the nonlinear total matter power spectrum to one-parameter
  variations around the fiducial cosmology at $z=0$. The upper row shows
  $P_{\mathrm{tot}}(k,z=0)$, and the lower row shows ratios to the fiducial
  prediction. Panels (a)--(d) vary $M_{\nu}$, $w_{0}$, $w_{a}$, and $\Ok$,
  respectively, with 40 sampled values for each parameter.
  {Alt text: One-parameter variations change the amplitude and scale dependence
  of the nonlinear matter power spectrum relative to the fiducial model.}
  }
  \label{fig:demo_ext_pk}
\end{figure*}

\begin{figure*}[htbp]
  \centering
  \includegraphics[width=0.85\linewidth]{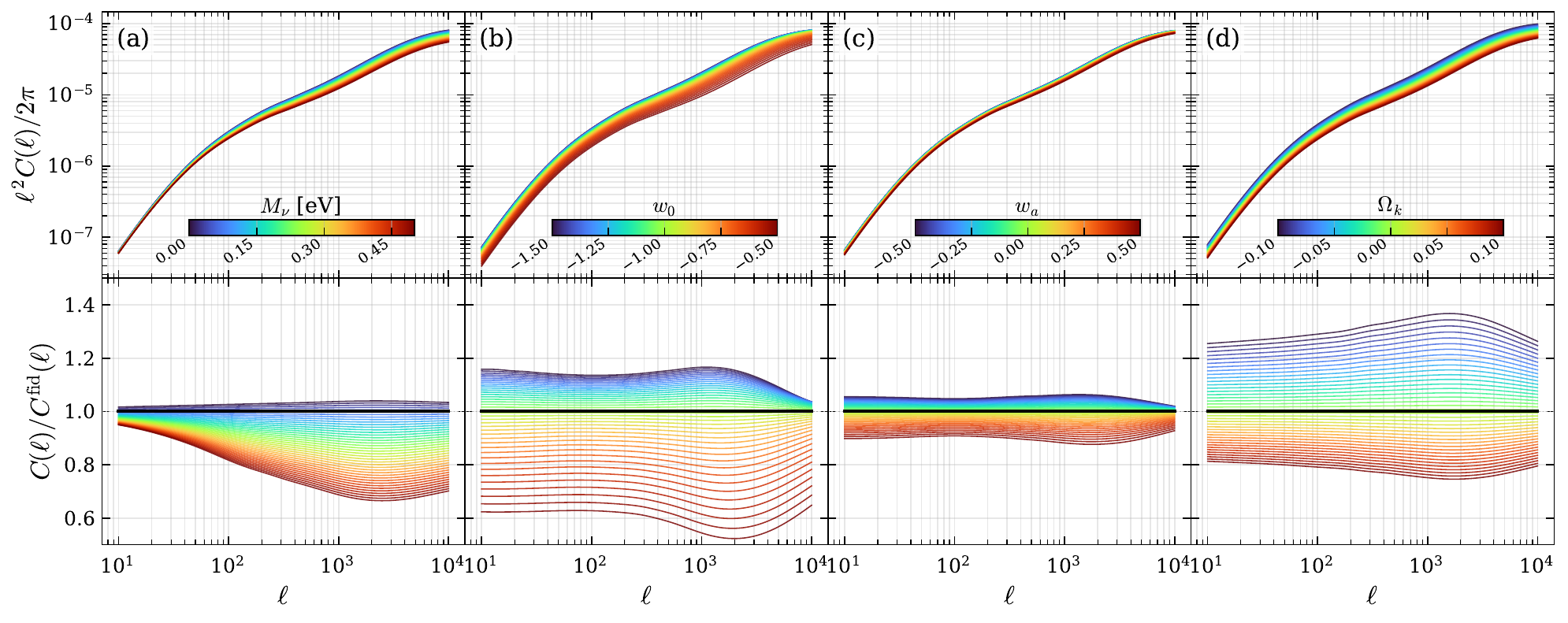}
  \caption{Same as figure~\ref{fig:demo_ext_pk}, but for the convergence power
  spectrum computed with a single source plane at $z_s=1$. The upper row shows
  $\ell^2 C_\ell/2\pi$, and the lower row shows $C_\ell/C_\ell^{\rm fid}$.
  {Alt text: The same parameter variations produce smooth changes in the
  convergence power spectrum after line-of-sight projection.}
  }
  \label{fig:demo_ext_cl}
\end{figure*}

\begin{table*}[htbp]
  \caption{DESI-based benchmark cosmologies used in figure~\ref{fig:demo_ext_desi}. Values are adopted from the posterior means reported in Table~5 of the DESI DR2 BAO cosmological analysis for the data combination listed in the second column. Parameters not specified by a given model are fixed to the fiducial \DE\ cosmology.}
  \label{table:demo_ext_cases}
  \centering
  \scalebox{1.0}{
  \begin{tabular}{llcccccccc}
    \hline
    \hline
    Model & Data combination
    & $\Om$ & $h$ & $w_0$ & $w_a$ & $\Ok$ & $\oc^*$ & $\sigma_8^*$ & $\Ode^*$ \\
    \hline
    fiducial
    & --
    & 0.3156 & 0.6740 & -1.000 & 0.00 & 0.0000 & 0.1198  & 0.8318 & 0.6844 \\
    DESI $\LCDM$
    & DESI+CMB
    & 0.3027 & 0.6817 & -1.000 & 0.00 & 0.0000 & 0.1178  & 0.8255 & 0.6973 \\
    DESI $o\LCDM$
    & DESI+CMB
    & 0.3034 & 0.6850 & -1.000 & 0.00 & 0.0023 & 0.1195 & 0.8321 & 0.6943 \\
    DESI $\wCDM$
    & DESI+CMB+Pantheon+
    & 0.3047 & 0.6797 & -0.995 & 0.00 & 0.0000 & 0.1179 & 0.8243 & 0.6953 \\
    DESI $w_0w_a\CDM$
    & DESI+CMB+Pantheon+
    & 0.3114 & 0.6751 & -0.838 & -0.62 & 0.0000 & 0.1190 & 0.8330 & 0.6886 \\
    DESI $o w_0w_a \CDM$
    & DESI+CMB+Pantheon+
    & 0.3117 & 0.6762 & -0.853 & -0.54 & 0.0011 & 0.1196 & 0.8337 & 0.6872 \\
    \hline
    \hline
  \end{tabular}}
  \begin{minipage}{0.95\linewidth}
    \footnotesize
    For the benchmark cases listed in this table, the nine-dimensional emulator input for each cosmology is specified by the unstarred values and by the fiducial \DE\ values $\ob=0.02225$, $\ns=0.9645$, $M_{\nu}=0.06\,\mathrm{eV}$, and $\lnAs=3.094$. The starred values are derived parameters recomputed from the same input specification and are listed for reference.
    The DESI $w_0w_a\CDM$ ($w_a=-0.62$) and DESI $ow_0w_a\CDM$ ($w_a=-0.54$) entries have $w_a$ values just outside the nominal emulator design range $w_a\in[-0.5,0.5]$, by $0.12$ and $0.04$ respectively, and thus correspond to mild extrapolations of the emulator.
  \end{minipage}
\end{table*}

\begin{figure*}[htbp]
  \centering
  \includegraphics[width=0.48\linewidth]{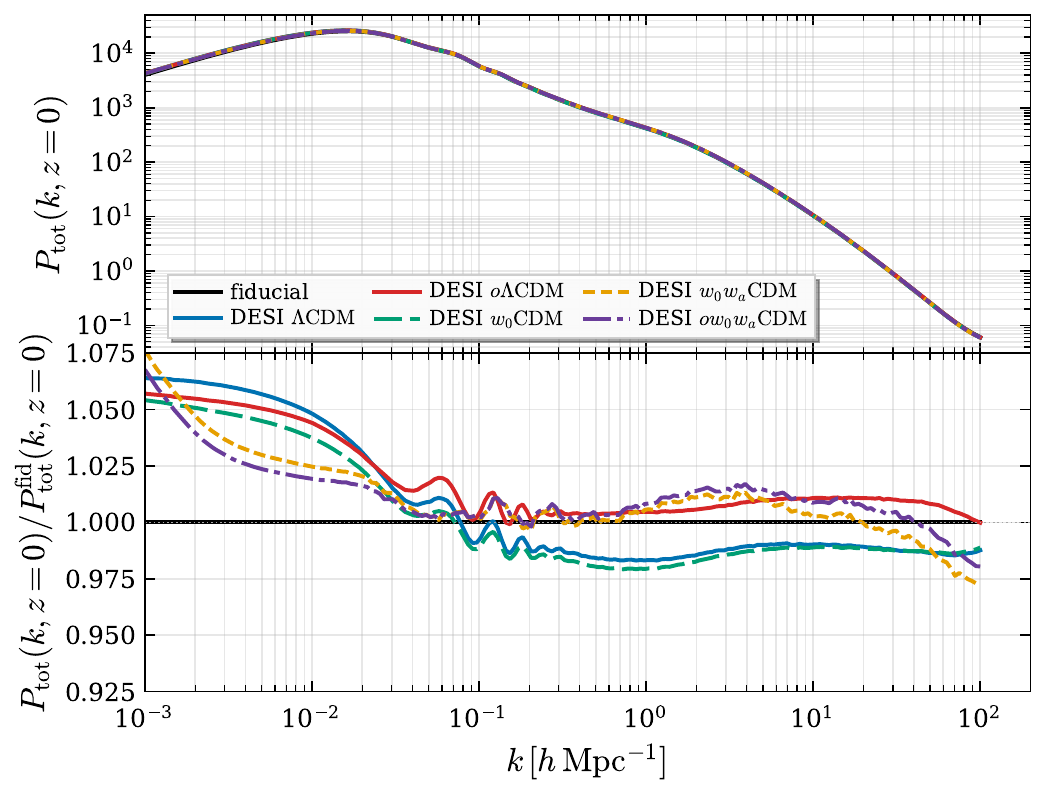}
  \includegraphics[width=0.48\linewidth]{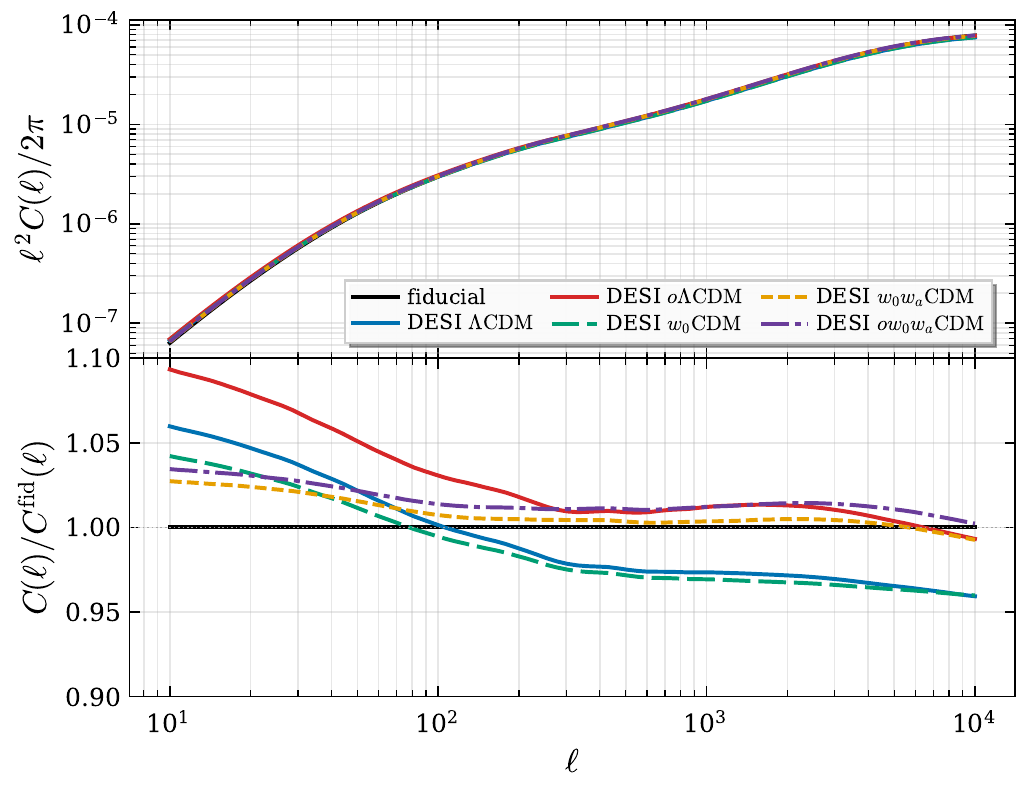}
  \caption{Clustering predictions for the DESI-based benchmark cosmologies
  listed in table~\ref{table:demo_ext_cases}. The left panel shows the nonlinear
  total matter power spectrum at $z=0$, and the right panel shows the convergence
  power spectrum computed with a single source plane at $z_s=1$. In each panel,
  the upper panels show the absolute prediction and the lower panels show
  the ratio to the fiducial \DE\ cosmology.
  {Alt text: DESI-based benchmark cosmologies give similar broadband spectra
  but visibly different ratios to the fiducial model.}
  }
  \label{fig:demo_ext_desi}
\end{figure*}

To illustrate the response of \DE\ across the extended cosmological parameter space, we evaluate the nonlinear total matter power spectrum and the corresponding
lensing convergence power spectrum for controlled parameter variations. This demonstration is not intended as a validation test, but rather as a visual check that the emulator predictions vary smoothly when parameters beyond the minimal $\LCDM$ model are changed.

We vary one extension parameter at a time around the fiducial cosmology:
$M_{\nu}$, $w_{0}$, $w_{a}$, and $\Ok$.  In the $M_{\nu}$ variation, we keep $\Om$ fixed and recompute the physical CDM density as
\be
\oc = \Om h^2 - \ob - \omega_{\nu}(M_{\nu}) ,
\ee
so that changing the neutrino mass does not inadvertently change either the total matter density or the curvature.  For all cases, the dark energy density
is recomputed from closure, $\Ode = 1-\Om-\Ok$.
This convention isolates the intended physical response: the neutrino mass sequence mainly changes the scale-dependent suppression of clustering, rather than introducing an artificial background density shift.

In addition to the one-parameter sequences, we evaluate DESI-based benchmark cosmologies selected from the posterior means reported in Table~5 of the DESI DR2 BAO cosmological analysis~\citep{DESI-DR2}. We use the $\LCDM$ and $o\LCDM$ posteriors from DESI+CMB, and the $\wCDM$,
$w_0w_a\CDM$, and $o w_0w_a\CDM$ posteriors from DESI+CMB+Pantheon+.
Parameters not specified by a given model are fixed to the fiducial \DE\ cosmology, as summarized in table~\ref{table:demo_ext_cases}. This setup illustrates how cosmologies selected primarily by background-expansion information can produce different clustering predictions.

This comparison is not intended as a parameter-inference analysis, but as a controlled demonstration of the ability of \DE\ to propagate DESI-motivated extended cosmologies to nonlinear matter power spectra and weak-lensing observables.

\section{Prediction time}
\label{sec:est_time}

\begin{table*}[hbtp]
  \caption{Wall-clock time per cosmology to evaluate the total matter power spectrum $P_{\mathrm{tot}}(k)$. The single redshift entry uses one input redshift per call, while the 50 redshifts entry reports the total over 50 redshifts spanning $0\le z\le 1$. Depending on the implementation, the 50 redshifts entry can reflect batched evaluation or repeated single redshift calls. Each entry is the mean over 20 randomly sampled cosmologies. The DE2 cost ratio is the ratio of the single redshift time to that of \DE. The $50/1$-$z$ cost is the ratio of the 50 redshifts time to the single redshift time.}
  \label{table:est_time}
  \centering
  \scalebox{1.0}{
  \begin{tabular}{lcccc}
    \hline
    \hline
    Estimator        & single redshift [s] & 50 redshifts [s] & DE2 cost ratio & $50/1$-$z$ cost \\
    \hline
    \DE              & 0.0169              & 0.1006           & 1.00           & 5.95            \\
    \DE\ + direct \textsc{Class}     & 21.3455             & 21.5335          & 1263.05        & 1.01            \\
    \EE              & 0.9182              & 1.0332           & 54.33          & 1.13            \\
    \BCE             & 0.0293              & 1.0536           & 1.73           & 35.96           \\
    \MTU             & 0.0117              & 0.2644           & 0.69           & 22.60           \\
    \AEMU            & 1.9274              & 94.2646          & 114.05         & 48.91           \\
    \GOKUN           & 0.0039              & 0.0042           & 0.23           & 1.08            \\
    \ALETHEIA        & 1.0490              & 52.1157          & 62.07          & 49.68           \\
    \CSST            & 0.0603              & 0.0619           & 3.57           & 1.03            \\
    \FEMU            & 0.0067              & 0.3390           & 0.40           & 50.60           \\
    \PA              & 0.0257              & 1.2068           & 1.52           & 46.96           \\
    \textsc{Halofit} & 7.2766              & 7.4387           & 430.57         & 1.02            \\
    \textsc{HMcode}  & 8.0016              & 8.0575           & 473.47         & 1.01            \\
    \hline
    \hline
  \end{tabular}
  }
\end{table*}

Table~\ref{table:est_time} reports the wall-clock time per cosmology to evaluate
the nonlinear total matter power spectrum $P_{\mathrm{tot}}(k)$ for a set of emulators
and fitting formulas. All timings were measured on an Intel Xeon Gold~6230 CPU
using 8 cores (8 OpenMP threads), and each entry is the mean over 20 randomly sampled
cosmologies.

\DE\ can either use its internal emulator for the linear power spectrum (see appendix~\ref{append:lin_pk_emu}) or compute $P_{\mathrm{lin}}(k)$ directly with the Boltzmann solver \textsc{Class}~\citep{class1,class2}. We therefore report two configurations: the default \DE\ setting and a diagnostic ``\DE\ + direct \textsc{Class}'' setting. In the latter case, we use the same high-accuracy \textsc{Class} configuration as in the simulation pipeline~\citep{Ginkaku26}. The relevant settings, with the corresponding
\textsc{Class} defaults in parentheses, are
\begin{quote}
   \scriptsize\ttfamily \raggedright
  P\_k\_max\_h/Mpc = 100 \quad (default: 1),\\
  tol\_perturbations\_integration = 1e-8 \quad (default: 1e-5),\\
  perturbations\_sampling\_stepsize = 1e-4 \quad (default: 0.1),\\
  tol\_ncdm\_bg = 1e-10 \quad (default: 1e-5),\\
  background\_Nloga = 4620 \quad (default: 3000).
\end{quote}
For \textsc{Halofit}~\citep{Takahashi12} and
\textsc{HMcode}~\citep{Mead2021-og}, we use the default precision settings of \textsc{Class}, except that the linear spectrum is computed up to $k_{\max}=10\,\hMpci$. Because the cost of a Boltzmann solver call depends on both the precision settings and the requested $k$ range, the absolute runtimes of Boltzmann-based pipelines in table~\ref{table:est_time} should be interpreted only for the settings specified here. The \DE\ + direct \textsc{Class} row is therefore a diagnostic timing for replacing the internal auxiliary emulators by the high-accuracy Boltzmann calculation used in our pipeline, not a default \textsc{Class} timing.

The wall-clock times reported in table~\ref{table:est_time} include any auxiliary computations performed internally by each public implementation to map cosmological parameters to $P_{\mathrm{tot}}(k)$, including the evaluation of $P_{\mathrm{lin}}(k)$ when required. In the \EE\ interface, $P_{\mathrm{lin}}(k)$ is computed by calling \textsc{Class} at its default accuracy settings, and the emulator then applies its nonlinear correction. In \ALETHEIA, the baseline linear power spectrum is computed with \textsc{Camb}. In \BCE, the nonlinear prediction is written as a boost factor multiplied by a linear spectrum, and by default the linear spectrum is evaluated by its internal linear emulator. \CSST\ emulates a correction relative to a \textsc{HMcode2020}-based baseline; we time the public matter-power-spectrum module as provided by its interface. Because \GOKUN\ directly emulates nonlinear power spectra at multiple resolutions, an explicit $P_{\mathrm{lin}}(k)$ calculation is not part of its core regression, although a separate linear power spectrum emulator is also provided. \MTU\ similarly does not rely on an explicit linear spectrum computation. The runtime of \AEMU\ includes not only the final mapping to $P_{\mathrm{tot}}(k)$ but also the evaluation of its HEFT module~\footnote{\url{https://github.com/AemulusProject/aemulus_heft}}. These implementation choices contribute to the spread in runtime among methods seen in table~\ref{table:est_time}.

The absolute timings should therefore not be interpreted as a direct measure of the regression model complexity alone. Very fast entries largely reflect public interfaces that avoid a fresh Boltzmann-solver call and evaluate compact interpolation or emulator representations directly, whereas slower entries may include baseline linear-spectrum calculations, additional internal modules, or redshift-by-redshift evaluations. The timings should therefore be interpreted together with the supported cosmological parameter space, the validated $(k,z)$ range, and the target quantity of each method.

In practical applications, one rarely needs $P(k)$ at a single redshift. Instead, $P(k)$ is typically required over a set of redshifts, for example when computing angular power spectra $C(\ell)$. To reflect this use case, we also measure the time to evaluate $P(k)$ on a grid of 50 redshifts spanning $0 \le z \le 1$. The last column of table~\ref{table:est_time}, labeled ``$50/1$-$z$ cost'', reports the ratio of the ``50 redshifts'' time to the ``single redshift'' time.
For methods dominated by the Boltzmann solver, this ratio is close to unity because the linear calculation is performed once and the incremental cost of evaluating additional redshifts is small. For emulators, the scaling depends on whether the implementation batches redshifts within a single call or evaluates them one by one. In \DE, supplying a list of redshifts triggers a single neural network evaluation for the full list. As a result, evaluating 50 redshifts costs only a factor of $\simeq 5$ more than a single redshift call, rather than scaling linearly with the number of redshifts.

Although emulators substantially accelerate nonlinear $P(k)$ evaluations, the per-call cost still matters because MCMC analyses may require tens of thousands of likelihood evaluations. In this regime, the differences in table~\ref{table:est_time} translate directly into substantial savings in total runtime.

\section{Summary and discussion}
\label{sec:summary}

In this work, we achieved subpercent-level agreement with the DQ2 simulation targets over the validated range by training a feed-forward neural network emulator on 1,000 cosmologies spanning a nine dimensional parameter space,
while keeping the evaluation time per cosmology practical for repeated calls in sampling-based inference pipelines, including MCMC (see table~\ref{table:est_time}).
This performance is enabled by three design choices that share a common motivation: in addition to the cosmological parameter vector, we supplement the neural network's inputs with three families of physically motivated auxiliary quantities that are expected to improve generalization across the parameter space. These are the linear matter power spectrum (which simplifies the regression target and encompasses the conventional boost-factor approach as a special case), descriptors of the simulation resolution, and a low-dimensional summary of the initial Gaussian random field; the latter two carry information about the actual spectrum realized in each finite-resolution, single-realization simulation that is not contained in the cosmological parameters alone.

As shown in figure~\ref{fig:pk_p9vsp9lp}, providing the linear power spectrum as an input improves the effective $k$-space resolution of the training targets and helps preserve BAO-scale oscillatory features.
Figure~\ref{fig:pk_mixed_emu} illustrates that mixed resolution training allows the emulator to leverage a small set of high resolution simulations
while maintaining broad coverage with lower resolution runs. 
Finally, figure~\ref{fig:pk_rand_check} shows that averaging the measured spectra over an ensemble of realizations reduces cosmic variance in the training targets, which stabilizes training and simplifies the accuracy assessment.

To place these results in context, we compared \DE\ with other public emulators and found that the agreement is not uniform across scales, redshifts, and model extensions (see figures~\ref{fig:comp_fid_emu}--\ref{fig:comp_nlpk_emu3}).
In practice, the choice of emulator should be guided by parameter coverage,
demonstrated accuracy over the relevant $(k,z)$ range, and the evaluation time required by the intended inference pipeline.

These cross-emulator differences also clarify how percent-level accuracy claims for nonlinear matter power spectrum emulators should be interpreted. A quoted accuracy is necessarily conditional: it refers to a specific simulation suite, measurement pipeline, parameter domain, redshift and wavenumber range, preprocessing procedure, and validation metric, so two emulators each reporting subpercent accuracy under their own validation criteria can disagree at the few-percent level on a common cosmology without internal contradiction.

Equally, the disagreement we observe should not be interpreted solely as a failure of the surrogate regression. It also combines differences in the underlying $N$-body solvers, treatment of massive neutrinos and other non-cold components, time-stepping and force-resolution choices, power spectrum estimation, conversion between cold-plus-baryon and total-matter spectra, and extrapolation or regularization prescriptions outside the directly simulated $(k,z)$ domain. As one illustration, when an emulator's regression target is defined relative to a separate fitting formula rather than to a simulation-derived quantity, residual mismatches between the fitting formula and the underlying $N$-body spectra are partially folded back into the emulator output, so what is sometimes read as ``emulator error'' is partly inherited from the analytical baseline.

The accuracy of an emulator cannot be separated from the accuracy and definition of the data on which it is trained. The companion \textsc{Ginkaku} paper~\citep{Ginkaku26} documents the upstream numerical ingredients --- the $N$-body solver, gauge treatment, linear-response implementation of massive neutrinos and other external sources, convergence tests, and post-processing pipeline --- that determine the simulation products on which \DE\ is trained, and the present paper propagates those products through a resolution-aware and realization-aware neural network and on to projected observables such as the cosmic-shear convergence spectrum. The two papers therefore document the full chain from $N$-body solver to weak-lensing observable, including the systematic-uncertainty contributions identified at each stage.

The parameter space covered by \DE\ is not tied to any specific cosmological observable, which allows it to be used across multiple observational programs and for extensions beyond $\LCDM$. This portability also makes \DE\ suitable for integration into modular parameter inference frameworks, such as \textsc{CosmoSIS}~\citep{Cosmosis} and \textsc{Cobaya}~\citep{Cobaya,Cobaya2}.  As an example of its use in cosmological inference, \DE\ has already been applied to an HSC-Y3 cosmic shear analysis~\citep{Terasawa2025-ar}.

In practical cosmological parameter inference, predictions for the matter power spectrum alone are often insufficient, and one typically requires halo and galaxy summary statistics that are more directly connected to observables. 
We therefore plan to extend the training framework of \DE\ to emulate halo and galaxy summary statistics, including observables based on HOD models and effects such as halo assembly bias. Such extensions will require reliable predictions across a wide halo mass range. This will likely require simulations with larger box sizes to reduce finite volume effects for rare massive halos, as well as higher mass resolution (smaller particle masses) to model the low-mass end accurately. 
Because our framework already takes inputs that describe numerical resolution, it provides a practical basis for combining simulations that vary in both box size and particle number. 

Beyond these extensions to more observable-level statistics, additional physical degrees of freedom remain outside the current model space, including $N_{\rm eff}$, the dark energy sound speed $c_s$, and baryonic or hydrodynamic effects.
Expanding both the physical model space and the numerical configurations will increase the dimensionality of the emulator inputs, making brute-force space-filling designs increasingly inefficient. In high-dimensional settings, fully predetermined space filling designs such as LHDs can become inefficient, whereas our WSD framework is straightforward to augment by appending new cosmologies.
Building on this flexibility, future simulation campaigns should be guided by sequential design criteria that use emulator uncertainty, observational posterior distributions, and estimated error contributions from numerical fidelity and realization scatter to decide which simulation should be added next.

These ideas fall under active learning, also termed sequential design, in the computer experiment literature, where a surrogate model is used to select new simulations that efficiently improve predictive performance~\citep{Sacks1989-ab,Santner2010-fa,Settles2009-gh}.
In cosmological applications, Bayesian optimization has been used to iteratively refine emulator training sets under a limited simulation budget~\citep{Rogers2019-xs,Takhtaganov2021-kk}.
Depending on the scientific goal, acquisition rules can prioritize regions of large predictive uncertainty to improve global accuracy, or concentrate new simulations in regions favored by observations to improve inference efficiency~\citep{Gutmann2016-va,Alsing2019-uc}.

For future extensions of \DE, however, the relevant design decision is broader than selecting only new cosmological parameter points. A more quantitative strategy would treat the next simulation as a joint choice over cosmology, numerical fidelity, and initial-condition realization: for example, whether to add a new low-resolution cosmology, promote an existing cosmology to a higher resolution or larger box, or generate additional random realizations at fixed cosmology. This problem is closely related to multi-fidelity computer experiments, where simulations at different levels of accuracy and cost are combined to improve predictions of the highest-fidelity response~\citep{Kennedy2001-nk,Ho2021-nu,Ho2023-ls}, and to stochastic simulation metamodeling, where the response is separated into an ensemble-mean function and an intrinsic realization-dependent component~\citep{Ankenman2010-qg}. In the latter setting, sequential design must decide whether the next run should explore a new input location or replicate an existing one to reduce stochastic noise and improve the local mean estimate~\citep{Binois2019-xn}. Our present implementation does not optimize these choices with an explicit acquisition function. Developing an error- and cost-aware acquisition criterion for this joint cosmology--fidelity--realization space is therefore a natural next step toward a more quantitative resolution-aware and realization-aware emulator construction.

Looking further ahead, the conditional nature of accuracy claims described above also frames how users should select and interpret an emulator in practice. A subpercent number on an emulator's own validation set is one piece of information, but it does not by itself answer the question of how accurately a given analysis can expect predictions in its own target cosmology, scale range, and observable. The structure of cross-emulator residuals --- where disagreement is located in the joint parameter, scale, and redshift space --- is in many cases more directly actionable for a specific application than a headline accuracy number, and we have aimed in this paper to expose that structure rather than smooth it away.

Cross-emulator comparison is itself a non-trivial exercise. Public emulators do not share a common parameter space: they differ in which extended parameters such as $w_a$, $\Omega_K$, $M_{\nu}$, $N_{\mathrm{eff}}$, or $\alpha_s$ are included, and even when the parameter list coincides, the supported ranges per dimension are not the same. A fair comparison must therefore restrict to a common parameter region, which typically lies near the center of the most permissive design and near the boundary of the most restrictive one, where each emulator's sampling density and extrapolation behavior also differ. Common conventions for the power-spectrum estimator, the $P_{\mathrm{cb}}\to P_{\mathrm{tot}}$ conversion, and the high-wavenumber closure used in downstream integrals are part of the same problem, since these choices can themselves contribute at the percent level once propagated to integrated observables such as $\xi(r)$ or $C(\ell)$.

Some elements of the community-level validation we would advocate are already partially assembled in the literature, including phase-matched initial conditions and explicit code-level convergence and code-to-code tests, parts of which are documented in the present paper and in the companion \textsc{Ginkaku} paper~\citep{Ginkaku26}. Natural next steps include explicit cross-code $N$-body comparisons on shared cosmologies with matched initial conditions, standardized high-wavenumber and projection prescriptions for downstream observables, blind cross-emulator challenges on a common set of test cosmologies, and treatment of the residual envelope between independently constructed emulators as an empirical input to theory-error covariances in survey likelihoods, rather than as a nuisance to be removed only at the presentation stage. Efforts of this kind would help separate surrogate-regression errors from simulation and pipeline systematics, and would provide a firmer empirical basis for theory-error models in next-generation survey analyses.

\section*{Funding}
This work is supported in part by MEXT/JSPS KAKENHI Grant Number JP19H00677, JP20H05861, JP21H01081, JP22K03634, JP24H00215, JP24H00221 and JP26H00402. We also acknowledge financial support from Japan Science and Technology Agency (JST) AIP Acceleration Research Grant Number JP20317829.

\begin{ack}
We thank Ryo Terasawa and Keitaro Ishikawa for useful discussions. 
Numerical simulations were carried out on Cray XC50 at the Center for Computational Astrophysics, National Astronomical Observatory of Japan. Code development and tests were partly carried out at the Yukawa Institute Computer Facility, Cray XC40 and Yukawa-21.
This research was supported by MEXT as ``Program for Promoting Researches on the Supercomputer Fugaku'' (Toward a unified view of the universe: from large scale structures to planets, JPMXP1020200109, project ID hp220173; Structure and Evolution of the Universe Unraveled by Fusion of Simulation and AI, JPMXP1020230406, project ID hp230204; Multi-wavelength Cosmological Simulations for Next-generation Surveys, JPMXP1020230407, project ID hp230202, hp240200 and hp250219) and used computational resources of supercomputer Fugaku provided by the RIKEN Center for Computational Science.
\end{ack}

\section*{Data and code availability}
The \DE\ code is publicly available at
\url{https://github.com/DarkQuestCosmology/dark_emulator2_public/}.
Supplementary material, including additional emulator-comparison results, is available at
\url{https://darkquestcosmology.github.io/dark_emulator2_supplement/}.
For extreme regions of the extended parameter space, particularly those involving large excursions in the dark energy sector, users are encouraged to replace the internal linear-power-spectrum emulator with a direct call to a Boltzmann solver, since linear-theory features associated with dark energy clustering and implementation choices in Boltzmann solvers can propagate into the nonlinear prediction.


\appendix

\section{Amplitude and linear power spectrum emulators}
\label{append:lin_pk_emu}

The nonlinear power spectrum emulator developed in this work uses the linear power spectrum in two ways. First, sampled values of the linear spectrum are used as additional input features together with the nine cosmological parameters. Second, on sufficiently large scales, where linear theory is
accurate, the nonlinear prediction is replaced by the linear spectrum. Because the nonlinear emulator uses $\sigma_{8}$ as its amplitude parameter, we also need to convert $\As$ to the corresponding $\sigma_{8}$ when the input amplitude is specified by $\As$. Both operations would otherwise require repeated calls to a Boltzmann solver. We therefore construct two auxiliary emulators: one for the amplitude conversion, $\As \leftrightarrow \sigma_{8}$, and one for the linear power spectrum. These auxiliary emulators allow the nonlinear emulator to be evaluated without invoking the Boltzmann solver. A single evaluation of the nonlinear emulator takes a few milliseconds, whereas a \textsc{Class} call with the high-precision settings and $k_{\max}=100\,\hMpci$ adopted in this work takes of order ten seconds per cosmology. These cost considerations motivate the auxiliary emulators described below.

\begin{table}[htbp]
  \caption{Parameter ranges used to generate the auxiliary training catalog for the amplitude and linear power spectrum emulators. The first nine rows list the independent input parameters. The remaining rows show dependent quantities derived from the accepted models. The nonlinear emulator ranges in table~\ref{table:params} are fully enclosed by the ranges listed here.}
  \label{table:lin_params}
  \centering
  \begin{tabular}{ccc}
    \hline
    \hline
    Parameter    & Range         & Fiducial value \\
    \hline
    $\Om$        & [0.03, 0.64]  & 0.3156         \\
    $\ob$        & [0.01, 0.04]  & 0.02225        \\
    $M_{\nu}$    & [0.0, 0.6]    & 0.06           \\
    $\Ok$        & [-0.12, 0.12] & 0              \\
    $\ns$        & [0.91, 1.02]  & 0.9645         \\
    $\sigma_{8}$ & [0.46, 1.36]  & 0.831          \\
    $h$          & [0.4, 1.0]    & 0.674          \\
    $w_{0}$      & [-1.6, -0.4]  & -1             \\
    $w_{a}$      & [-0.6, 0.6]   & 0              \\
    \hline
    $\oc$        & [0.005, 0.35] & 0.1198         \\
    $\Ode$       & [0.41, 0.94]  & 0.6843         \\
    $\lnAs$      & [0.53, 6.93]  & 3.0911         \\
    $S_{8}$      & [0.59, 0.96]  & 0.85233        \\
    \hline
    \hline
  \end{tabular}
\end{table}

The two auxiliary emulators are trained on an independently generated catalog of linear-theory calculations, rather than on a subset of the nonlinear simulation cosmologies. The enlarged ranges in table~\ref{table:lin_params} are introduced as a numerical buffer around the nonlinear emulator domain. They are not an additional physical prior and should not be interpreted as extending the calibrated or validated domain of the nonlinear emulator.

In the $\Om$--$\sigma_{8}$ plane, the auxiliary catalog uses the same spline-defined support region as the nonlinear simulation design, following the HSC-Y1-motivated curved degeneracy direction~\citep{Hamana2020-bh} shown in figure~\ref{fig:9d_params}. Within this support, the auxiliary models are drawn independently and are then accepted only after the additional physical and numerical cuts are satisfied. For the linear power spectrum emulator, we relax the distance cut in equation~\ref{eq:p_dist} from $D<5$ to $D<10$. This looser cut improves the stability of the auxiliary interpolation near the boundary of the nonlinear emulator domain, but it does not enlarge the region over which the nonlinear emulator has been calibrated or validated.

We generate 50,000 accepted cosmological models with \textsc{Class}~\citep{class1,class2}. Of these, 49,000 models are used for training and validation, and 1,000 models are reserved as an independent test set. The following subsections describe the amplitude conversion emulator and the linear power spectrum emulator separately.

\subsection{Amplitude emulator}

\begin{figure}[htbp]
  \centering
  \includegraphics[width=\linewidth]{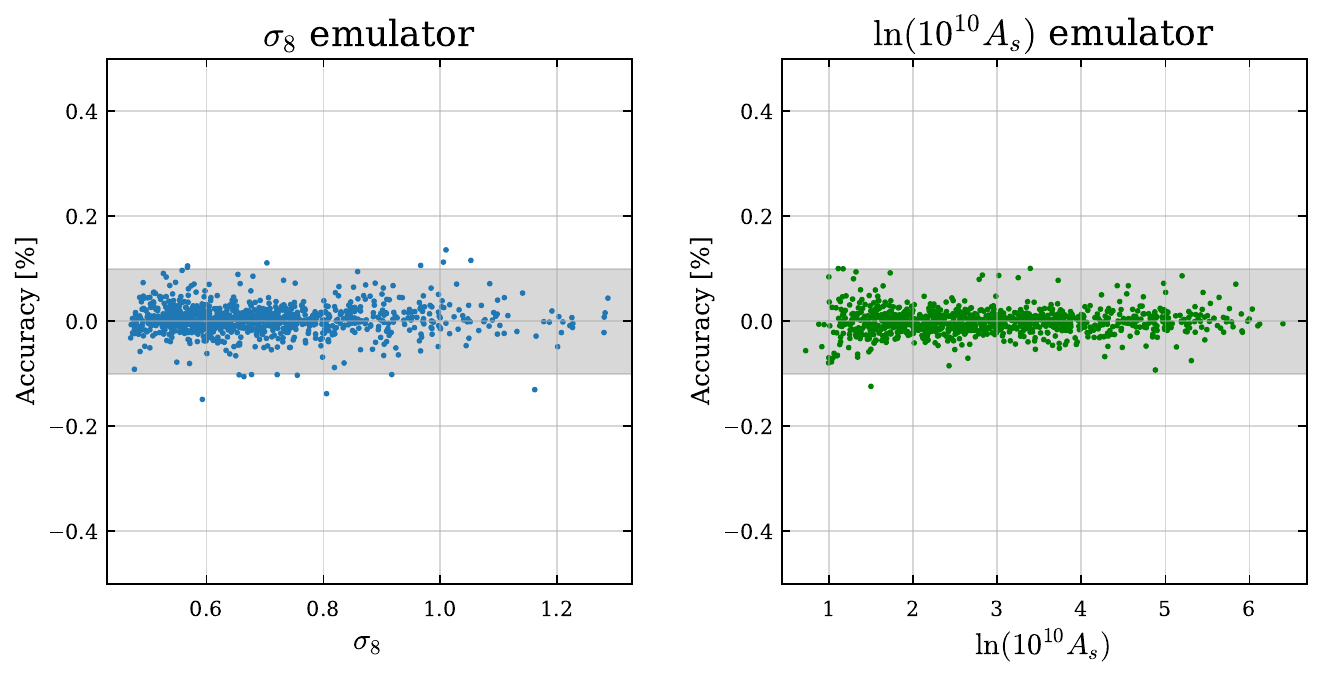}
  \caption{Accuracy of the amplitude emulator on 1,000 independent test cosmologies. The left panel shows the fractional error in the recovered $\sigma_{8}$ for inputs specified by $\As$. The right panel shows the fractional error in the recovered $\ln(10^{10}\As)$ for inputs specified by $\sigma_{8}$. The shaded band indicates a $0.1\%$ error range.
  {Alt text: Residuals for both amplitude conversions are tightly clustered within the displayed accuracy band, with no broad systematic trend across the test cosmologies.}
  }
  \label{fig:amp_emu}
\end{figure}

The amplitude parameter $\As$ is defined at the fixed pivot scale $k_{\mathrm{pivot}}=0.05\,\mathrm{Mpc}^{-1}$ adopted throughout this work. At fixed values of the remaining cosmological parameters, the linear power spectrum scales linearly with $\As$, and hence $\sigma_{8}$ scales as $\sqrt{\As}$. However, the coefficient relating $\sqrt{\As}$ to $\sigma_{8}$ depends on the shape and growth of the linear spectrum, and therefore on parameters such as $\Om$, $\ob$, $h$, $n_s$, $M_\nu$, $w_0$, $w_a$, and $\Ok$. Evaluating this coefficient directly would require computing the linear spectrum with a Boltzmann solver for each input cosmology. The amplitude emulator is introduced to replace this Boltzmann solver evaluation with a fast neural network prediction.

Figure~\ref{fig:amp_emu} shows the accuracy of the amplitude emulator on the 1,000 independent test cosmologies. The network architecture and training settings are listed in table~\ref{table:net_params}. The errors are centered close to zero and have a scatter well below the $0.1\%$ level. The relative errors remain within $0.1\%$ for both $\As \rightarrow \sigma_{8}$ and $\sigma_{8} \rightarrow \As$. When the input cosmology is specified with $\As$ rather than $\sigma_{8}$, the amplitude emulator first predicts the corresponding $\sigma_{8}$, which is then used as the $\sigma_{8}$ component of the nine-dimensional input vector for the nonlinear power spectrum emulator.
The sub-$0.1\%$ conversion error is expected to be subdominant relative to the target accuracy of the nonlinear emulator. We therefore do not treat the amplitude conversion error as a separate contribution to the nonlinear power spectrum error budget.

\subsection{Linear power spectrum emulator}

\begin{figure}[htbp]
  \centering
  \includegraphics[width=\linewidth]{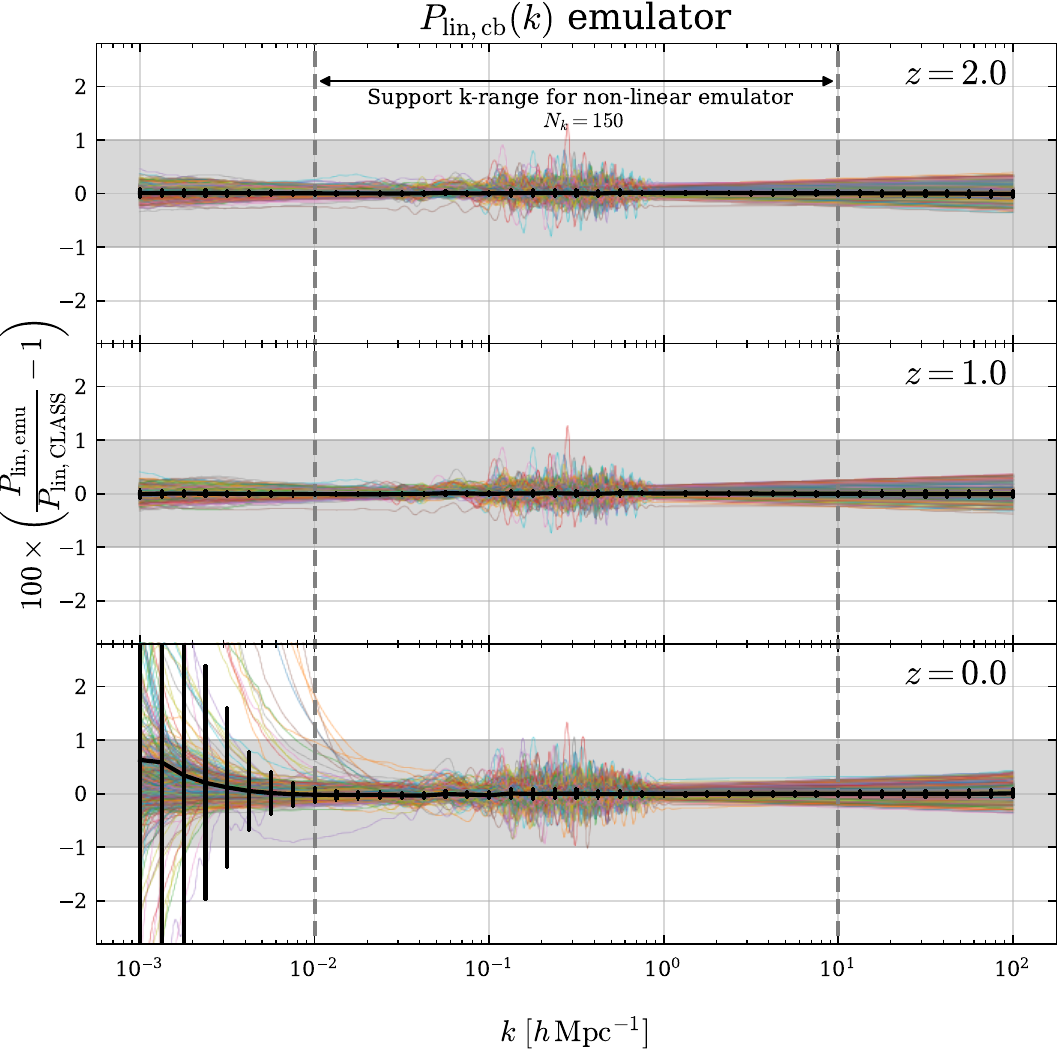}
  \caption{Accuracy of the linear power spectrum emulator for $P_{\mathrm{lin,cb}}(k)$ evaluated against \textsc{Class}. Colored curves show the residuals for 1,000 independent test cosmologies, and black curves show the mean and $1\sigma$ scatter. The shaded band indicates a $1\%$ error range. The vertical dashed lines enclose the 150 $k$ bins used as inputs to the nonlinear power spectrum emulator.
  {Alt text: Most residuals in the linear spectrum remain near zero over the main wavenumber range, while a small number of curves show larger deviations at very low wavenumber.}
  }
  \label{fig:lin_pk_emu}
\end{figure}

We first assess the overall accuracy of the linear power spectrum emulator on the independent test set. We then isolate a low-$k$ boundary case associated with extreme phantom dark energy models, and finally use two-dimensional accuracy maps to locate this degradation in parameter space.

The linear power spectrum emulator is trained as an explicit function of redshift, because massive-neutrino free streaming induces scale-dependent growth that cannot be represented by a simple rescaling of the $z=0$ spectrum. For each accepted cosmology, the target spectra are evaluated at 20 redshifts
between $z=0$ and 6, namely $z=6, 5.92, 5.31, 4.75, \ldots, 0.32, 0.2, 0.1,$ and $0$. The network architecture and training settings are listed in table~\ref{table:net_params}.

Figure~\ref{fig:lin_pk_emu} shows the accuracy against \textsc{Class} for the 1,000 independent test cosmologies. Over the 150 logarithmically spaced $k$ bins used as inputs to the nonlinear power spectrum emulator, the emulator reproduces the overall shape of $P_{\mathrm{lin,cb}}(k,z)$ with high precision. The typical scatter is below $0.1\%$, although a small number of test cases show visible residual oscillations around the BAO scale. The residuals remain within the shaded $1\%$ band over the displayed range, except for the low-$k$ boundary behavior discussed below.

The main limitation appears at very small wavenumbers ($k<0.01\,\hMpci$) and low redshift for a small number of extreme phantom models, primarily with $w_0<-1.4$. In this corner, the \textsc{Class} spectra can show a sharp large-scale suppression associated with the treatment of dark energy perturbations. By default, \textsc{Class} uses the Parametrized Post-Friedmann (PPF) prescription~\citep{Hu2007-ss} to maintain numerical stability across the phantom divide. The resulting feature is most visible in the $N$-body gauge~\citep{2015PhRvD..92l3517F}, which is the gauge convention used for consistency with the Newtonian $N$-body simulations. Such sharp large-scale features are difficult for a smooth emulator to reproduce exactly. We therefore regard the corresponding low-$k$ degradation as a limitation of the auxiliary linear spectrum interpolation in a gauge-sensitive boundary region, rather than as a failure of the nonlinear emulator calibration on the scales used for validation.

\begin{figure}[htbp]
  \centering
  \includegraphics[width=\linewidth]{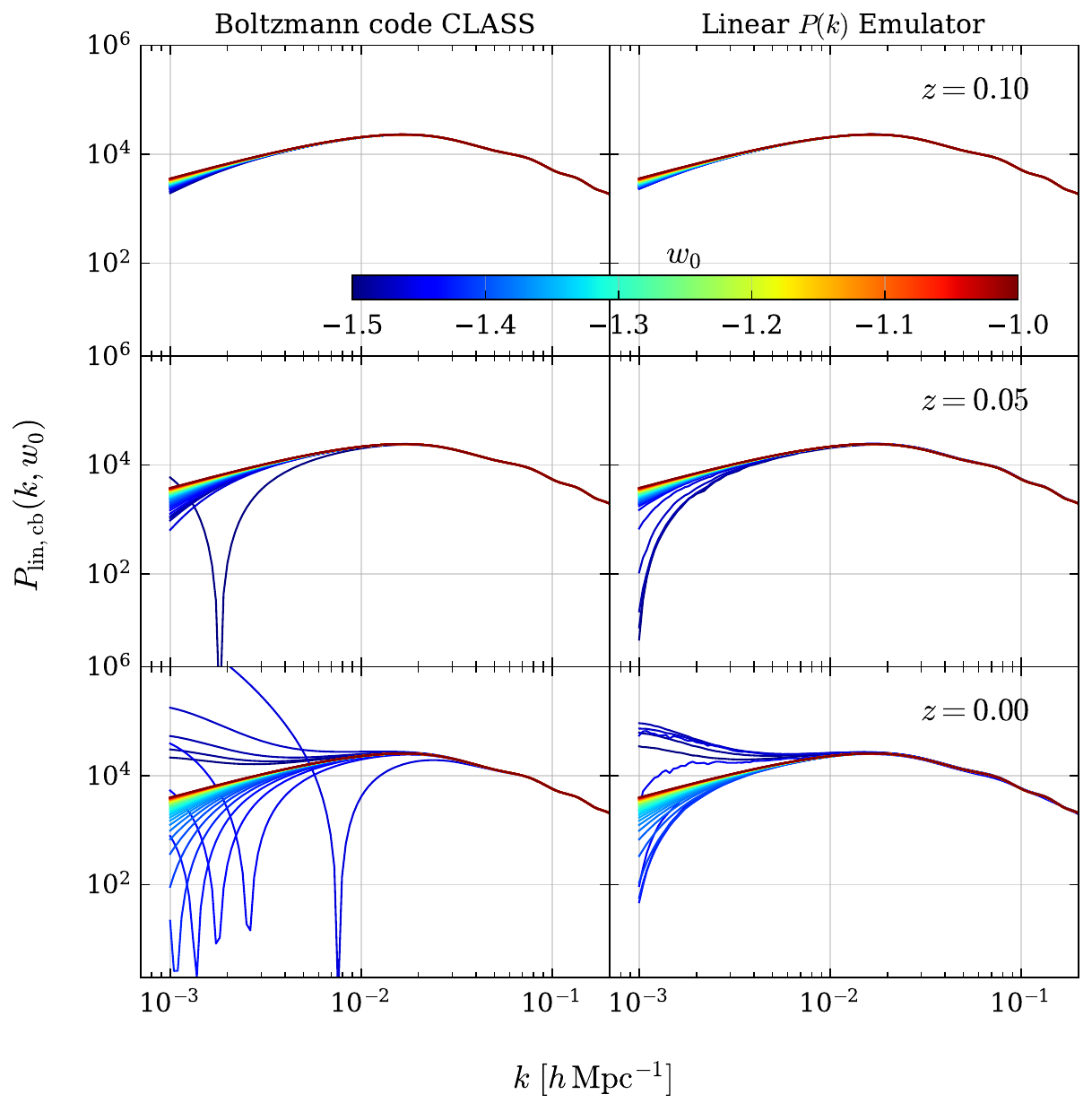}
  \caption{One-parameter diagnostic for an extreme dark energy model. The value of $w_0$ is varied while the remaining cosmological parameters are fixed at their fiducial values. For $w_0<-1.4$, the \textsc{Class} spectrum can show
  sharp large-scale damping at very low $k$ and low redshift. Such non-smooth features in the target spectra are difficult for the emulator to reproduce.
  {Alt text: The emulator follows the Class spectrum over most scales but does not fully reproduce the sharp large-scale damping that appears for the extreme dark energy model.}}
  \label{fig:lin_pk_w0}
\end{figure}

Figure~\ref{fig:lin_pk_w0} illustrates this behavior with a one-parameter sequence in which $w_0$ is varied while all other cosmological parameters are held fixed at the fiducial values. This figure is intended as a diagnostic of the low-$k$ boundary case, not as a representative example of the typical linear emulator accuracy over the training domain.

\begin{figure}[htbp]
  \centering
  \includegraphics[width=0.49\linewidth]{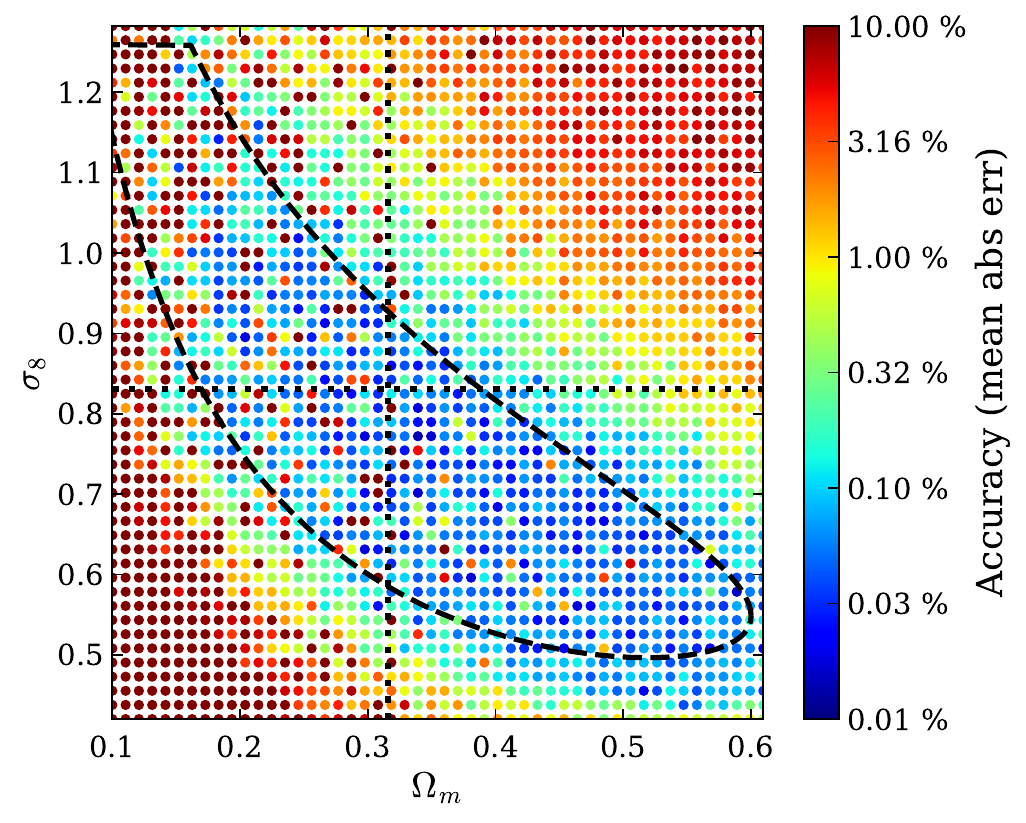}
  \includegraphics[width=0.49\linewidth]{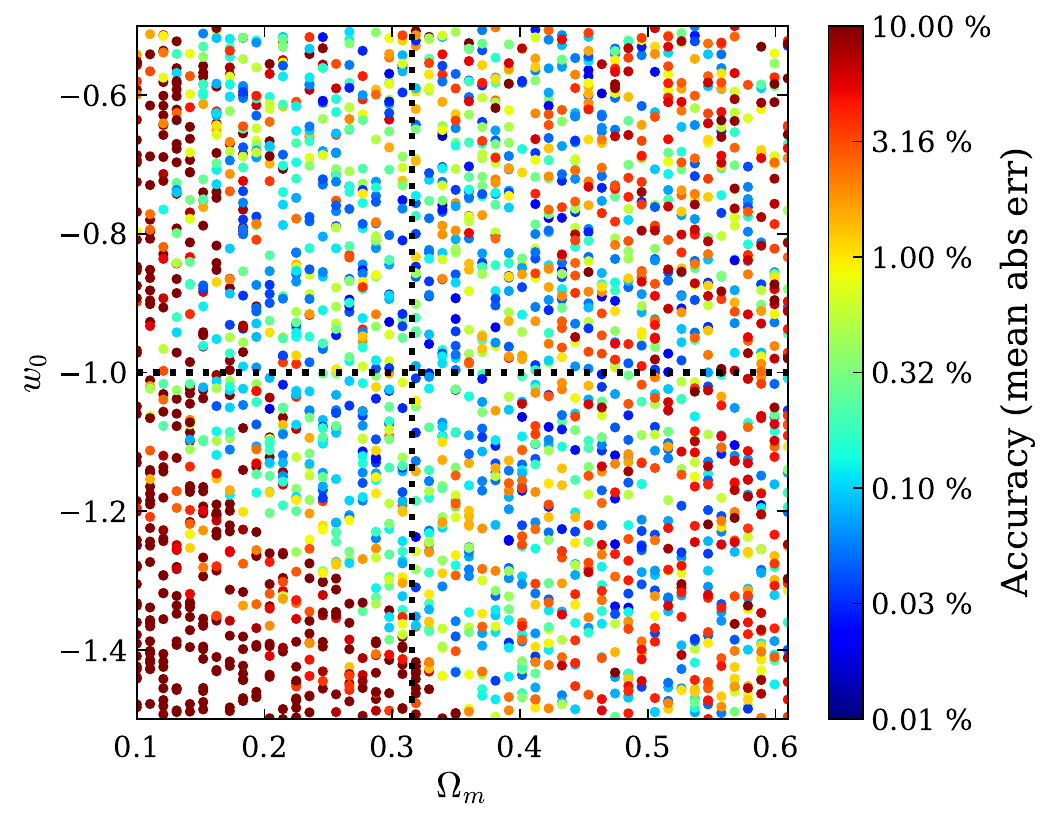}
  \caption{Two-dimensional accuracy of the linear power spectrum emulator in the $\wwnoCDM$ model. The left and right panels show the $\Om$--$\sigma_8$ and $\Om$--$w_0$ planes, respectively. Accuracy is quantified as the mean absolute percentage error relative to \textsc{Class}. Larger errors appear in the strongly negative-$w_0$ and low-$\Om$ region and near parts of the boundary of the support region, including the high-$\sigma_8$ side. The black dashed line delineates the support region used by the nonlinear $P(k)$ emulator, and the dotted lines mark the fiducial values.
  {Alt text: Larger errors appear mainly for strongly negative dark energy equation of state, low matter density, and parts of the support boundary, while much of the sampled domain remains accurate.}}
  \label{fig:Om_sig8_w0_acc}
\end{figure}

To localize the low-$k$ degradation in parameter space,
figure~\ref{fig:Om_sig8_w0_acc} shows two-dimensional accuracy maps on the $\Om$--$\sigma_8$ and $\Om$--$w_0$ planes. The maps are based on 2,500 evaluation points, with the remaining cosmological parameters sampled within their supported ranges. The largest errors are not controlled by $w_0$ alone.
They are enhanced when strongly negative $w_0$ is combined with low $\Om$, consistent with the sharp large-scale damping in the \textsc{Class} spectra discussed above. The $\Om$--$\sigma_8$ projection also shows that larger errors can occur near parts of the boundary of the support region, including the high-$\sigma_8$ side. Since the emulated linear spectra enter the nonlinear emulator both as input features and as large-scale replacements, the accuracy degradation in these regions can propagate to the final nonlinear prediction. We therefore treat these regions as localized caveats of the full prediction pipeline in boundary applications.

\begin{figure}[htbp]
  \centering
  \includegraphics[width=0.5\linewidth]{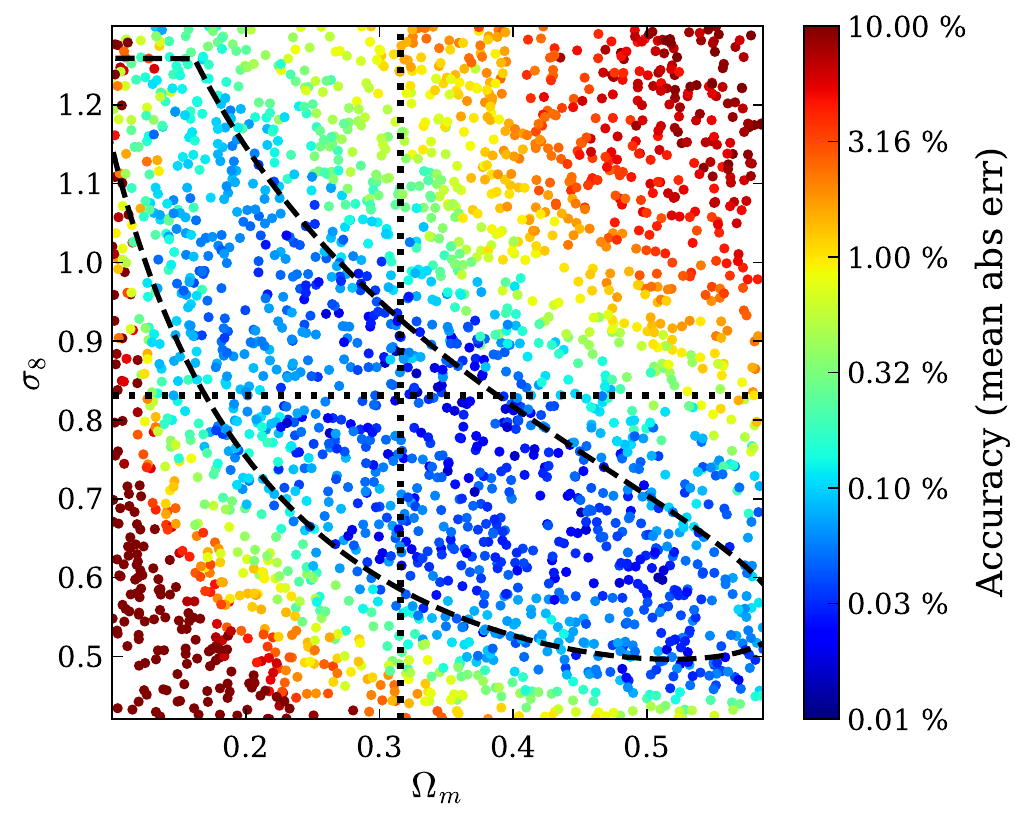}
  \caption{Two-dimensional accuracy of the linear power spectrum emulator for $\LCDM$ cosmologies projected onto the $\Om$--$\sigma_8$ plane. The sampled region is motivated by the HSC-Y3 cosmic shear parameter priors of \citet{Terasawa2025-ar} and restricted to the banana-shaped support used by the nonlinear emulator, which was designed to cover the HSC-Y1-motivated $\Om$--$\sigma_8$ posterior region.
  {Alt text: The accuracy map is broadly uniform across the displayed Lambda cold dark matter parameter region, without a strong localized degradation pattern.}}
  \label{fig:Om_sig8_acc_LCDM}
\end{figure}

As a reference check closer to conventional $\LCDM$ weak lensing analyses, figure~\ref{fig:Om_sig8_acc_LCDM} repeats the same diagnostic for $\LCDM$ cosmologies in the $\Om$--$\sigma_8$ plane. The sampled region is motivated by the cosmological parameter priors adopted in the HSC-Y3 cosmic shear analysis of \citet{Terasawa2025-ar}, as summarized in table~I of that paper, and is restricted to the banana-shaped support used by the nonlinear emulator. This support was designed to cover the HSC-Y1-motivated $\Om$--$\sigma_8$ posterior region described in section~\ref{subsec:sampling}. Within this relevant support region, the accuracy is approximately uniform and shows no localized degradation comparable to that seen in the extreme $w_0<-1.4$ boundary case.

The degradation seen in figure~\ref{fig:Om_sig8_w0_acc} is therefore best interpreted as a boundary caveat associated with a specific combination of low $\Om$ and strongly negative $w_0$, rather than as a generic limitation within the $\LCDM$ reference region considered above. This interpretation is also consistent with the DESI-based illustrative cases used in section~\ref{subsec:demo_cosmo}, which are based on the DESI DR2 cosmological analysis~\citep{DESI-DR2} and lie near $w_0\simeq -1$ or on the $w_0>-1$ side, rather than in the $w_0<-1.4$ corner. We therefore regard the localized low-$k$ degradation in the extended $\wwnoCDM$ diagnostic as a boundary caveat of the full prediction pipeline, not as representative behavior in the $\LCDM$ reference region or in the DESI-motivated examples considered in this work.

\section{Conversion from $P_{\mathrm{cb}}$ to $P_{\mathrm{tot}}$}
\label{append:cb_to_tot}

The \DE network emulates the cold components power spectrum $P_{\mathrm{cb}}$. 
Some cosmological applications, most notably weak lensing and related projected statistics, require the total matter power spectrum $P_{\mathrm{tot}}$, which includes the contribution of massive neutrinos. 
We therefore map $P_{\mathrm{cb}}$ to $P_{\mathrm{tot}}$ using a scale- and redshift-dependent ratio calibrated in linear theory.

We define the linear theory total-to-cb ratio as
\be 
\mathcal{R}^\mathrm{lin}_\mathrm{tot/cb}(k,z)
\equiv \frac{P_{\mathrm{lin,tot}}(k,z)}{P_{\mathrm{lin,cb}}(k,z)}\,, 
\label{eq:Rlin_def}
\ee 
where $P_{\mathrm{lin,cb}}$ and $P_{\mathrm{lin,tot}}$ are computed with \textsc{Class}
for identical cosmological parameters. Equivalently, 
\be 
P_\mathrm{lin,tot}(k,z)
= \mathcal{R}^\mathrm{lin}_\mathrm{tot/cb}(k,z)\,P_\mathrm{lin,cb}(k,z)\,. 
\label{eq:lin_map}
\ee

\begin{figure}[htbp]
  \centering
  \includegraphics[width=\linewidth]{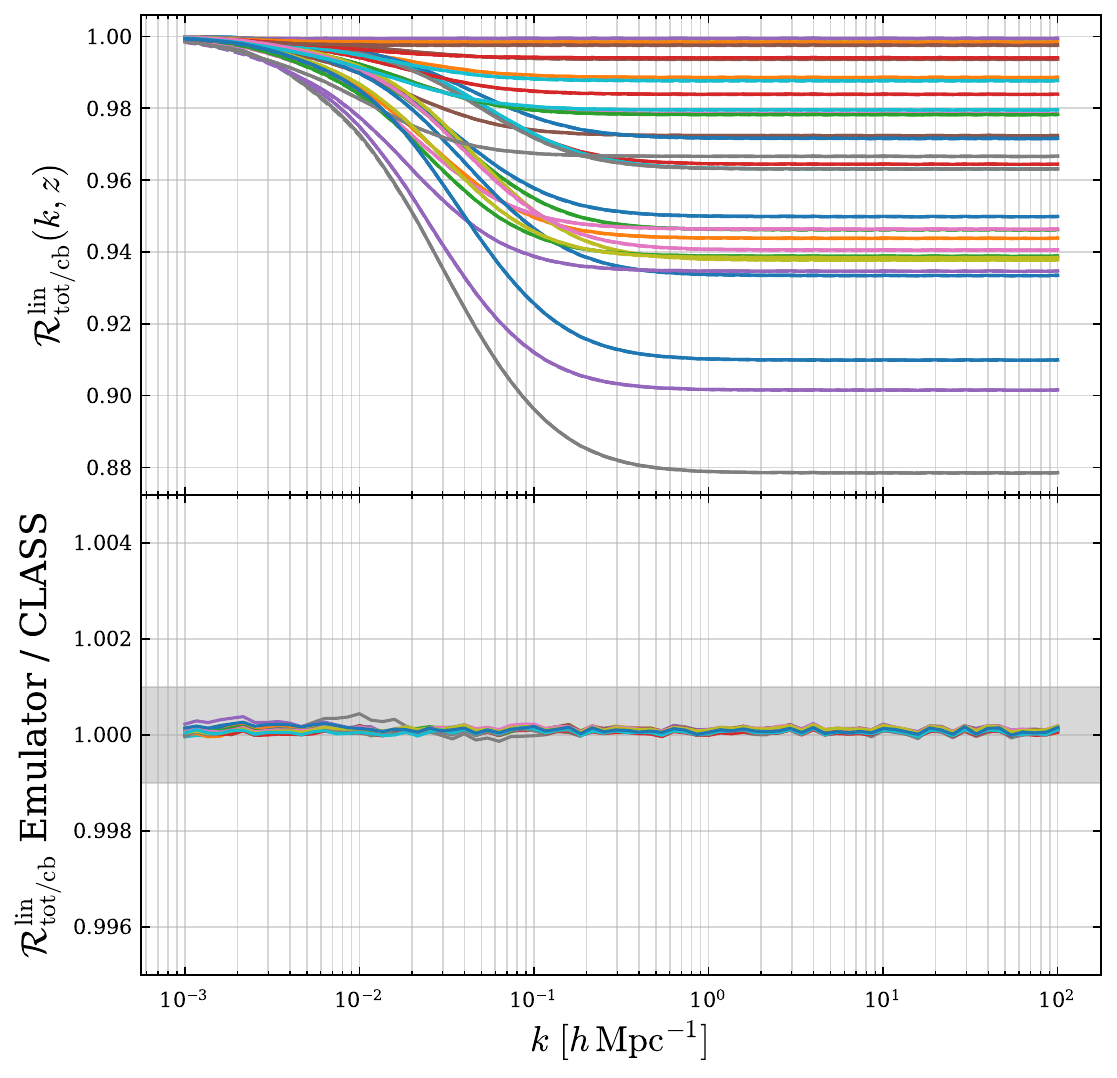}
  \caption{Linear theory total-to-cb ratio
  $\mathcal{R}^{\mathrm{lin}}_\mathrm{tot/cb}(k,z)$ used to convert
  $P\mathrm{cb}$ to $P_{\mathrm{tot}}$. The upper panel compares representative
  cosmologies, with solid curves from the emulator and dotted curves from
  \textsc{Class}. The lower panel shows the emulator-to-\textsc{Class} ratio;
  the shaded band indicates a $0.1\%$ range. 
  {Alt text: The emulator curves closely overlap the Class reference curves, and the residual panel shows deviations confined within the narrow displayed accuracy band.}
  }
  \label{fig:cb_to_tot_factor}
\end{figure}

In the DQ2 setup, which adopts a linear response treatment of massive neutrinos~\citep{2013MNRAS.428.3375A}, we train a compact neural network (see table~\ref{table:net_params}) to emulate $\mathcal{R}^{\mathrm{lin}}_{\mathrm{tot/cb}}(k,z)$ across our cosmological parameter space, using \textsc{Class} predictions as training targets. 
Figure~\ref{fig:cb_to_tot_factor} shows that the emulator reproduces the \textsc{Class}-derived ratio to better than 0.1\% over the $k$ range and cosmologies considered. 
For nonlinear predictions, we apply this linear theory ratio multiplicatively to the nonlinear $P_{\mathrm{cb}}$, 
\be 
\label{eq:nonlinear_map} P_\mathrm{tot}(k,z)
\approx \mathcal{R}^\mathrm{lin}_\mathrm{tot/cb}(k,z)\,P_\mathrm{cb}(k,z)\,. 
\ee
This factorization is consistent with the linear response picture. 
The dominant scale dependence associated with neutrino free-streaming is already present at linear order, while the leading nonlinear evolution is expected to arise mainly from the cb sector.
Within the DQ2 domain, the residual error from this mapping remains below the intrinsic error budget of the nonlinear emulator.

\section{Sensitivity diagnostics}
\label{append:sensitivity}

This appendix presents sensitivity diagnostics for the trained neural network emulator. We perform this analysis for the fixed-phase low-resolution emulator discussed in Section~\ref{subsec:9d_param_linpk}, for which a sufficiently large independent test set is available. Specifically, we use the 30 \textsc{TLF} test cosmologies at $z=0$. When the sampled linear power spectrum is included as an input, it is recomputed from the modified cosmological parameter vector and is not treated as an independently varied feature.

\subsection{Permutation feature importance}

Permutation feature importance (PFI; \cite{Breiman2001-RF}) is a model-agnostic method for quantifying how much each input feature contributes to the prediction accuracy of a trained model. 
In the present analysis, we quantify the importance of the nine cosmological parameters. 
When the sampled linear power spectrum is included in the inputs, it is regenerated from the perturbed cosmological parameter vector and is not treated as an independently permuted feature.

For a given feature $j$, we compute PFI as follows. 
We first evaluate the emulator on the test set using the original unshuffled features and compute the mean absolute percentage error, denoted by $\overline{\mathrm{MAPE}}_{\mathrm{truth}}$.
We then randomly permute the values of feature $j$ across the test set, reconstruct the full input vector for each perturbed sample, and recompute the MAPE. 
When the sampled linear power spectrum is included in the inputs, we recompute $\mathbf{p}_{\mathrm{lin}}$ from the perturbed cosmological parameter vector so that the inputs remain physically consistent.

Repeating this procedure several times and averaging over the permutations yields $\overline{\mathrm{MAPE}}_{\mathrm{PFI},j}$. The importance of feature $j$ is defined as 
\be 
I_j \equiv
\frac{\overline{\mathrm{MAPE}}_{\mathrm{PFI},j}}{\overline{\mathrm{MAPE}}_{\mathrm{truth}}}.
\label{eq:pfi_1d}
\ee 
If $I_{j}\simeq 1$, permuting the feature has little impact on the prediction error and the feature is effectively irrelevant for the emulator, whereas larger values indicate more important features.

\begin{figure}[htpb]
  \centering
  \includegraphics[width=0.7\linewidth]{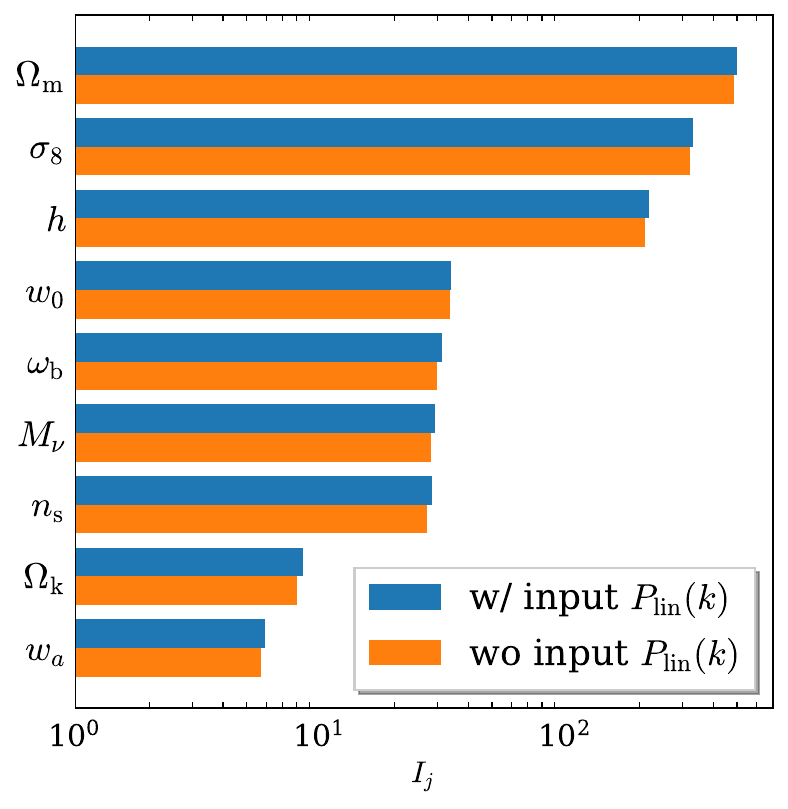}
  \caption{Permutation feature importance (PFI) for the nine cosmological input parameters, evaluated with and without the sampled linear power spectrum included in the inputs. The parameters are ordered from most important (top) to least important (bottom). The importance metric is defined in equation~\ref{eq:pfi_1d}.
  {Alt text: The longest bars correspond to matter density, sigma eight, and the Hubble parameter. Curvature and dark energy evolution have much smaller importance values.}
  }
  \label{fig:pfi_1d}
\end{figure}

Figure~\ref{fig:pfi_1d} shows the permutation importance of the nine cosmological parameters, evaluated both without and with the sampled linear power spectrum included in the inputs.
We estimate the importance using the matter power spectrum in the range $0.001 < k < 10 \, \hMpci$ at $z=0$, using 50 independent random permutations of feature $j$ across the test set, and computing $\overline{\mathrm{MAPE}}_{\mathrm{PFI},j}$ as the mean of the resulting per-cosmology MAPE values over the 30 test cosmologies and the 50 permutations. In this metric, among the cosmological parameters, $\Om$, $\sigma_{8}$, and $h$ are the most important features.
These parameters mainly control the overall growth and amplitude of the power spectrum and thus affect $P(k)$ over the full $k$ range, so permuting them leads to a large degradation in emulator accuracy. 
The parameters $w_{0}$, $\ob$, $M_{\nu}$, and $n_{s}$ also have non-negligible importance, as they modify the spectral tilt or induce localized scale-dependent features around characteristic scales. 
We find that adding the linear power spectrum to the inputs has little impact on the relative importance ranking of the cosmological parameters.

\begin{figure}[htpb]
  \centering
  \includegraphics[width=0.7\linewidth]{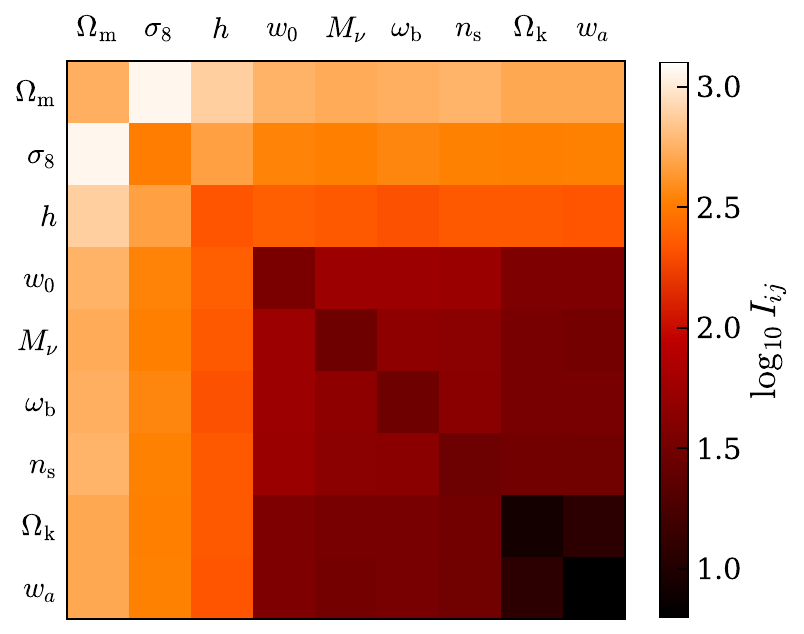}
  \caption{Grouped permutation feature importance for all pairs of input cosmological parameters. The grouped importance is defined in equation~\ref{eq:pfi_2d}, and the plotted color scale shows $\log_{10}I_{ij}$. 
  {Alt text: Cells with large importance cluster around pairs containing the individually important parameters. No pair made only of weak individual parameters becomes dominant.}
  }
  \label{fig:pfi_2d}
\end{figure}

The emulator can also exploit interactions among multiple features. Rather than exhaustively exploring all combinations, we compute a grouped PFI for pairs of features by permuting two features simultaneously. 
The corresponding grouped importance is defined as
\be 
I_{ij} \equiv 
\frac{\overline{\mathrm{MAPE}}_{\mathrm{PFI},ij}}{\overline{\mathrm{MAPE}}_{\mathrm{truth}}}
. \label{eq:pfi_2d} 
\ee 
For visualization, figure~\ref{fig:pfi_2d} shows $\log_{10} I_{ij}$.
Overall, the grouped importance is largely associated with parameters that already have large one-dimensional PFIs, and we do not find a pair of individually unimportant features that becomes highly important only when combined. The largest grouped importance is found for the $\Om$--$\sigma_8$ pair, whose impact exceeds that of either parameter considered individually. This result indicates that the emulator is particularly sensitive to the joint variation of $\Om$ and $\sigma_8$, a dependence often summarized in weak-lensing analyses by the amplitude combination $S_8=\sigma_8\sqrt{\Om/0.3}$. Because $\Om$ and $\sigma_8$ also affect the shape and nonlinear evolution of $P(k)$ differently, this should not be interpreted as sensitivity to $S_8$ alone.

We emphasize that PFI characterizes the sensitivity of the neural network emulator to its inputs and should not be interpreted as a statement about causal relationships among the cosmological parameters.

\subsection{Partial dependence}

Partial dependence (PD; \cite{Hastie2009-PD}) is a standard diagnostic for interpreting machine learning models by varying a selected feature while averaging over the empirical distribution of the remaining features. Here we use this diagnostic to examine how the emulator error changes when one cosmological parameter is varied while the remaining parameters are kept at the test set values. For each grid value of a selected feature, we replace that feature by the chosen value for all test cosmologies, keep the remaining features fixed, and recompute the emulator prediction. When the sampled linear power spectrum is included in the inputs, we also recompute $\mathbf{p}_{\mathrm{lin}}$ from the modified cosmological parameter vector so that the input remains internally consistent.

Let $\boldsymbol{x}_{n}$ denote the input vector of the $n$-th test cosmology, and let $x_j$ be the selected feature. For a grid value $a$ of this feature, we construct a modified input vector $\boldsymbol{x}_{n}^{(j \leftarrow a)}$ by replacing $x_{n,j}$ with $a$ while keeping the remaining features fixed.
For compactness, we denote $P_{\mathrm{DE2},ni}^{(j,a)}
\equiv P_{\mathrm{DE2}}(k_i,z;\boldsymbol{x}_{n}^{(j \leftarrow a)})$ and $P_{\mathrm{sim},ni} \equiv P_{\mathrm{sim}}(k_i,z;\boldsymbol{x}_{n})$.
Since the modified inputs generally do not correspond to simulated test cosmologies, this diagnostic should not be interpreted as the emulator error at the modified parameter value. Instead, we first define the relative deviation from the original simulation spectrum as
\be
\Delta_{ni}^{(j,a)}
&\equiv&
\frac{P_{\mathrm{DE2},ni}^{(j,a)}-P_{\mathrm{sim},ni}}
     {P_{\mathrm{sim},ni}},
\nonumber \\
\mathrm{MAPE}_{\mathrm{PD},n}(a;j)
&=&
\frac{100}{N_k}
\sum_{i=1}^{N_k}
\left|\Delta_{ni}^{(j,a)}\right|
\quad [\%].
\label{eq:mape_pd}
\ee

\begin{figure*}[htbp]
  \centering
  \includegraphics[width=0.8\linewidth]{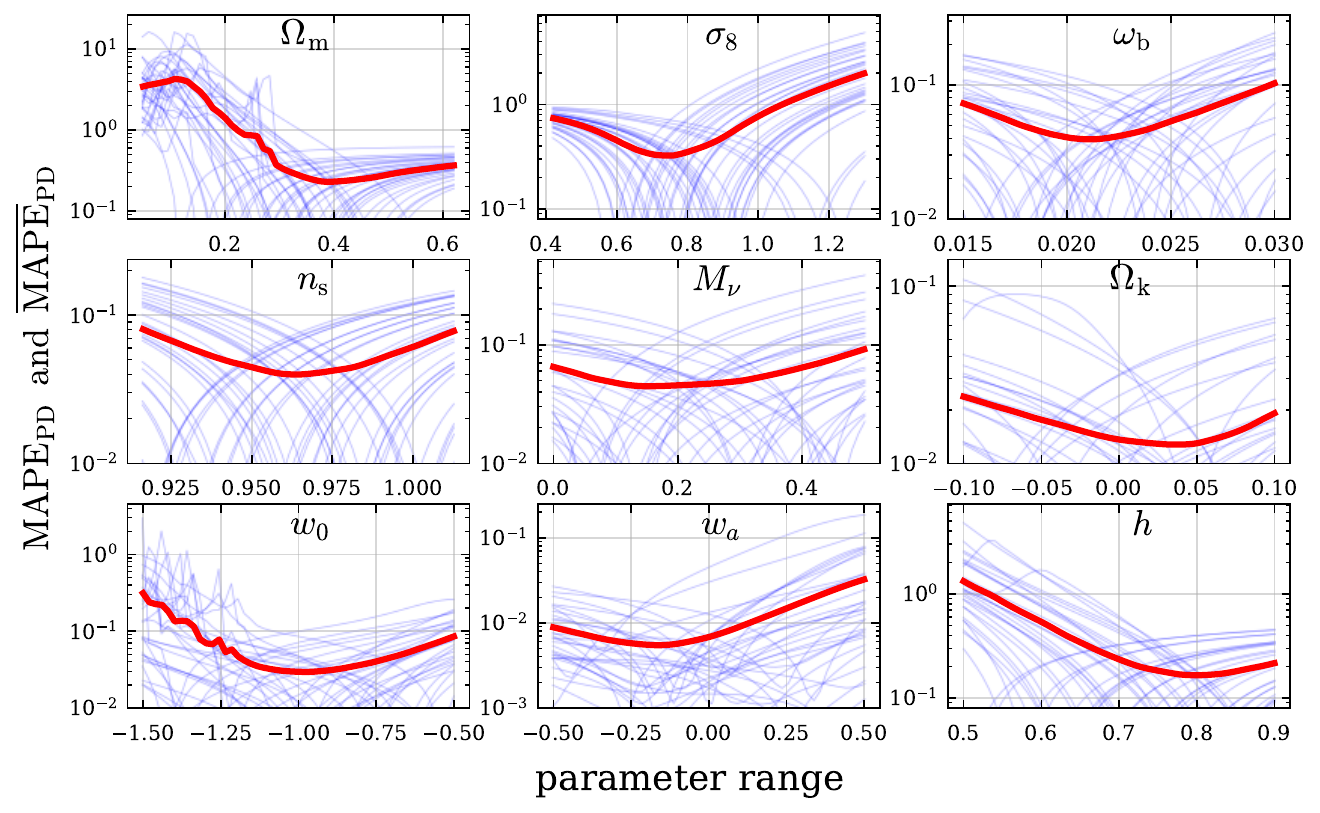}
  \caption{One-dimensional partial dependence plots for each input parameter at $z=0$. The blue lines in each panel show $\mathrm{MAPE}_{\mathrm{PD},n}(a;j)$ for individual instances (test cosmologies), and the red line shows the average $\overline
  {\mathrm{MAPE}}_{\mathrm{PD}}(a;j)$ over the 30 test cosmologies. The dips in the log-scale MAPE for each blue curve occur near the original feature values of the corresponding test cosmology. 
  {Alt text: Average error curves are generally lower near the interior of the sampled ranges and rise toward several parameter boundaries, especially for matter density and sigma eight.}
  }
  \label{fig:pd_1d}
\end{figure*}

Figure~\ref{fig:pd_1d} shows the one-dimensional PD-style diagnostics at $z=0$. The vertical axis measures the relative deviation induced by varying the selected feature, rather than the emulator error at the modified parameter value. The overall magnitude of $\mathrm{MAPE}_{\mathrm{PD}}$ in each panel is broadly consistent with the PFI results for this emulator.

For each individual test cosmology, the blue curve typically reaches its minimum near the original value of the selected feature, as expected from the definition above. The averaged curves show that, for most features, the induced deviation is smaller in the interior of the sampled range and increases toward the boundaries. For $\ob$, $n_s$, $M_{\nu}$, $\Ok$, and $w_a$, the dependence on the selected feature is relatively weak. By contrast, smaller values of $\Om$ and larger values of $\sigma_8$ induce larger deviations. This trend reflects the design of the training set: for these two parameters, the sampling density is higher near the fiducial model, and many extreme combinations lie outside the high-density, banana-shaped region sampled by the training cosmologies.

The increase in $\mathrm{MAPE}_{\mathrm{PD}}$ at small $w_0$ is partly attributable to inaccuracies in the linear power spectrum emulator for models with clustering dark energy (see appendix~\ref{append:lin_pk_emu}). The increase at small $h$ is not directly predictable from the design and may arise from couplings with other input features.

While PD summarizes the average trend as a function of each feature, the behavior is still instance-specific when we focus on individual test cosmologies (the blue $\mathrm{MAPE}_{\mathrm{PD}}$ curves in figure~\ref{fig:pd_1d}).
In principle, PD can also be extended to investigate the effects of combinations of features, analogously to the grouped PFI, but we leave such multi-dimensional PD analyses for future work. 
As with PFI, PD should be interpreted as describing the sensitivity of the emulator to its inputs, rather than as evidence for causal relationships among the cosmological parameters.

\section{Behavior outside the calibrated domain}
\label{append:extrapolation}

This appendix documents the behavior of \DE\ outside the calibrated ranges in effective resolution and redshift. These tests are not intended to extend the validated domain of the emulator, but to clarify how the prediction behaves when the model is queried beyond the supported range.

\begin{figure}[htbp]
  \centering
  \includegraphics[width=\linewidth]{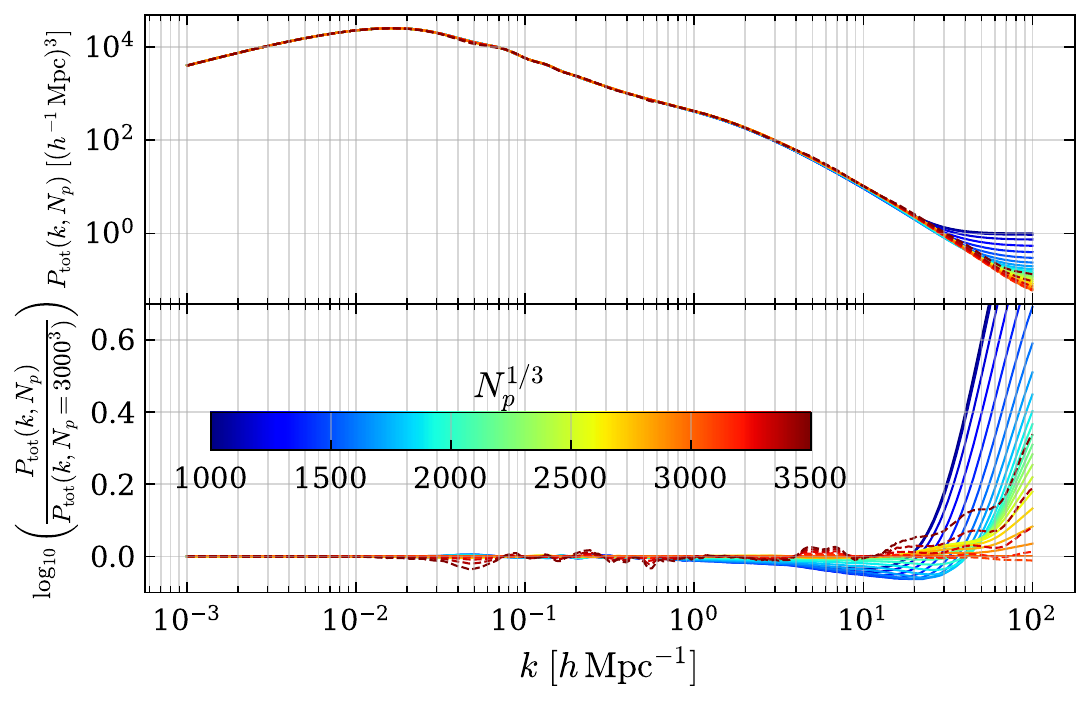}
  \caption{Resolution extrapolation of \DE at the DQ2 fiducial cosmology. Curves
  are shown for $N_{\mathrm{p}}^{1/3}=1500$--$3500$ in steps of 100. Solid
  segments indicate the supported range, and dotted segments show extrapolation.
  The bottom panel shows ratios relative to the $N_{\mathrm{p}}^{1/3}=3000$
  prediction.
  {Alt text: Predictions remain smooth near the supported resolution but develop oscillatory artifacts at high wavenumber when the effective particle resolution is extrapolated too far beyond the training range.}
  }
  \label{fig:app_emu_reso}
\end{figure}

Figure~\ref{fig:app_emu_reso} shows how the predicted spectrum changes as we vary
the effective resolution parameter at the DQ2 fiducial cosmology. The
predictions remain stable up to $N_{\mathrm{p}}^{1/3}=3100$, whereas for $N_{\mathrm{p}}
^{1/3}\ge 3200$ the spectrum develops unphysical features at high $k$. We also
observe oscillatory artifacts at specific wavenumbers in the intermediate-$k$
regime, as well as at high $k$. We therefore do not recommend using \DE beyond
the supported resolution. The mixed-resolution training strategy is nevertheless compatible with future updates that incorporate additional higher-resolution simulations.

\begin{figure}[htbp]
  \centering
  \includegraphics[width=\linewidth]{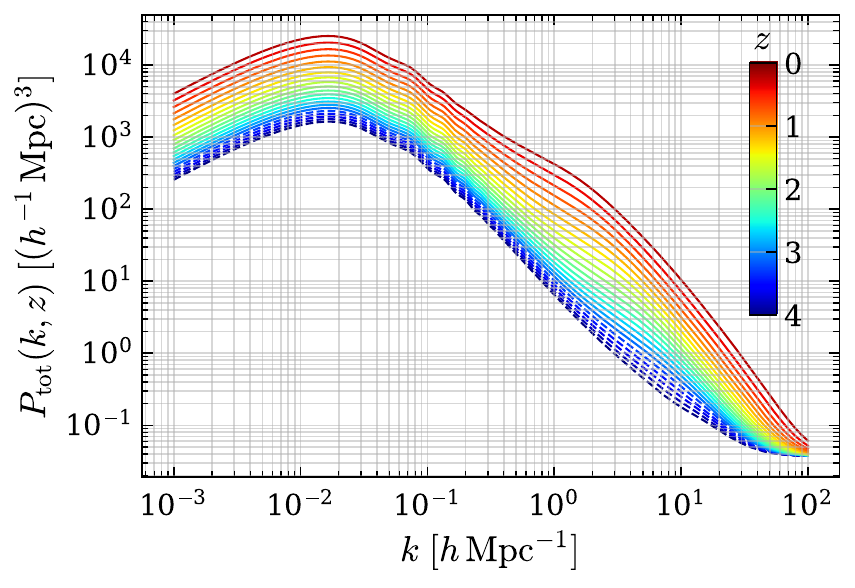}
  \caption{Redshift extrapolation of \DE at the DQ2 fiducial cosmology. Curves
  are shown for $z=0$--4 in steps of 0.2. Solid segments indicate the redshift
  range supported by the nonlinear emulator, and dotted segments show
  extrapolation.
  {Alt text: The spectra evolve smoothly across the trained redshift range and remain qualitatively smooth under mild extrapolation, while the tail at high wavenumber approaches a floor similar to shot noise at higher redshift.}
  }
  \label{fig:app_emu_z}
\end{figure}

Figure~\ref{fig:app_emu_z} shows the redshift dependence of the \DE prediction
at the DQ2 fiducial cosmology, including mild extrapolation beyond the training
range. \DE is trained on 11 snapshots spanning $z=0$--3 for the nonlinear power
spectrum. The spectra remain smooth up to $z\simeq 4$, suggesting that the qualitative
approach toward the particle shot-noise floor is at least captured. The
emulator can be used beyond the supported redshift range to some extent, but it
should not be used for precision analyses in this regime.

We provide an optional high-$k$ extension with shot-noise plateau suppression because
integrals for the correlation function and the convergence power spectrum
require $P(k)$ beyond $k\simeq 100\,\hMpci$. Even when the emulator is queried
at high effective resolution, the spectrum can approach a near-constant shot-noise
floor around $k\simeq 100\,\hMpci$, depending on the input parameters. As
shown in figure~\ref{fig:app_emu_z}, this shot-noise floor becomes prominent at
high redshift and is also visible, to a lesser extent, at low redshift. A flat
high-$k$ tail leads to an unphysical contribution to $\xi(r)$ and $C(\ell)$
and related integrals. We therefore impose a decaying high-$k$ tail in the extrapolated
regime.

\begin{figure}[htbp]
  \centering
  \includegraphics[width=\linewidth]{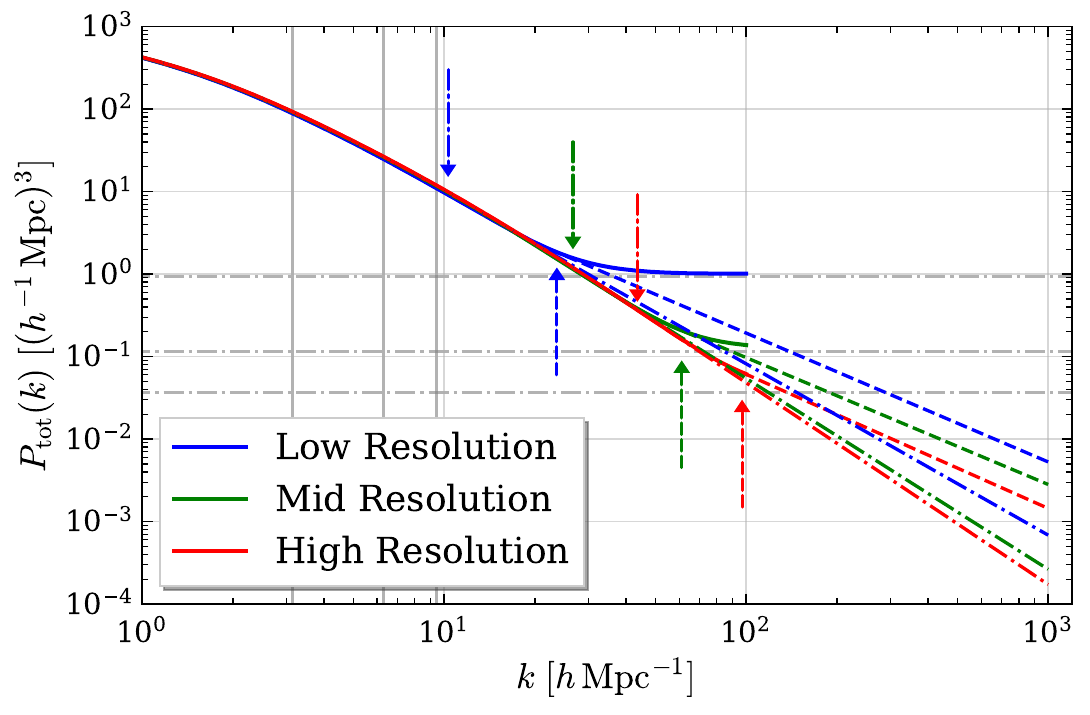}
  \caption{Shot-noise plateau suppression and high-$k$ extension at the DQ2 fiducial
  cosmology. Solid curves show the emulator output for $k\le 100\,\hMpci$.
  Dashed curves show slope transition determined by the slope method. Dash-dotted
  curves show transition at $P(k)=10\,P_{\mathrm{sn}}$. Vertical lines mark
  the Nyquist wavenumbers for the low-, mid-, and high-resolution settings, and
  horizontal lines mark the corresponding shot-noise levels. Arrows indicate
  the transition points used by each method. 
  {Alt text: The original spectra flatten toward shot-noise plateaus at high wavenumber, while both extension prescriptions replace the plateau with declining tails beyond their transition points.}
  }
  \label{fig:app_reduce_shotnoise}
\end{figure}

For $k>100\,\hMpci$, \DE provides two heuristic options. The first is a slope method
that detects the onset of a shot-noise plateau in log--log space. The second is
a threshold method that transitions once the spectrum reaches a fixed multiple
of the shot-noise level in equation~\ref{eq:pksn}. figure~\ref{fig:app_reduce_shotnoise}
compares these two prescriptions for suppressing the shot-noise plateau and
extending the spectrum. In the slope method, we switch to a linear
extrapolation in log-log space at the first point where the local slope
reaches $-1.5$, and then extend the spectrum to higher $k$. In the threshold
method, we transition at the wavenumber where $P(k)=10\,P_{\mathrm{sn}}$. Both
methods yield a monotonically decreasing high-$k$ tail. The slope method retains
more of the original emulator output, whereas the threshold method provides a
more stable transition point. Because the slope threshold requires tuning and the
current \DE accuracy at very high $k$ is limited, we adopt the
$10\,P_{\mathrm{sn}}$ threshold in the $C(\ell)$ calculations below.

In the convergence power spectrum calculations in section~\ref{subsec:cl}, we apply the same high-$k$ extension to \DE\ and to other public emulators. If an emulator does not reach the $10\,P_{\mathrm{sn}}$ level within its valid range, we extrapolate linearly in log--log space from a point near its maximum wavenumber $k_{\max}$.

\section{Hyperparameter optimization with CMA-ES}
\label{append:cmaes}

\begin{figure*}[htbp]
  \centering
  \includegraphics[width=0.7\linewidth]{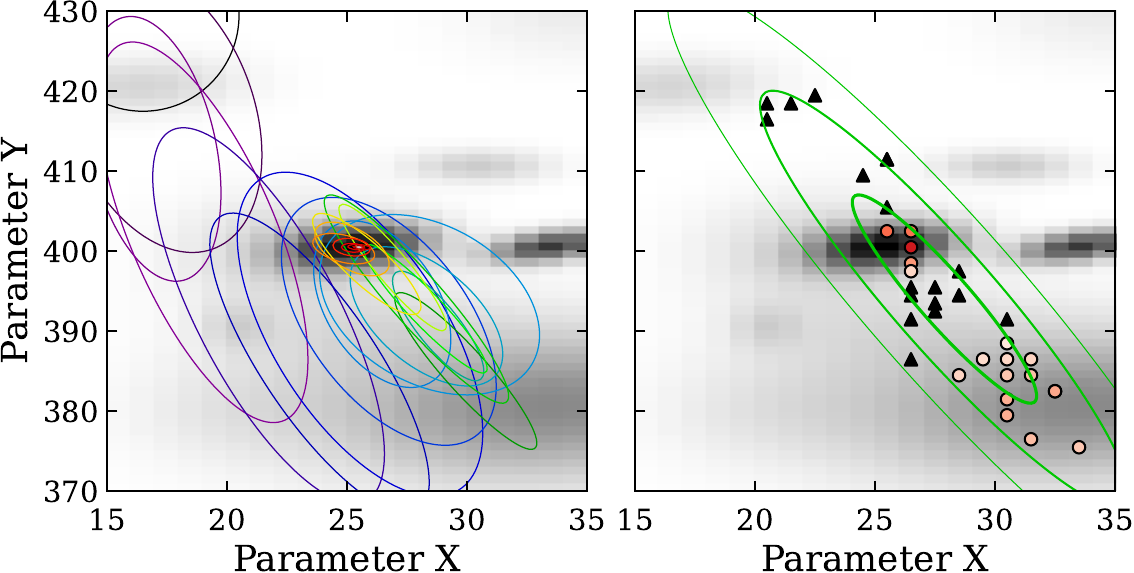}
  \caption{ Demonstration of CMA-ES with $\sigma=0.4$, $\lambda=36$, and $\mu=1
  8$. The left panel shows the $1\sigma$ contour of the multivariate normal
  sampling distribution over successive generations. Contours are colored by generation,
  and the plot shows up to 32 generations. The right panel shows the $1\sigma$,
  $2\sigma$, and $3\sigma$ contours together with the sampling points from one
  representative generation. Red points receive positive recombination weights
  and are used to update the next sampling distribution. White points receive zero
  weight because of their high objective values. 
  {Alt text: The sampling distribution contracts and changes orientation over generations, while selected candidates with low objective values determine the next update of the distribution.}
  }
  \label{fig:cmaes}
\end{figure*}

\begin{table*}[htbp]
  \caption{Hyperparameter search ranges explored in the network optimization
  survey.}
  \label{table:survey_params}
  \centering
  \scalebox{0.95}{
  \begin{tabular}{ll}
    \hline
    \hline
    Hyperparameter                                  & Values or range tested                                               \\
    \hline
    Number of hidden layers ($N_{\mathrm{hidden}}$) & $\{1, 2, 3, \dots, 10, 11\}$                                         \\
    Neurons per layer ($N_{\mathrm{neuron}}$)       & $\{400, 600, 800, \dots, 2800, 3000\}$                               \\
    Training epochs ($N_{\mathrm{epoch}}$)          & $\{1000, 1500, 2000, \dots, 9000, 10000\}$                           \\
    Batch size ($N_{\mathrm{batch}}$)               & $\{50, 100, 150, \dots , 350, 400\}$                                 \\
    Activation function                             & RReLU, LeakyReLU, GELU, SELU, ELU, Softsign                          \\
    Optimizer                                       & SGD, ASGD, LAMB, ADAM, ADAMAX, ADAMW                                 \\
    Learning-rate scheduler                         & LinearLR, StepLR, ExponentialLR, ReduceLROnPlateau,                  \\
                                                    & OneCycleLR, CyclicLR, CosineAnnealingLR, CosineAnnealingWarmRestarts \\
    Loss function                                   & RMSE (validation)                                                    \\
    \hline
    \hline
  \end{tabular}
  }
\end{table*}

Training the neural network requires selecting hyperparameters such as the number of hidden layers, the number of neurons in each hidden layer, the batch size, the number of epochs, the activation function, the optimizer, the learning rate, and the learning-rate schedule.
With a fixed training data set, poor hyperparameter choices can substantially degrade emulator accuracy. 
Because our search space mixes discrete and continuous variables, the validation loss is a noisy and non-convex black-box function of these choices. We therefore treat hyperparameter selection as a global black-box optimization problem rather than a gradient-based one~\citep{Rios2013-wb}.

Classical grid search scales poorly with dimensionality and can waste evaluations on uninformative directions. Pure random search is often more efficient than grid search, but it does not adaptively concentrate evaluations around promising regions~\citep{Bergstra2011-qj,Bergstra2012-ey,Bergstra2013-mj}. 
We therefore employ the Covariance Matrix Adaptation Evolution Strategy (CMA-ES)~\citep{Hansen1996-me,Hansen2003-ev,Hansen2016-jk}, using the mixed discrete and continuous extension with box constraints described by \citet{Hamano2022-fr}. CMA-ES is invariant under rank-preserving transformations of the objective and is well suited to multimodal and noisy response surfaces~\citep{Hansen2003-ev,Hansen2016-jk}. 
At each generation, CMA-ES proposes a population of $\lambda$ candidate configurations, so the $\lambda$ objective evaluations are embarrassingly parallel. 
We also considered derivative-free local search methods such as Nelder-Mead~\citep{Nelder1965-xy} and Gaussian-process-based Bayesian optimization~\citep{Rasmussen2005-yd}. 
In our setting, however, the mixed discrete and continuous space, strong parameter interactions, stochastic noise, and the need for batch-parallel evaluations make CMA-ES a more practical choice.

In the present application, the CMA-ES update cycle is summarized as follows.
\begin{enumerate}
  \item Draw $\lambda$ candidate points from a multivariate normal distribution, $\mathcal{N}(m_{t},\sigma_{t}^{2}C_{t})$.

  \item Evaluate the objective $f(x)$ (the validation RMSE in our case) for each candidate and select the best $\mu$ points, typically $\mu \approx \lambda/2$.

  \item Update the mean $m_{t+1}$ by weighted recombination of the selected points. Update the global step size $\sigma_{t+1}$ using cumulative step-size adaptation (CSA), and update the covariance $C_{t+1}$ with rank-one and rank-$\mu$ updates~\citep{Hansen2003-ev,Hansen2016-jk}.

  \item Repeat steps 1--3 until a stopping criterion is met, for example when $\sigma_{t}$ becomes small, when the objective stops improving for several generations, or when the evaluation budget is exhausted.
\end{enumerate}
Mixed discrete and continuous hyperparameters and box constraints are handled by projection and repair at evaluation time; for categorical hyperparameters encoded as integer variables, we additionally apply small randomized perturbations before projection/repair to promote exploration in the discrete subspace~\citep{Hamano2022-fr}.
See \citet{Hansen2016-jk} for details of CMA and the covariance update rules.

Figure~\ref{fig:cmaes} provides a schematic illustration of CMA-ES on an idealized black-box objective. 
The sampling distribution rapidly contracts toward a high-performance region while its covariance adapts to the local anisotropy of the objective. 
The sequence of iso-density ellipses in the left panel shows that CMA-ES explores broadly at early generations and then exploits by aligning and shrinking the covariance along the valley toward the optimum. 
In the right panel, the selected candidates (red points) concentrate near the current mean and receive positive recombination weights.
These selected points determine the subsequent update of $(m_{t},\sigma_{t},C_{t})$.

In the production search, we used $(\lambda,\mu)=(48,24)$ and $\sigma_{0}=0.4$, and ran CMA-ES for 25 generations, for a total of 1,200 objective evaluations. 
The exploration ranges for all hyperparameters are listed in table~\ref{table:survey_params}.
We ran CMA-ES for a fixed evaluation budget rather than until a strict convergence criterion was met. 
Although the search did not formally converge, it efficiently concentrated evaluations in a high-performing region of the hyperparameter space. 
Because the validation RMSE is noisy under stochastic training, several configurations can become effectively indistinguishable near the optimum. 
We therefore re-evaluated a small set of top candidates and performed a targeted local sweep before selecting the final configuration reported in table~\ref{table:net_params}.

\bibliographystyle{apj}
\bibliography{lssref}

\end{document}